\documentclass[12pt]{article}
\usepackage{authblk}

\usepackage[square,numbers]{natbib}
\usepackage{caption}
\usepackage{booktabs}
\usepackage[english]{babel}
\usepackage[utf8]{inputenc}
\usepackage{fancyhdr}
\usepackage{amsmath,amssymb}
\usepackage{parskip}
\usepackage{graphicx}
\usepackage{hyperref}
\usepackage{mathtools}
\usepackage{tikz}
\usepackage{wrapfig}
\usepackage{float}
\usepackage{array}

\usepackage{rotating} 

\usepackage[top=1in, left=1in, right=1in, bottom=1in]{geometry}
\usepackage{amssymb,amsmath}%

\makeatletter
\def\ps@pprintTitle{%
 \let\@oddhead\@empty
 \let\@evenhead\@empty
 \def\@oddfoot{\hfil\thepage\hfil}%
 \let\@evenfoot\@oddfoot}
\makeatother

\begin{document}

\title{\textbf{Multiple Scattering of Elastic Waves in Polycrystals}}

\author[1]{{\small Anubhav Roy}\thanks{Corresponding author: \texttt{aero.aroy@stanford.edu}}}
\author[2]{\small Christopher M. Kube}

\affil[1]{\small Department of Aeronautics and Astronautics, Stanford University\\ \newline \small Durand Building, 496 Lomita Mall, Stanford, CA 94305, USA}
\affil[2]{\small Department of Engineering Science and Mechanics, The Pennsylvania State University\\ \newline Earth and Engineering Sciences Building, University Park, PA 16802, USA}

\date{}
\maketitle

\begin{abstract}
Elastic waves that propagate in polycrystalline materials attenuate due to scattering of energy out of the primary propagation direction in addition to becoming dispersive in their group and phase velocities. Attenuation and dispersion are modeled through multiple scattering theory to describe the mean displacement field or the mean elastodynamic Green's function. The Green's function is governed by the Dyson equation and was solved previously \cite{weaver1990diffusivity} by truncating the multiple scattering series at first-order, which is known as the first-order smoothing approximation (FOSA). FOSA allows for multiple scattering but places a restriction on the scattering events such that a scatterer can only be visited once during a particular multiple scattering process. In other words, recurrent scattering between two scatterers is not permitted. In this article, the Dyson equation is solved using the second-order smoothing approximation (SOSA). The SOSA permits scatterers to be visited twice during the multiple scattering process and, thus, provides a more complete picture of the multiple scattering effects on elastic waves. The derivation is valid at all frequencies spanning the Rayleigh, stochastic, and geometric scattering regimes without additional approximations that limit applicability in strongly scattering cases (like the Born approximation). The importance of SOSA is exemplified through analyzing specific weak and strongly scattering polycrystals. Multiple scattering effects contained in SOSA are shown to be important at the beginning of the stochastic scattering regime and are particularly important for transverse (shear) waves. This step forward opens the door for a deeper fundamental understanding of multiple scattering phenomena in polycrystalline materials. 
\end{abstract}


\begin{figure}
    \centering
    \includegraphics[width=1\linewidth]{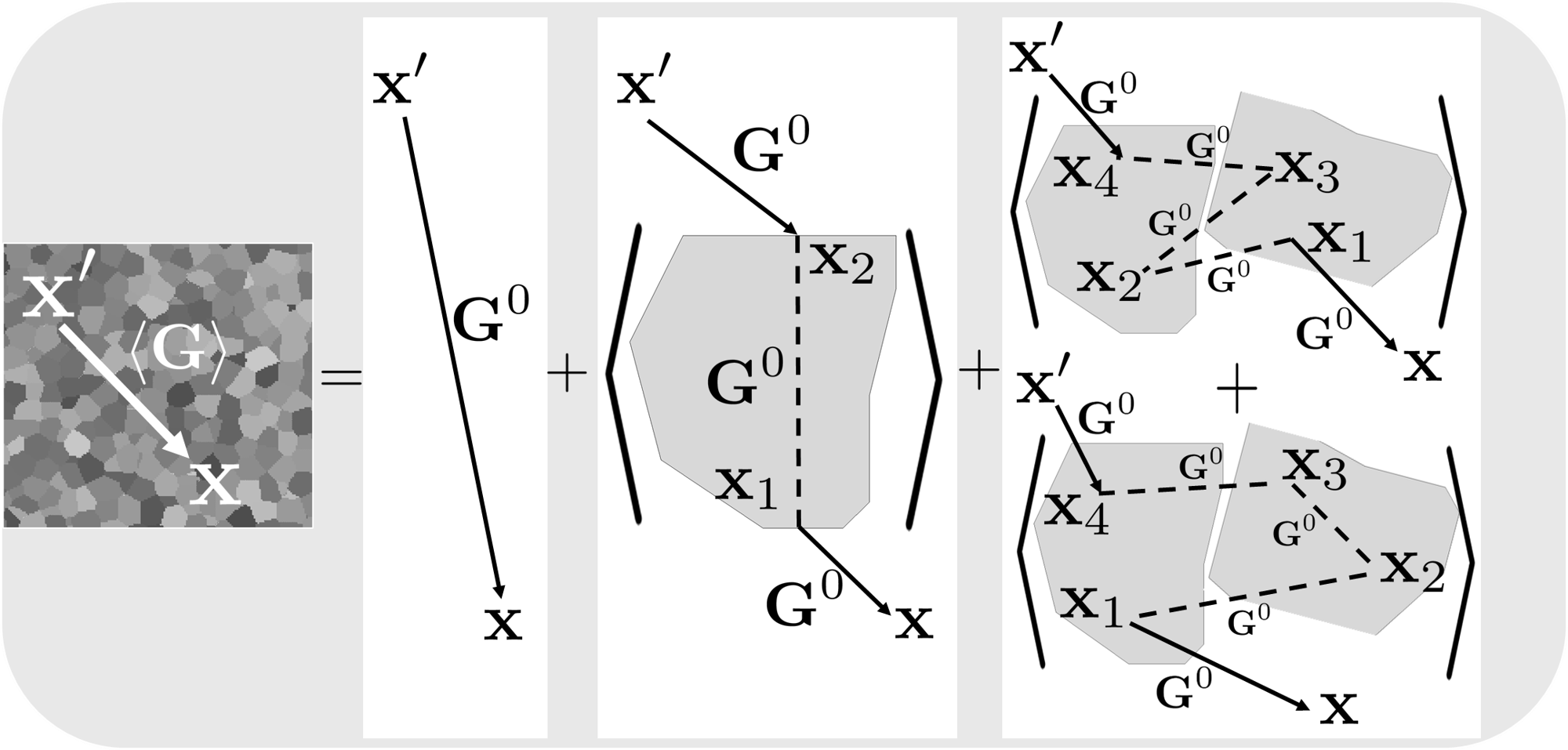}
    \captionsetup{labelformat=empty} 
    \caption*{\textbf{Graphical Abstract:} Pictorial representation for the mean Green's function with first- and second-order scattering effects, which includes recurrent scattering from grains.}
\end{figure}


\begin{center}
\end{center}


\section{Introduction}
\label{intro}
\textit{Authors' note: The first part of the introduction discusses the foundational elastostatic homogenization theories for polycrystalline materials that helped motivate the authors and expanded their foundation toward elastodynamic scattering. They are close analogs to the elastodynamic multiple scattering theory to follow. Readers mostly interested in elastodynamics may skip the first paragraph without loss of generality.} 

\par Polycrystalline metals are completely composed of grains that can spatially vary. Each grain is assumed crystalline and, thus, has a crystal symmetry and associated crystallographic axes that can be differently oriented from other grains. The polycrystal has no matrix phase as any particular grain is surrounded by other grains. In this work, statistical isotropy refers to the situation in which all volume fractions of crystallographic orientations are equally probable in the polycrystal. Homogenization theories seek to leverage the statistical character of polycrystals for the prediction of properties. Statistical isotropy is a simplifying assumption that often forms the starting point of homogenization theories. Elastostatic homogenization has a rich history spanning several decades. In the review paper, Hashin \cite{hashin1983analysis} subdivided the homogenization theories of composites into broad categories: the direct approach, variational bounding, and approximation methods, including self-consistent schemes. Voigt \cite{voigt1910lehrbuch} and Reuss \cite{reuss1929berechnung} provided early homogenization approaches that could be applied to bound the effective properties of polycrystals, assuming uniform strain and uniform stress, respectively. The uniform stress and strain assumptions permit volume-averaged quantities to be established. For polycrystals, in particular, ergodicity assumptions are applied, allowing volume averaging to be approximated by ensemble and orientation averages. The Voigt average, in particular, is important in the present communication as the orientation average of stiffness, denoted as $\langle\textbf{C}\rangle$, is used to represent a non-scattering statistically homogeneous reference medium with stiffness $\textbf{C}^0=\langle\textbf{C}\rangle$. Hill \cite{hill1952elastic} presented the classical extremum principles on elasticity, the bounds of which reduce to the Voigt \cite{voigt1910lehrbuch} and Reuss \cite{reuss1929berechnung} estimates. Later, a more generalized relationship between the effective elastic moduli and the elastostatic strain energy was established by Hill \cite{hill1963elastic} for statistically homogeneous and isotropic materials.  Kr{\"o}ner \cite{kroner1958berechnung,kroner1967elastic} developed an approximation-based approach \cite{hashin1983analysis} involving self-consistent schemes and utilized Green's functions for the statistical homogenization of polycrystals.  The foundational work by Hashin and Shtrikman \cite{hashin1962variational, hashin1963variational} uses a variational approach to provide bounds on elastic moduli for polycrystals \cite{hashin1962variational} and multi-phase materials such as composites \cite{hashin1963variational}. Walpole \cite{walpole1966bounds, walpole1966boundsII} utilized Green's functions to extend the Hashin and Shtrikman's variational approach \cite{hashin1963variational} and provide effective property estimates for composites with inclusions and arbitrary anisotropy. The self-consistent scheme by Budiansky \cite{budiansky1965elastic} utilizes Eshelby's analytical solutions \cite{eshelby1957determination} to provide effective property estimates for isotropic composites with spherical inclusions and was later extended by Willis \cite{willis1977bounds} for anisotropic composites. Hill \cite{hill1965self} employed a self-consistent scheme to incorporate plasticity into the model, inspiring the later-developed crystal plasticity theories \cite{lubarda2001elastoplasticity,lebensohn1993self}. Kr{\"o}ner \cite{kroner1977bounds} combined a variational approach with the previously established self-consistent scheme \cite{kroner1958berechnung}. Furthermore, in \cite{kroner1967elastic,kroner1977bounds}, Kr{\"o}ner derived predictions of effective elastic stiffness in the form of a Lippmann-Schwinger equation with successive terms involving the $N$-point spatial correlation functions of the polycrystal. The Lippmann-Schwinger equation seen in \cite{kroner1967elastic,kroner1977bounds}, and derived for elastostatics, is the closest to our present treatment for elastodynamics involving the Dyson equation. Additionally,  Kr{\"o}ner proved \cite{kroner1977bounds} that truncation of the Lippmann-Schwinger equation returned $N$-th order bounds on the effective stiffness. For example, truncating at terms involving two-point correlation functions provides the Hashin-Shtrikman bounds \cite{hashin1962variational, hashin1963variational}. Kube and co-workers \cite{kube2016bounds,kube2016elasticM} exemplified \cite{kroner1967elastic,kroner1977bounds} and provided the ability to calculate bounds of any order through an iterative scheme with increasing order converging to self-consistent estimates. 
\par Instead of using volume-averaged stress or strain to formulate effective stiffness properties, elastodynamic homogenization and scattering theories usually seek to predict the volume-averaged displacement fields that account for scattering in the polycrystal. Most scattering models assume each grain is homogeneous on its interior and share boundaries with neighboring grains. Thus, spatial heterogeneity is the stiffness and/or density variation from grain to grain, yet it is uniform within each grain. Physically, discontinuities such as dislocations and grain boundary decorations like precipitates can scatter elastic waves. However, in many settings, such as ultrasonic nondestructive evaluation, grain boundary scattering dominates measurements. The grain boundary scattering leads to attenuation of the propagating elastic wave in which its amplitude decays exponentially. Additionally, the phase and group velocity of the wave become dispersive as these quantities vary from grain to grain, leading to a non-uniform phase across wave fronts. Thus, velocity dispersion and attenuation are the primary quantities of interest when modeling elastic waves in polycrystals. Depending on the specific treatment, the elastodynamic scattering theories recover the effective static properties in the long wavelength limit (see Stanke and Kino \cite{stanke1984unified} comparison to Hashin-Shtrikman estimates in their Table II). 
 \par Pekeris \cite{pekeris1947note} presented an early model for the radiative scattering of sound waves in media with spatial wavespeed heterogeneity. The spatially-dependent average of the variation in wavespeed was analytically performed through introducing a now often used exponentially decaying two-point spatial correlation function \cite{pekeris1947note}. Mason and McSkimin \cite{mason1947attenuation, mason1948energy} formulated scattering-based attenuation formulae for elastic waves by considering spatial heterogeneity in the anisotropic elastic stiffness constants and mass density. Formulae for both longitudinal and shear wave attenuations were established for low (Rayleigh scattering) and high frequency (geometric scattering) based on scattering cross-section forms given by Rayleigh \cite{rayleigh1896theory}. Huntington \cite{huntington1950ultrasonic} derived formulae for the intermediate or stochastic scattering regime. 
\par As the three scattering regimes are important in later theories in addition to the current article, we make a brief note on their terminology and definition  (see also Section \ref{resultsSec}). The three scattering regimes include the Rayleigh, intermediate (stochastic), and geometric regimes, where the Rayleigh regime has a fourth-order frequency dependence and is appropriate when the wavelength is around two orders of magnitude greater than the mean scatterer size. The intermediate (stochastic) has a second-order frequency dependence when the wavelength is around an order of magnitude greater than the mean scatterer size. While the geometric regime exhibits an attenuation that is independent of frequency and occurs when the wavelength is smaller than the mean scatterer size. 
\par It is noted here that the strict frequency independence in the geometric regime is a result of modeling limitations that do not include multiple scattering and diffraction effects from grain shape. The present article addresses existing modeling limitations related to multiple scattering effects over all frequency ranges and scattering regimes.
\par While Mason and McSkimin \cite{mason1947attenuation, mason1948energy} arrived at attenuation expressions for both longitudinal and shear waves, they did not incorporate mode conversion scattering effects. In this context, mode conversion refers to a longitudinal (or transverse) mode that scatters into a transverse (or longitudinal) mode during the multiple scattering process, which is prominent in polycrystals. Lifshits and Parkhamovskii \cite{lifshits1950theory} incorporated mode conversion and arrived at the same scattering integrals as seen in more recent theories of Stanke and Kino (UT) \cite{stanke1984unified} and Weaver \cite{weaver1990diffusivity}. Similarly, Bhatia and Moore \cite{bhatia1959scattering,bhatia1959scatteringII} investigated the effects of crystallite anisotropy on the mode-converted scattering at the grain boundaries. Papadakis \cite{papadakis1961grain} provided attenuation formulae based on a distribution of grain sizes, while Hirsekorn \cite{hirsekorn1982scattering} modeled the scattering from grains assuming point scatterers. 
\par In the late 1950s and early 1960s, the mathematician Joseph B. Keller contributed to modeling elastic waves in polycrystalline media \cite{karal1959elastic,karal1964elasticTry2}. In particular, Karal and Keller \cite{karal1964elasticTry2} formulated a perturbative expansion incorporating various orders of scattering effects. Keller was at New York University at the time  Refs. \cite{karal1959elastic} and \cite{karal1964elasticTry2} were published. He was later on Fred Stanke's dissertation committee at Stanford University and must have influenced the UT of Stanke and Kino \cite{stanke1984unified}. The UT model \cite{stanke1984unified} incorporated first-order scattering effects on the mean plane wave displacement field $\langle\textbf{u}\rangle$ through the use of Keller's scattering framework \cite{karal1964elasticTry2}. Weaver \cite{weaver1990diffusivity} derived the same scattering integral as Stanke and Kino \cite{stanke1984unified} (see Equation 94 from \cite{stanke1984unified} and Equation 3.15 from \cite{weaver1990diffusivity} and the discussion in \cite{kube2017iterative}). Weaver \cite{weaver1990diffusivity} approached the scattering problem through the use of the Dyson equation borrowed from quantum field theory and adapted from Frisch \cite{frisch1968wave}. The framework in Frisch \cite{frisch1968wave} and other resources for the Dyson equation \cite{rytov1989principles, dal2022waves} is based on scalar theory. The extension to the vector theory needed for elastodynamics \cite{weaver1990diffusivity} was a significant leap forward. All orders of multiple scattering are conveniently organized within the Dyson series using diagrams. Truncating the diagrammatic series at first-order is referred to as the first-order smoothing approximation (FOSA) \cite{weaver1990diffusivity} and is equivalent to the UT. The FOSA is also known as the bilocal approximation in some contexts \cite{tsang2004scattering}. Both the UT \cite{stanke1984unified} and FOSA \cite{weaver1990diffusivity} include multiple scattering effects on the mean displacement field, in which each scatterer can only be visited once. The statement that each scatterer can only be visited once does not stipulate strict single scattering. For example, within FOSA, a wave can scatter from multiple grains in succession, where each grain is visited only once. The condition of lone visits to scatterers leads to FOSA involving only the two-point correlation function to describe the statistical spatial layout of the scatterers. Higher-order effects where scatterers or grains can be visited more than once, which is known as recurrent scattering, are accounted for in diagrams involving \textit{N}-point correlation functions when \textit{N}$>2$. The case of $N=4$ is the second-order smoothing approximation (SOSA) and includes two strongly connected diagrams that represent all possibilities of grains being visited twice. These multiple scattering effects have not been evaluated for elastic waves in polycrystals beyond the low-frequency Rayleigh limit \cite{roy2024improved}, which was only recently considered by the present authors  (the authors refer to SOSA as the third-order smoothing approximation in Ref. \cite{roy2024improved}). Multiple scattering effects within the SOSA of the Dyson equation is the focus of this article.
\par This article addresses multiple scattering effects on the mean displacement field representing elastic waves in polycrystals in which grains can be visited more than once. Our approach is based on the evaluation of higher-order diagrams in the Dyson series \cite{weaver1990diffusivity,frisch1968wave, rytov1989principles}. In Section 2, the Dyson equation governing the mean elastodynamic Green's function is formulated with an emphasis on the higher-order diagrams representing up to four-point correlations. At this order, scattering paths between two grains are accounted for. Extending scattering models from single scattering to multiple scattering is a major challenge to execute analytically. To the best of our knowledge, the present work marks the first instance of analytically going beyond first-order multiple scattering effects on elastic waves in polycrystals.

\par The article is organized as follows. Section \ref{Secstathom} introduces a derivation of the Dyson equation with a clear delineation between the Feynman diagrams that contribute to the appropriate self-energies of FOSA and SOSA. The SOSA would be equivalent to a so-called quadrilocal approximation if one preferred the bilocal approximation terminology. In Section \ref{SecSFTDomain}, we introduce the solution procedure in the wave number domain and derive the dispersion equations governing the squared effective wavenumbers in terms of the scaled self-energies. The derivation in Section \ref{SecSFTDomain} is applicable to either FOSA or SOSA. In Section \ref{FOSA_Sec}, the scaled self-energies for FOSA are derived, which is then followed by Section \ref{SOSA_Sec} where SOSA is considered. At the conclusion of Section \ref{SecSFTDomain}, the scaled self-energies are provided in integral forms. Sections \ref{SecResultsAll} and \ref{SecResultsAllSOSA} are the results sections that describe the specific integration steps for FOSA and SOSA, respectively. In addition, Section \ref{SecResultsAll} gives results for FOSA in all frequency and scattering regimes for the first time. Section \ref{SecResultsAllSOSA} provides the primary results of this article, including quantitative comparisons between SOSA and FOSA predicted attenuation and phase velocity dispersion. Lastly, it is noted that the Dyson equation is commonly used in many applications in physics and scattering theory embodied in a wealth of reference material. For clarity and conciseness, this article includes references to only those that had an influence on the present work. Readers interested in other applications or finer detail of the Dyson equation are recommended to consult \cite{frisch1968wave, rytov1989principles, dal2022waves, stefanucci2013nonequilibrium, sheng2007introduction, sato2012seismic}.

\section{Elastodynamic Dyson equation}

\label{Secstathom}
We consider the propagation of elastic waves in a space-filling grainy medium appropriate for modeling propagation and scattering in polycrystals. The starting point is that of the heterogeneous medium in which we assume the spatial heterogeneity from grain to grain is related to the elastic stiffness $\textbf{C}(\textbf{x})=C_{ijkl}(\textbf{x})$ only. The Green's function $\textbf{G}(\textbf{x},\textbf{x}',t)$ is of primary importance and describes the elastic wave propagation from $\textbf{x}'$ to $\textbf{x}$ due to a space-time impulse at $\textbf{x}'$. The balance of linear momentum dictates the governing equation of motion of the Green's function,
\begin{align}
\label{indGov}
\left[\boldsymbol{\mathcal{L}}^0(\textbf{x})+\boldsymbol{\mathcal{L}}(\textbf{x})\right]\boldsymbol{\cdot}\textbf{G}\left(\textbf{x},\textbf{x}',t\right)=\textbf{I}\delta^3\left(\textbf{x}-\textbf{x}'\right)\delta\left(t\right),
\end{align}
where $\textbf{I}=\delta_{ij}$ is the Kronecker delta tensor, and the two differential operators were introduced,
\begin{align}
\mathcal{L}_{ik}^0\left(\textbf{x}\right)=C_{ijkl}^{0}\nabla_{x_j}\nabla_{x_l}-\rho\delta_{ik}\partial^2/\partial t^2,
\end{align}
\begin{align}
\label{heteroL}
\mathcal{L}_{ik}\left(\textbf{x}\right)=\nabla_{x_j}\left(\gamma_{ijkl}\left(\textbf{x}\right)\nabla_{x_l}\right),
\end{align}
where the gradient $\nabla_{x_i}=\partial/\partial x_i$, $\rho$ is the mass density (assumed constant), and $\textbf{C}^0$ is a reference stiffness tensor in a corresponding homogeneous and non-scattering medium. Within the operator $\boldsymbol{\mathcal{L}}(\textbf{x})$ is the stiffness fluctuation,
\begin{align} 
\label{cHetero}
\boldsymbol{\gamma}(\textbf{x})=\textbf{C}(\textbf{x}) -\textbf{C}^0.
\end{align}  
that provides for the spatial heterogeneity in stiffness. $\boldsymbol{\mathcal{L}}^0$ is the operator that governs the Green's function of a non-scattering reference medium,
\begin{align}
\label{L0G0}
\boldsymbol{\mathcal{L}}^0\cdot\textbf{G}^0\left(\textbf{x},\textbf{x}',t\right)=\textbf{I}\delta^3\left(\textbf{x}-\textbf{x}'\right)\delta\left(t\right).
\end{align}
Combining Equations (\ref{indGov}) and (\ref{L0G0}), leads to the Lippmann-Schwinger equation, 
\begin{align}
    \label{iterm1}
    \textbf{G}(\textbf{x},\textbf{x}')=\textbf{G}^0(\textbf{x},\textbf{x}')-\int \textbf{G}^0(\textbf{x},\textbf{x}_1)\boldsymbol{\cdot}\boldsymbol{\mathcal{L}}\left(\textbf{x}_1\right)\boldsymbol{\cdot}\textbf{G}(\textbf{x}_1,\textbf{x}')d^3\textbf{x}_1,
\end{align}
which can be iterated to form the Born series,
\begin{align}
\label{full^g}
    \textbf{G}(\textbf{x},\textbf{x}')&=\textbf{G}^0(\textbf{x},\textbf{x}')-\int \textbf{G}^0(\textbf{x},\textbf{x}_1)\boldsymbol{\cdot}\boldsymbol{\mathcal{L}}\left(\textbf{x}_1\right)\boldsymbol{\cdot}\textbf{G}^0(\textbf{x}_1,\textbf{x}')d^3\textbf{x}_1\notag\\
    \quad & +\int\int \textbf{G}^0(\textbf{x},\textbf{x}_1)\boldsymbol{\cdot}\boldsymbol{\mathcal{L}}\left(\textbf{x}_1\right)\boldsymbol{\cdot}\textbf{G}^0(\textbf{x}_1,\textbf{x}_2)\boldsymbol{\cdot}\boldsymbol{\mathcal{L}}\left(\textbf{x}_2\right)\boldsymbol{\cdot}\textbf{G}^0(\textbf{x}_2,\textbf{x}')d^3\textbf{x}_1d^3\textbf{x}_2\notag\\
    \quad & -\int\int\int \textbf{G}^0(\textbf{x},\textbf{x}_1)\boldsymbol{\cdot}\boldsymbol{\mathcal{L}}\left(\textbf{x}_1\right)\boldsymbol{\cdot}\textbf{G}^0(\textbf{x}_1,\textbf{x}_2)\boldsymbol{\cdot}\boldsymbol{\mathcal{L}}\left(\textbf{x}_2\right)\boldsymbol{\cdot}\textbf{G}^0(\textbf{x}_2,\textbf{x}_3)\notag\\ \quad & \boldsymbol{\cdot}\boldsymbol{\mathcal{L}}\left(\textbf{x}_3\right)\boldsymbol{\cdot}\textbf{G}^0(\textbf{x}_3,\textbf{x}')d^3\textbf{x}_1d^3\textbf{x}_2d^3\textbf{x}_3\notag\\
    \quad & +\int\int\int\int \textbf{G}^0(\textbf{x},\textbf{x}_1)\boldsymbol{\cdot}\boldsymbol{\mathcal{L}}\left(\textbf{x}_1\right)\boldsymbol{\cdot}\textbf{G}^0(\textbf{x}_1,\textbf{x}_2)\boldsymbol{\cdot}\boldsymbol{\mathcal{L}}\left(\textbf{x}_2\right)\boldsymbol{\cdot}\textbf{G}^0(\textbf{x}_2,\textbf{x}_3)\notag\\
    \quad &  \boldsymbol{\cdot}\boldsymbol{\mathcal{L}}\left(\textbf{x}_3\right)\boldsymbol{\cdot}\textbf{G}^0(\textbf{x}_3,\textbf{x}_4)\boldsymbol{\cdot}\boldsymbol{\mathcal{L}}\left(\textbf{x}_4\right)\boldsymbol{\cdot}\textbf{G}^0(\textbf{x}_4,\textbf{x}')d^3\textbf{x}_1d^3\textbf{x}_2d^3\textbf{x}_3d^3\textbf{x}_4+\cdots,
\end{align}
where $\textbf{x}_1$, $\textbf{x}_2$, $\textbf{x}_3$ and $\textbf{x}_4$ are spatial vectors corresponding to the probable scattering locations. Analytical solutions to Equation (\ref{full^g}) are difficult. However, the Born series provides the needed pathway toward the Dyson equation that governs the mean Green's function, i.e., the averaged displacement field due to an impulse in the polycrystal. It is usually the mean field that is experimentally accessible when considering elastic waves in polycrystals, as laboratory elastic wave sources and receivers are usually much greater than the scatterer (grain) sizes, and the propagating wave propagates through volumes containing a larger number of scatterers (grains). To move toward the Dyson equation, we introduce averaged quantities on the stiffness and its fluctuations. The ensemble average stiffness is denoted as the stiffness of the reference medium,
\begin{align}
\label{gamma_CRV}
\langle \textbf{C} \left(\textbf{x}\right)\rangle = \textbf{C}^0,
\end{align}
where the angle bracket $\langle\boldsymbol{\cdot}\rangle$ represents the operator for ensemble averaging. We assume that the stiffness fluctuation is a centered Gaussian random quantity such that the ensemble average of the fluctuation vanishes,
\begin{align}
\label{gamma_CRV_N}
\langle \boldsymbol{\gamma} \left(\textbf{x}\right)\rangle= 0.
\end{align}
Additionally, multiple scattering effects must include higher-order $N$-point averages on the fluctuations, which can be represented as,
\begin{flalign}
\label{gamma_CRV_prop}
\langle\gamma\left(\textbf{x}_1\right)\gamma\left(\textbf{x}_2\right)\boldsymbol{\cdot}\boldsymbol{\cdot}\boldsymbol{\cdot}\gamma\left(\textbf{x}_{2n}\right)\rangle &= \sum_{p.p.} \prod_{\substack{r={1}\\k={2}\\r<k}}^{2n} \langle \gamma\left(\textbf{x}_r\right)\gamma\left(\textbf{x}_k\right)\rangle\notag\\
\quad =&\langle\gamma\left(\textbf{x}_1\right)\gamma\left(\textbf{x}_2\right)\rangle+\langle\gamma\left(\textbf{x}_1\right)\gamma\left(\textbf{x}_2\right)\rangle\langle\gamma\left(\textbf{x}_3\right)\gamma\left(\textbf{x}_4\right)\rangle\notag\\
\quad+&\langle\gamma\left(\textbf{x}_1\right)\gamma\left(\textbf{x}_3\right)\rangle\langle\gamma\left(\textbf{x}_2\right)\gamma\left(\textbf{x}_4\right)\rangle+\langle\gamma\left(\textbf{x}_1\right)\gamma\left(\textbf{x}_4\right)\rangle\langle\gamma\left(\textbf{x}_2\right)\gamma\left(\textbf{x}_3\right)\rangle+\cdots
\end{flalign}
For Gaussian fluctuations \cite{rytov1989principles, tsang2004scattering}, all the odd-moment correlations vanish \cite{frisch1968wave} as,
\begin{align}
\label{gamma_CRVG_prop}
\langle\gamma\left(\textbf{x}_1\right)\gamma\left(\textbf{x}_2\right)\boldsymbol{\cdot}\boldsymbol{\cdot}\boldsymbol{\cdot}\gamma\left(\textbf{x}_{2n+1}\right)\rangle = 0. 
\end{align}
Equation (\ref{gamma_CRVG_prop}) is a precursor to the vanishing of odd-order diagrams in the self-energy to be shown shortly. As indicated in Equations (\ref{gamma_CRV_prop}) and (\ref{gamma_CRVG_prop}), the two-point covariance of the stiffness correlations is of primary importance. A standard ergodic assumption is utilized that states the ensemble average can be set equal to the volumetric average. With this in mind, the covariance is decomposed into a purely tensorial part $\Xi_{ijkl}^{\alpha\beta\gamma\delta}$ and a spatial part represented by the two point correlation function $\eta(\textbf{x}_1,\textbf{x}_2)$,
\begin{align}
\label{<gamma_Redef>}
\langle \gamma_{\alpha\beta\kappa\delta} \left(\textbf{x}_1\right) \gamma_{ijkl} \left(\textbf{x}_2\right) \rangle =\Xi^{\alpha\beta\kappa\delta}_{ijkl}\eta\left(\textbf{x}_1,\textbf{x}_2\right) .
\end{align}
The tensor shown on the right-hand side of Equation (\ref{<gamma_Redef>}) can be written in terms of the local stiffness tensor $\textbf{C}$ and the homogeneous reference stiffness tensor $\textbf{C}^0$,
\begin{align}
\label{covdef}
\Xi^{\alpha\beta\kappa\delta}_{ijkl} = \left\langle \left(C_{\alpha\beta\kappa\delta}-C^0_{\alpha\beta\kappa\delta}\right) \left(C_{ijkl}-C^0_{ijkl}\right) \right\rangle.
\end{align}
In Equation (\ref{covdef}), the angle brackets can now be interpreted as orientation averages over the rotations SO(3) since the averages are being applied to purely tensorial stiffness tensors. The aforementioned averaging operations were similarly applied in Refs. \cite{weaver1990diffusivity,stanke1984unified}. Here, we repeat some details on the averages to help make the present article self-contained.
\par We will constrain the focus to grains that have crystallographic symmetry belonging to the cubic Laue class. By virtue of (Franz) Neumann's principle, the stiffness tensor $\textbf{C}$ and products formed by pairs of $\textbf{C}$ must obey the symmetry class of the underlying crystal symmetry (cubic in this work). Thus, the interior of each grain is represented by the stiffness tensor,
\begin{align}
\label{intGrain}
C_{ijkl}=C_{12}\delta_{ij}\delta_{kl}+C_{44}\left(\delta_{ik}\delta_{jl}+\delta_{il}\delta_{jk}\right)+\nu a_{iu}a_{ju}a_{ku}a_{lu},
\end{align}
where $C_{11}$, $C_{12}$, $C_{44}$ are the three independent stiffness constants specific to cubic symmetry and $\nu=C_{11}-C_{12}-2C_{44}$ is an unnormalized measure of elastic anisotropy for materials with cubic crystallographic symmetry. In Equation (\ref{intGrain}), the repeated index $u$ implies summation over the terms $u=1,2,3$. The matrices $\textbf{a}$ are rotation matrices that enable one to write the stiffness $\textbf{C}$ that has been rotated through the matrices $\textbf{a}$ from a reference stiffness. For the present purpose, the crystallographic axes of a reference stiffness are of no concern since we only need orientation averages of $\textbf{C}$ and we assume the polycrystal is untextured. In other words, there is no preferential orientations of the grain's crystallographic axes, which implies that the polycrystalline media considered here are statistically isotropic and all orientational averages produce isotropic tensors. The orientation average on Equation (\ref{intGrain}) gets applied onto the rotation matrices,
\begin{align}
\label{a_ensem}
\langle a_{iu}a_{ju}a_{ku}a_{lu}\rangle=\frac{1}{5}(\delta_{ij}\delta_{kl}+\delta_{ik}\delta_{jl}+\delta_{il}\delta_{jk}).
\end{align}
With Equations (\ref{intGrain}) and (\ref{a_ensem}), we can easily establish that the two independent elastic stiffnesses for the non-scattering reference medium are $C_{12}^0=C_{12}+\nu/5$ and $C_{44}^0=C_{44}+\nu/5$ while the isotropic symmetry dictates that $C_{11}^0=C_{12}^0+2C_{44}^0$. The average needed to construct $\langle C_{ijkl}C_{\alpha\beta\gamma\delta}\rangle$ follows a similar procedure, but requires averages over products of eight rotation matrices. These have been evaluated \cite{kube2015stress},
\begin{align}
\label{latingreekEq}
    \langle a_{iu}a_{ju}a_{ku}a_{lu}a_{\alpha \lambda}a_{\beta\lambda}a_{\gamma\lambda}a_{\delta\lambda}\rangle=\frac{4}{105}I_{LLLLGGGG}+\frac{1}{420}I_{LGLGLGLG}+\frac{13}{35}I_{LLGGLGLG}
\end{align}
where both $u$ and $\lambda$ are summed over the values from 1, 2, and 3. The eighth rank tensors on the right-hand side of Equation (\ref{latingreekEq}) are taken from Ref. \cite{kube2015stress} and also given in Equations (\ref{ILG1} - \ref{ILG3}) of Appendix \ref{covTrans}. Following this procedure, the entire tensor $\Xi_{ijkl}^{\alpha\beta\gamma\delta}$ can be constructed as seen in Ref. \cite{weaver1990diffusivity} and written out in full in Equation (\ref{covfin2}) of Appendix \ref{covTrans}.
\par The orientation averages enter into the tensorial part $\Xi_{ijkl}^{\alpha\beta\gamma\delta}$ of the covariance. The spatial part is decoupled through the two-point spatial correlation function as seen in Equation (\ref{<gamma_Redef>}). In Refs. \cite{weaver1990diffusivity,stanke1984unified}, the two-point correlation (TPC) function was chosen to be a single decaying exponential with a correlation length $\ell$. The exponential was chosen in Refs. \cite{weaver1990diffusivity,stanke1984unified} to promote analytical modeling while having a strong historical basis in describing Debye random media. Furthermore, the single exponential has worked remarkably well to capture the microstructure influence on attenuation and scattering of elastic waves in real metallic polycrystalline microstructures. Like previous work \cite{weaver1990diffusivity,stanke1984unified}, we similarly adopt the exponential two-point correlation function for its mathematical simplicity. The TPC as a function of two spatial vectors $\textbf{x}_1$ and $\textbf{x}_2$ is,
\begin{align}
\label{eta_exp}
\eta\left(\textbf{x}_1,\textbf{x}_2\right) = e^{-|\textbf{x}_1-\textbf{x}_2|/\ell},
\end{align}
where $\ell$ is the correlation length, which is a measure of the average distance in which two grains have different crystallographic orientations and, thus, a measure of scatterer (grain) size. In Section \ref{SecSFTDomain}, the spatial Fourier transform of the TPC to the wave vector domain $\textbf{p}$ will be needed and is 
\begin{align}
\label{TPC_fosatransform}
\tilde{\eta}\left(\textbf{p}\right)=\frac{1}{\left(2\pi\right)^3}\int d^3\textbf{r} \eta(\textbf{r})e^{-i\textbf{p}\boldsymbol{\cdot}\textbf{r}} =\frac{1}{\left(2\pi\right)^3}\int d^3\textbf{r} e^{-|\textbf{r}|/\ell}e^{-i\textbf{p}\boldsymbol{\cdot}\textbf{r}}=\frac{\ell^3}{\pi^2\left(1+\ell^2|\textbf{p}|^2\right)^2}.
\end{align}
\par The Green's function $\left\langle\textbf{G}\right\rangle$ describes the coherent propagation of an elastic wave whose amplitude and phase velocity are modified through grain scattering effects \cite{weaver1990diffusivity}. The covariance and two-point statistics just described enter into the Dyson equation, which will be the next focus. Substituting the operator $\mathcal{L}$ from Equation (\ref{heteroL}), the Dyson series for the mean field arrives from the ensemble average of Equation (\ref{full^g}),

\begin{align}
\label{GmeanF}
       \langle G_{ij}\left(\textbf{x},\textbf{x}'\right) \rangle &=G^0_{ij}\left(\textbf{x},\textbf{x}'\right)\notag\\
       \quad &-\int\int G^0_{ii_1}\left(\textbf{x},\textbf{x}_1\right)\nabla_{x_{1_{j_1}}} \left[\vphantom{\frac{1}{2}}\right. \langle  \gamma_{i_1j_1k_1l_1}\left(\textbf{x}_1\right)\gamma_{j_2k_2i_2l_2}\left(\textbf{x}_2\right)\rangle G^{0''}_{k_1j_2}\left(x_{1_{l_1}},x_{2_{k_2}}\right) \left.\vphantom{\frac{1}{2}}\right]\notag\\
 \quad & \nabla_{x_{2_{l2}}} G^0_{i_2j}\left(\textbf{x}_2,\textbf{x}'\right)d^3\textbf{x}_1d^3\textbf{x}_2\notag\\
    \quad & -\int \int\int\int G^0_{ii_1}\left(\textbf{x},\textbf{x}_1\right) \nabla_{x_{1_{j_1}}} \left[\vphantom{\frac{1}{2}}\right. \langle \gamma_{i_1j_1k_1l_1}\left(\textbf{x}_1\right)\gamma_{j_2k_4k_2l_2}\left(\textbf{x}_2\right)\rangle\langle\gamma_{j_3j_5k_3l_3}\left(\textbf{x}_3\right)\notag\\
    \quad &\gamma_{j_4j_6i_2l_4}\left(\textbf{x}_4\right)\rangle G^{0''}_{k_1j_2}\left(\textbf{x}_{1_{l_1}},\textbf{x}_{2_{k_4}}\right)G^{0''}_{k_2j_3}\left(\textbf{x}_{2_{l_2}},\textbf{x}_{3_{j_5}}\right)G^{0''}_{k_3j_4}\left(\textbf{x}_{3_{l_3}},\textbf{x}_{4_{j_6}}\right)\left.\vphantom{\frac{1}{2}}\right]\nabla_{x_{4_{l_4}}}\notag\\
    \quad & G^0_{i_2j}\left(\textbf{x}_4,\textbf{x}'\right)d^3\textbf{x}_1d^3\textbf{x}_2d^3\textbf{x}_3d^3\textbf{x}_4\notag\\
    \quad & -\int \int\int\int G^0_{ii_1}\left(\textbf{x},\textbf{x}_1\right) \nabla_{x_{1_{j_1}}} \left[\vphantom{\frac{1}{2}}\right. \langle \gamma_{i_1j_1k_1l_1}\left(\textbf{x}_1\right)\gamma_{j_2k_4k_2l_2}\left(\textbf{x}_3\right)\rangle\langle\gamma_{j_3j_5k_3l_3}\left(\textbf{x}_2\right)\notag\\
    \quad &\gamma_{j_4j_6i_2l_4}\left(\textbf{x}_4\right)\rangle G^{0''}_{k_1j_2}\left(\textbf{x}_{1_{l_1}},\textbf{x}_{2_{k_4}}\right) G^{0''}_{k_2j_3}\left(\textbf{x}_{2_{l_2}},\textbf{x}_{3_{j_5}}\right)G^{0''}_{k_3j_4}\left(\textbf{x}_{3_{l_3}},\textbf{x}_{4_{j_6}}\right)\left.\vphantom{\frac{1}{2}}\right]\nabla_{x_{4_{l_4}}}\notag\\
    \quad & G^0_{i_2j}\left(\textbf{x}_4,\textbf{x}'\right)d^3\textbf{x}_1d^3\textbf{x}_2d^3\textbf{x}_3d^3\textbf{x}_4\notag\\
    \quad & -\int \int\int\int G^0_{ii_1}\left(\textbf{x},\textbf{x}_1\right) \nabla_{x_{1_{j_1}}} \left[\vphantom{\frac{1}{2}}\right. \langle \gamma_{i_1j_1k_1l_1}\left(\textbf{x}_1\right)\gamma_{j_2k_4k_2l_2}\left(\textbf{x}_4\right)\rangle\langle\gamma_{j_3j_5k_3l_3}\left(\textbf{x}_2\right)\notag\\
    \quad &\gamma_{j_4j_6i_2l_4}\left(\textbf{x}_3\right)\rangle G^{0''}_{k_1j_2}\left(\textbf{x}_{1_{l_1}},\textbf{x}_{2_{k_4}}\right) G^{0''}_{k_2j_3}\left(\textbf{x}_{2_{l_2}},\textbf{x}_{3_{j_5}}\right)G^{0''}_{k_3j_4}\left(\textbf{x}_{3_{l_3}},\textbf{x}_{4_{j_6}}\right)\left.\vphantom{\frac{1}{2}}\right]\nabla_{x_{4_{l_4}}}\notag\\
    \quad & G^0_{i_2j}\left(\textbf{x}_4,\textbf{x}'\right)d^3\textbf{x}_1d^3\textbf{x}_2d^3\textbf{x}_3d^3\textbf{x}_4+\cdot \cdot \cdot
\end{align}

 where $G^{0''}_{ij}\left(x_{r_k}, x_{s_l}\right)=\nabla_{x_{r_k}}\nabla_{x_{s_l}}G_{ij}^0\left(\textbf{x}_{r},\textbf{x}_{s}\right)$. Equation~(\ref{GmeanF}) explicitly displays the complexity associated with the increasing order of multiple scattering events. Feynman diagrams are convenient for representing the Dyson series shown in Equation~(\ref{GmeanF}), 
\begin{equation}\label{eqDiag}
\begin{tikzpicture}[scale=1.5,>=latex,baseline=0.75ex]

\draw[line width=2.5pt] (0,0) -- (1,0) node[right] {$=$};

\draw (1.5,0) -- node[below] {$0$} (2,0) node[right] {$+$};
\draw (2.5,0) -- node[below] {$2$} (3.5,0) node[right] {$+$};
\draw (4,0) -- node[below] {$4_{I}$} (5.75,0);

\draw (6.25,0) node[left] {$+$}  -- node[below] {$4_{II}$} (7.5,0);
\draw (8,0) node[left] {$+$} -- node[below] {$4_{III}$} (9.5,0) node[right]{$+\boldsymbol{\cdot}\boldsymbol{\cdot}\boldsymbol{\cdot}\\$};

 


\draw[densely dashed] (2.75,0) arc [start angle=180, end angle=0, x radius=0.25, y radius=0.25];

\draw[densely dashed] (4.25,0) arc [start angle=180, end angle=0, x radius=0.25, y radius=0.25];
\draw[densely dashed] (5,0) arc [start angle=180, end angle=0, x radius=0.25, y radius=0.25];

\draw[densely dashed] (6.5,0) arc [start angle=180, end angle=0, x radius=0.25, y radius=0.25];
\draw[densely dashed] (6.75,0) arc [start angle=180, end angle=0, x radius=0.25, y radius=0.25];

\draw[densely dashed] (8.25,0) arc [start angle=180, end angle=0, x radius=0.5, y radius=0.5];
\draw[densely dashed] (8.5,0) arc [start angle=180, end angle=0, x radius=0.25, y radius=0.25];

\draw[densely dashed] (2.75,0) arc [start angle=180, end angle=0, x radius=0.25, y radius=0.25];

\draw[densely dashed] (4.25,0) arc [start angle=180, end angle=0, x radius=0.25, y radius=0.25];
\draw[densely dashed] (5,0) arc [start angle=180, end angle=0, x radius=0.25, y radius=0.25];

\draw[densely dashed] (6.5,0) arc [start angle=180, end angle=0, x radius=0.25, y radius=0.25];
\draw[densely dashed] (6.75,0) arc [start angle=180, end angle=0, x radius=0.25, y radius=0.25];

\draw[densely dashed] (8.25,0) arc [start angle=180, end angle=0, x radius=0.5, y radius=0.5];
\draw[densely dashed] (8.5,0) arc [start angle=180, end angle=0, x radius=0.25, y radius=0.25];


\filldraw[black] (2.75,0) circle (0.03);
\filldraw[black] (3.25,0) circle (0.03);

\filldraw[black] (4.25,0) circle (0.03);
\filldraw[black] (4.75,0) circle (0.03);
\filldraw[black] (5,0) circle (0.03);
\filldraw[black] (5.5,0) circle (0.03);

\filldraw[black] (6.5,0) circle (0.03);
\filldraw[black] (6.75,0) circle (0.03);
\filldraw[black] (7,0) circle (0.03);
\filldraw[black] (7.25,0) circle (0.03);

\filldraw[black] (8.25,0) circle (0.03);
\filldraw[black] (8.5,0) circle (0.03);
\filldraw[black] (9,0) circle (0.03);
\filldraw[black] (9.25,0) circle (0.03);
\end{tikzpicture}
\end{equation}
where, 

a) the circle represents the spatial location of the scatterers:
\begin{tikzpicture}[scale=1.5,>=latex,baseline=-0.5ex]
\filldraw[black] (0,0) circle (0.03) ;
\end{tikzpicture}

b) the solid line represent the homogeneous propagator $\textbf{G}^0$: \begin{tikzpicture}[scale=1.5,>=latex,baseline=-0.5ex]
\draw (1.5,0) -- (2,0) ;
\end{tikzpicture}

c) the thick solid line represents the mean Green's function $\langle\textbf{G}\rangle$: \begin{tikzpicture}[scale=1.5,>=latex,baseline=-0.5ex]
\draw[line width=2.5pt] (1.5,0) -- (2,0) ;
\end{tikzpicture}

 d) the dashed loop without circles represents the statistical correlations, $\langle\boldsymbol{\gamma}\boldsymbol{\gamma}\rangle$: \begin{tikzpicture}[scale=1.5,>=latex]
\draw[densely dashed] (0,0) arc [start angle=180, end angle=0, x radius=0.25, y radius=0.25];
\end{tikzpicture}

The first term of the diagrammatic Dyson series shown in Equation (\ref{eqDiag}) represents the homogeneous propagation $\textbf{G}^0$ unimpaired by scattering effects. Similarly, the following term corresponds to $2-$point statistics. For $4$-point statistics, there can be three possible correlations, of which the first term, $4_I$, is a higher-order repetition of the two-point statistics demonstrated by the diagrammatic term $2$. Thus, the diagrammatic term $4_I$ corresponds to the connection index \cite{rytov1989principles} of 2 and can therefore be recognized as a weakly connected or reducible diagram. On the other hand, diagrammatic terms $2$, $4_{II}$, and $4_{III}$ are strongly connected or irreducible with connection index $\left(CI\right)=1$.
\par The sum of all strongly connected diagrams with $CI=1$, including the terms $2, 4_{II}, 4_{III}$ can be shown analytically as
\begin{align}
\label{Gstrong}
\langle G_{ij}\rangle_{CI=1} &=-\int\int G^0_{ii_1}\left(\textbf{x},\textbf{x}_1\right)\nabla_{x_{1_{j_1}}} \left[\vphantom{\frac{1}{2}}\right. \langle  \gamma_{i_1j_1k_1l_1}\left(\textbf{x}_1\right)\gamma_{j_2k_2i_2l_2}\left(\textbf{x}_2\right)\rangle G^{0''}_{k_1j_2}\left(x_{1_{l_1}},x_{2_{k_2}}\right) \left.\vphantom{\frac{1}{2}}\right]\notag\\
 \quad & \nabla_{x_{2_{l2}}} G^0_{i_2j}\left(\textbf{x}_2,\textbf{x}'\right)d^3\textbf{x}_1d^3\textbf{x}_2\notag\\
    \quad & -\int \int\int\int G^0_{ii_1}\left(\textbf{x},\textbf{x}_1\right) \nabla_{x_{1_{j_1}}} \left[\vphantom{\frac{1}{2}}\right. \langle \gamma_{i_1j_1k_1l_1}\left(\textbf{x}_1\right)\gamma_{j_2k_4k_2l_2}\left(\textbf{x}_3\right)\rangle\langle\gamma_{j_3j_5k_3l_3}\left(\textbf{x}_2\right)\notag\\
    \quad &\gamma_{j_4j_6i_2l_4}\left(\textbf{x}_4\right)\rangle G^{0''}_{k_1j_2}\left(\textbf{x}_{1_{l_1}},\textbf{x}_{2_{k_4}}\right) G^{0''}_{k_2j_3}\left(\textbf{x}_{2_{l_2}},\textbf{x}_{3_{j_5}}\right)G^{0''}_{k_3j_4}\left(\textbf{x}_{3_{l_3}},\textbf{x}_{4_{j_6}}\right)\left.\vphantom{\frac{1}{2}}\right]\nabla_{x_{4_{l_4}}}\notag\\
    \quad & G^0_{i_2j}\left(\textbf{x}_4,\textbf{x}'\right)d^3\textbf{x}_1d^3\textbf{x}_2d^3\textbf{x}_3d^3\textbf{x}_4\notag\\
    \quad & -\int \int\int\int G^0_{ii_1}\left(\textbf{x},\textbf{x}_1\right) \nabla_{x_{1_{j_1}}} \left[\vphantom{\frac{1}{2}}\right. \langle \gamma_{i_1j_1k_1l_1}\left(\textbf{x}_1\right)\gamma_{j_2k_4k_2l_2}\left(\textbf{x}_4\right)\rangle\langle\gamma_{j_3j_5k_3l_3}\left(\textbf{x}_2\right)\notag\\
    \quad &\gamma_{j_4j_6i_2l_4}\left(\textbf{x}_3\right)\rangle G^{0''}_{k_1j_2}\left(\textbf{x}_{1_{l_1}},\textbf{x}_{2_{k_4}}\right) G^{0''}_{k_2j_3}\left(\textbf{x}_{2_{l_2}},\textbf{x}_{3_{j_5}}\right)G^{0''}_{k_3j_4}\left(\textbf{x}_{3_{l_3}},\textbf{x}_{4_{j_6}}\right)\left.\vphantom{\frac{1}{2}}\right]\nabla_{x_{4_{l_4}}}\notag\\
    \quad & G^0_{i_2j}\left(\textbf{x}_4,\textbf{x}'\right)d^3\textbf{x}_1d^3\textbf{x}_2d^3\textbf{x}_3d^3\textbf{x}_4+\cdots.
\end{align}

In Equation (\ref{Gstrong}), pairs of homogeneous propagators, $\textbf{G}^0$, appear at the head and tail of each term.  The sum of all the strongly connected terms $2, 4_{II}, 4_{III}$ can be diagrammatically shown \cite{frisch1968wave, rytov1989principles} as
\begin{equation}
\begin{tikzpicture}[scale=1.5,>=latex]
\draw (7.5,-3)  node[left] {$\langle\textbf{G}_{CI=1}\rangle = \textbf{G}^0\boldsymbol{\cdot}\textbf{m}\boldsymbol{\cdot}\textbf{G}^0=$} -- (8,-3) ;
\draw (8,-3) .. controls (8.125,-2.75) and (8.875,-2.75) .. (9,-3);
\draw (9,-3) .. controls (8.875,-3.25) and (8.125,-3.25) .. (8,-3);
\draw (9,-3) --  (9.5,-3) node[right] {$,$};
\filldraw[black] (8,-3) circle (0.03);
\filldraw[black] (9,-3) circle (0.03);
\end{tikzpicture}
\end{equation}
where the horizontal propagators at the head and tail denote $\textbf{G}^0$, and the central lobe represents the so-called self-energy or the mass operator series \textbf{m} composed of only strongly connected diagrams. Thus, the diagrammatic mass operator series can be written by removing the propagators from the head and tail of all the strongly connected diagrams in Equation (\ref{eqDiag}), as

\begin{equation}\label{GausmassSeries}
\begin{tikzpicture}[scale=1.45,>=latex]

\draw (0,-3) .. controls (0.125,-2.75) and (0.875,-2.75) .. (1,-3)node[right] {$=$};
\draw (1,-3) .. controls (0.875,-3.25) and (0.125,-3.25) .. (0,-3);
\draw (0,-3) .. controls (0.125,-2.75) and (0.875,-2.75) .. (1,-3)node[right] {$=$};
\draw (1,-3) .. controls (0.875,-3.25) and (0.125,-3.25) .. (0,-3);
\filldraw[black] (0,-3) circle (0.03);
\filldraw[black] (1,-3) circle (0.03);

\draw (1.5,-3) -- node[below] {$_1\textbf{m}^{(1)}$} (2,-3) node[right] {$+$};
\draw[densely dashed] (1.5,-3) arc [start angle=180, end angle=0, x radius=0.25, y radius=0.25];
\draw[densely dashed] (1.5,-3) arc [start angle=180, end angle=0, x radius=0.25, y radius=0.25];
\filldraw[black] (1.5,-3) circle (0.03);
\filldraw[black] (2,-3) circle (0.03);

\draw (2.5,-3) -- node[below] {$_1\textbf{m}^{(2)}$} (3.25,-3) node[right] {$+$};
\draw[densely dashed] (2.5,-3) arc [start angle=180, end angle=0, x radius=0.25, y radius=0.25];
\draw[densely dashed] (2.75,-3) arc [start angle=180, end angle=0, x radius=0.25, y radius=0.25];

\draw[densely dashed] (2.5,-3) arc [start angle=180, end angle=0, x radius=0.25, y radius=0.25];
\draw[densely dashed] (2.75,-3) arc [start angle=180, end angle=0, x radius=0.25, y radius=0.25];
\filldraw[black] (2.5,-3) circle (0.03);
\filldraw[black] (2.75,-3) circle (0.03);
\filldraw[black] (3,-3) circle (0.03);
\filldraw[black] (3.25,-3) circle (0.03);

\draw (3.75,-3) -- node[below] {$_2\textbf{m}^{(2)}$} (4.75,-3) node[right] {$+$};
\draw[densely dashed] (3.75,-3) arc [start angle=180, end angle=0, x radius=0.5, y radius=0.5];
\draw[densely dashed] (4,-3) arc [start angle=180, end angle=0, x radius=0.25, y radius=0.25];
\draw[densely dashed] (3.75,-3) arc [start angle=180, end angle=0, x radius=0.5, y radius=0.5];
\draw[densely dashed] (4,-3) arc [start angle=180, end angle=0, x radius=0.25, y radius=0.25];

\filldraw[black] (3.75,-3) circle (0.03);
\filldraw[black] (4,-3) circle (0.03);
\filldraw[black] (4.5,-3) circle (0.03);
\filldraw[black] (4.75,-3) circle (0.03);

\draw (5.25,-3) -- node[below] {$_1\textbf{m}^{(3)}$} (6.5,-3);
\draw[densely dashed] (5.25,-3) arc [start angle=180, end angle=0, x radius=0.25, y radius=0.25];
\draw[densely dashed] (5.5,-3) arc [start angle=180, end angle=0, x radius=0.375, y radius=0.375];
\draw[densely dashed] (6,-3) arc [start angle=180, end angle=0, x radius=0.25, y radius=0.25];
\draw[densely dashed] (5.25,-3) arc [start angle=180, end angle=0, x radius=0.25, y radius=0.25];
\draw[densely dashed] (5.5,-3) arc [start angle=180, end angle=0, x radius=0.375, y radius=0.375];
\draw[densely dashed] (6,-3) arc [start angle=180, end angle=0, x radius=0.25, y radius=0.25];

\filldraw[black] (5.25,-3) circle (0.03);
\filldraw[black] (5.5,-3) circle (0.03);
\filldraw[black] (5.75,-3) circle (0.03);
\filldraw[black] (6,-3) circle (0.03);
\filldraw[black] (6.25,-3) circle (0.03);
\filldraw[black] (6.5,-3) circle (0.03);

\draw (7,-3) node[left] {$+$} -- node[below] {$_2\textbf{m}^{(3)}$} (8.5,-3) node[right] {$+$};
\draw[densely dashed] (7,-3) arc [start angle=180, end angle=0, x radius=0.25, y radius=0.25];
\draw[densely dashed] (7,-3) arc [start angle=180, end angle=0, x radius=0.25, y radius=0.25];

\draw[densely dashed] (7.25,-3) arc [start angle=180, end angle=0, x radius=0.625, y radius=0.625];
\draw[densely dashed] (7.25,-3) arc [start angle=180, end angle=0, x radius=0.625, y radius=0.625];

\draw[densely dashed] (7.75,-3) arc [start angle=180, end angle=0, x radius=0.25, y radius=0.25];
\draw[densely dashed] (7.75,-3) arc [start angle=180, end angle=0, x radius=0.25, y radius=0.25];

\filldraw[black] (7,-3) circle (0.03);
\filldraw[black] (7.25,-3) circle (0.03);
\filldraw[black] (7.5,-3) circle (0.03);
\filldraw[black] (7.75,-3) circle (0.03);
\filldraw[black] (8.25,-3) circle (0.03);
\filldraw[black] (8.5,-3) circle (0.03);

\draw (9,-3) -- node[below] {$_2\textbf{m}^{(3)}$} (10.25,-3) node[right] {$+ \cdots$};
\draw[densely dashed] (9,-3) arc [start angle=180, end angle=0, x radius=0.25, y radius=0.25];
\draw[densely dashed] (9,-3) arc [start angle=180, end angle=0, x radius=0.25, y radius=0.25];

\draw[densely dashed] (9.25,-3) arc [start angle=180, end angle=0, x radius=0.375, y radius=0.375];
\draw[densely dashed] (9.25,-3) arc [start angle=180, end angle=0, x radius=0.375, y radius=0.375];

\draw[densely dashed] (9.75,-3) arc [start angle=180, end angle=0, x radius=0.25, y radius=0.25];
\draw[densely dashed] (9.75,-3) arc [start angle=180, end angle=0, x radius=0.25, y radius=0.25];

\filldraw[black] (9,-3) circle (0.03);
\filldraw[black] (9.25,-3) circle (0.03);
\filldraw[black] (9.5,-3) circle (0.03);
\filldraw[black] (9.75,-3) circle (0.03);
\filldraw[black] (10,-3) circle (0.03);
\filldraw[black] (10.25,-3) circle (0.03);
\end{tikzpicture}
\end{equation}

\par The sum of all weakly connected diagrams with $CI=2$, including the terms like $4_{I}$ can be shown diagrammatically as,

\begin{equation}
\begin{tikzpicture}[scale=1.5,>=latex]

\draw (7.5,-3) node[left]{$\left\langle\textbf{G}_{CI=2}\right\rangle\equiv \textbf{G}^0\boldsymbol{\cdot}\textbf{m}\boldsymbol{\cdot}\textbf{G}^0\boldsymbol{\cdot}\textbf{m}\boldsymbol{\cdot}\textbf{G}^0=$} --  (8,-3) ;
\draw (8,-3) .. controls (8.125,-2.75) and (8.875,-2.75) .. (9,-3);
\draw (9,-3) .. controls (8.875,-3.25) and (8.125,-3.25) .. (8,-3);
\draw (9,-3) --  (9.5,-3) ;
\draw (9.5,-3) .. controls (9.625,-2.75) and (10.375,-2.75) .. (10.5,-3);
\draw (10.5,-3) .. controls (10.375,-3.25) and (9.625,-3.25) .. (9.5,-3);
\draw (10.5,-3) --  (11,-3) ;
\filldraw[black] (8,-3) circle (0.03);
\filldraw[black] (9,-3) circle (0.03);
\filldraw[black] (9.5,-3) circle (0.03);
\filldraw[black] (10.5,-3) circle (0.03);
\end{tikzpicture}
\end{equation}

Similarly, the sum of all the diagrams  with $CI=3$, can be shown as : 
\begin{equation}
\begin{tikzpicture}[scale=1.5,>=latex]

\draw (7.5,-3)node[left]{$\left\langle\textbf{G}_{CI=3}\right\rangle\equiv \textbf{G}^0\boldsymbol{\cdot}\textbf{m}\boldsymbol{\cdot}\textbf{G}^0\boldsymbol{\cdot}\textbf{m}\boldsymbol{\cdot}\textbf{G}^0\boldsymbol{\cdot}\textbf{m}\boldsymbol{\cdot}\textbf{G}^0=$} --  (8,-3) ;
\draw (8,-3) .. controls (8.125,-2.75) and (8.875,-2.75) .. (9,-3);
\draw (9,-3) .. controls (8.875,-3.25) and (8.125,-3.25) .. (8,-3);
\draw (9,-3) --  (9.5,-3) ;
\draw (9.5,-3) .. controls (9.625,-2.75) and (10.375,-2.75) .. (10.5,-3);
\draw (10.5,-3) .. controls (10.375,-3.25) and (9.625,-3.25) .. (9.5,-3);
\draw (10.5,-3) --  (11,-3) ;
\draw (11,-3) .. controls (11.125,-2.75) and (11.875,-2.75) .. (12,-3);
\draw (12,-3) .. controls (11.875,-3.25) and (11.125,-3.25) .. (11,-3);
\draw (12,-3) --  (12.5,-3) ;
\filldraw[black] (8,-3) circle (0.03);
\filldraw[black] (9,-3) circle (0.03);
\filldraw[black] (9.5,-3) circle (0.03);
\filldraw[black] (10.5,-3) circle (0.03);
\filldraw[black] (11,-3) circle (0.03);
\filldraw[black] (12,-3) circle (0.03);
\end{tikzpicture}
\end{equation}

Thus, in terms of connection index, Equation (\ref{GmeanF}) can be rewritten as,
\begin{align}
    \langle \textbf{G}\rangle=\textbf{G}^0+\langle \textbf{G}_{CI=1}\rangle+\langle \textbf{G}_{CI=2}\rangle+\langle \textbf{G}_{CI=3}\rangle+\cdots,
\end{align}
and, correspondingly, the diagrammatic Equation~(\ref{eqDiag}) can be rewritten as,
\begin{equation}\label{diag_CI}
\begin{tikzpicture}[scale=1.5,>=latex]
\draw[line width=2.5pt] (5,-3) -- (5.5,-3) node[right] {$=$};
\draw (6,-3) --  (6.5,-3) node[right] {$+$};
\draw (7,-3) --  (7.5,-3) ;
\draw (7.5,-3) .. controls (7.625,-2.75) and (8.375,-2.75) .. (8.5,-3);
\draw (8.5,-3) .. controls (8.375,-3.25) and (7.625,-3.25) .. (7.5,-3);
\draw (8.5,-3) --  (9,-3) node[right] {$+$};
\filldraw[black] (7.5,-3) circle (0.03);
\filldraw[black] (8.5,-3) circle (0.03);

\draw (9.5,-3) --  (10,-3) ;
\draw (10,-3) .. controls (10.125,-2.75) and (10.875,-2.75) .. (11,-3);
\draw (11,-3) .. controls (10.875,-3.25) and (10.125,-3.25) .. (10,-3);
\draw (11,-3) --  (11.5,-3) ;
\draw (11.5,-3) .. controls (11.625,-2.75) and (12.375,-2.75) .. (12.5,-3);
\draw (12.5,-3) .. controls (12.375,-3.25) and (11.625,-3.25) .. (11.5,-3);
\draw (12.5,-3) --  (13,-3) node[right] {$+\cdots$};
\filldraw[black] (10,-3) circle (0.03);
\filldraw[black] (11,-3) circle (0.03);
\filldraw[black] (11.5,-3) circle (0.03);
\filldraw[black] (12.5,-3) circle (0.03);
\end{tikzpicture}
\end{equation}
The above diagrammatic Equation (\ref{diag_CI}) is the elastodynamic Dyson equation
\begin{equation}\label{diag_Dyson}
\begin{tikzpicture}[scale=1.5,>=latex]
\draw[line width=2.5pt] (5,-3) -- (5.5,-3) node[right] {$=$};
\draw (6,-3) --  (6.5,-3) node[right] {$+$};
\draw (7,-3) --  (7.5,-3) ;
\draw (7.5,-3) .. controls (7.625,-2.75) and (8.375,-2.75) .. (8.5,-3);
\draw (8.5,-3) .. controls (8.375,-3.25) and (7.625,-3.25) .. (7.5,-3);
\draw [line width=2.5pt](8.5,-3) --  (9,-3);
\filldraw[black] (7.5,-3) circle (0.03);
\filldraw[black] (8.5,-3) circle (0.03);
\end{tikzpicture}
\end{equation}
which can be proven equal to Equation (\ref{diag_CI}) through repeated iteration of Equation (\ref{diag_Dyson}). In the preceding derivation, the weakly and strongly connected diagrams are sometimes referred to as reducible and irreducible diagrams, respectively. The Dyson equation is an exact mathematical representation of the mean Green's function of the random medium (polycrystal in the present case). It holds in weakly and strongly scattering situations. All orders of multiple scattering effects are included within the self-energy as seen in Equation (\ref{GausmassSeries}). In practice, the self-energy is truncated, which gives rise to the ordered approximations of FOSA and SOSA. The Dyson equation in non-diagrammatic form is,
\begin{align}
\label{Dyson}
    \langle \textbf{G}(\textbf{x},\textbf{x}')\rangle=\textbf{G}^{0}(\textbf{x},\textbf{x}') + \int\int \textbf{G}^{0}(\textbf{x},\textbf{x}_1)\boldsymbol{\cdot}\textbf{m}(\textbf{x}_1,\textbf{x}_r)\boldsymbol{\cdot}\langle\textbf{G}(\textbf{x}_r,\textbf{x}')\rangle d^3\textbf{x}_1d^3\textbf{x}_r,
\end{align}
which is seen in Weaver \cite{weaver1990diffusivity}. The derivation to arrive at Equation (\ref{Dyson}) is common to scattering problems in physics \cite{frisch1968wave, rytov1989principles}. The derivation of the Dyson equation here aims to provide a self-contained theory that remains in the context of elastodynamic scattering. Furthermore, the derivation is important to contextualize our contribution in including the higher-order scattering effects not present in previous models \cite{weaver1990diffusivity,stanke1984unified}. 

As will be shown in Section \ref{SecSFTDomain}, the self-energy is vital in solving the Dyson equation (and, inherently, the attenuation and velocity dispersion). To obtain analytical solutions, the mass operator series seen in Equation (\ref{GausmassSeries}) must be truncated. Truncation of the diagrammatic self-energy series shown in Equation (\ref{GausmassSeries}) to the first term corresponds to the bilocal \cite{tsang2004scattering} or FOSA. As a reminder, solutions for attenuation and velocity dispersion from FOSA \cite{weaver1990diffusivity} are equivalent to the Unified Theory of Stanke and Kino \cite{stanke1984unified}. The FOSA involves the evaluation of the operator
\begin{align}
\label{mspc1}
    _1m_{i_1i_2}^{(1)}(\textbf{x}_1,\textbf{x}_2)=-\nabla_{x_{1_{j_1}}} \left[\vphantom{\frac{1}{2}}\right. \langle  \gamma_{i_1j_1k_1l_1}\left(\textbf{x}_1\right)\gamma_{i_2j_2k_2l_2}\left(\textbf{x}_2\right)\rangle G^{0''}_{l_1l_2}\left(x_{1_{k_1}},x_{2_{k_2}}\right) \left.\vphantom{\frac{1}{2}}\right]\nabla_{x_{2_{j_2}}} \left[\vphantom{\frac{}{}}\right. , 
\end{align}
where the open left square bracket is a reminder of the operator nature of the self-energy. Thus, the FOSA of the Dyson series is
\begin{flalign}
\label{Dysoniter}
    \langle \textbf{G}(\textbf{x},\textbf{x}')\rangle_{FOSA}&=\textbf{G}^{0}(\textbf{x},\textbf{x}') + \int\int \textbf{G}^{0}(\textbf{x},\textbf{x}_1)\boldsymbol{\cdot}\textbf{m}^{(1)}(\textbf{x}_1,\textbf{x}_2) \boldsymbol{\cdot}\textbf{G}^0(\textbf{x}_2,\textbf{x}')d^3\textbf{x}_1d^3\textbf{x}_2\notag\\
    \quad & + \int\int \textbf{G}^{0}(\textbf{x},\textbf{x}_1)\boldsymbol{\cdot}\textbf{m}^{(1)}(\textbf{x}_1,\textbf{x}_2) \boldsymbol{\cdot}\textbf{G}^0(\textbf{x}_2,\textbf{x}_3)\boldsymbol{\cdot}\textbf{m}^{(1)}(\textbf{x}_3,\textbf{x}_4)  \notag\\
    \quad & \boldsymbol{\cdot}\textbf{G}^0(\textbf{x}_4,\textbf{x}')d^3\textbf{x}_1d^3\textbf{x}_2d^3\textbf{x}_3d^3\textbf{x}_4+ \cdots,
\end{flalign}
and in diagrammatic form
\begin{equation}\label{bilocalDiag}
\begin{tikzpicture}[scale=1.5,>=latex]

\draw[line width=2.5pt] (0,0) -- (1,0) node[right] {$=$};

\draw (1.5,0) -- node[below] {$0$} (2,0) node[right] {$+$};
\draw (2.5,0) -- node[below] {$2$} (3.5,0) node[right] {$+$};
\draw (4,0) -- node[below] {$4_{I}$} (6,0) node[right] {$+$};
\draw (6.5,0) -- node[below] {$6_{I}$} (9.5,0)node[right] {$+ \cdots.$};

\draw[densely dashed] (2.75,0) arc [start angle=180, end angle=0, x radius=0.25, y radius=0.25];
\draw[densely dashed] (4.25,0) arc [start angle=180, end angle=0, x radius=0.25, y radius=0.25];
\draw[densely dashed] (5.25,0) arc [start angle=180, end angle=0, x radius=0.25, y radius=0.25];
\draw[densely dashed] (6.75,0) arc [start angle=180, end angle=0, x radius=0.25, y radius=0.25];
\draw[densely dashed] (7.75,0) arc [start angle=180, end angle=0, x radius=0.25, y radius=0.25];
\draw[densely dashed] (8.75,0) arc [start angle=180, end angle=0, x radius=0.25, y radius=0.25];

\filldraw[black] (2.75,0) circle (0.03);
\filldraw[black] (3.25,0) circle (0.03);
\filldraw[black] (4.25,0) circle (0.03);
\filldraw[black] (4.75,0) circle (0.03);
\filldraw[black] (5.25,0) circle (0.03);
\filldraw[black] (5.75,0) circle (0.03);
\filldraw[black] (6.75,0) circle (0.03);
\filldraw[black] (7.25,0) circle (0.03);
\filldraw[black] (7.75,0) circle (0.03);
\filldraw[black] (8.25,0) circle (0.03);
\filldraw[black] (8.75,0) circle (0.03);
\filldraw[black] (9.25,0) circle (0.03);
\end{tikzpicture}    
\end{equation}

The Dyson equation within the FOSA is
\begin{equation}
\begin{tikzpicture}[scale=1.5,>=latex]
\label{bilocalDyson}
\draw[line width=2.5pt] (4.5,-3) -- (5.5,-3) node[right] {$=$};
\draw (6,-3) --  (7,-3)  node[right] {$+$};
\draw (7.5,-3) --  (8.5,-3); 
\draw (8.5,-3) --  (9,-3);
\draw[densely dashed] (8.5,-3) arc [start angle=180, end angle=0, x radius=0.25, y radius=0.25];
\draw[densely dashed] (8.5,-3) arc [start angle=180, end angle=0, x radius=0.25, y radius=0.25];
\draw [line width=2.5pt](9,-3) --  (10,-3)node[right] {$,$};
\filldraw[black] (8.5,-3) circle (0.03);
\filldraw[black] (9,-3) circle (0.03);
\end{tikzpicture}
\end{equation}
which can easily be shown to reduce to Equation (\ref{bilocalDiag}) through repeated iteration. Equation (\ref{bilocalDiag}) clearly shows multiple scattering events and, thus, solutions to the FOSA include multiple scattering. However, the limitation of FOSA is that all scatterers can only be visited once, and recurrent scattering is not included.
\begin{table}[H]
  \centering
  \renewcommand{\arraystretch}{1.6}
  \caption{Density $\rho$ (kg/$\text{m}^3$), monocrystalline elastic constants $C_{11}$, $C_{12}$, and $C_{44}$ (GPa), elastic constants of the non-scattering reference medium $C_{11}^0$, $C_{44}^0$ (GPa), anisotropy parameter $\nu$ (GPa), Zener's anisotropy parameter $A=(C_{11}-C_{12})/(2C_{44})=\nu/(2C_{44})+1$, and degree of heterogeneity $\epsilon_L$ and $\epsilon_T$ for polycrystalline aluminum, iron, and lithium.}
  \label{TabMat}
    \begin{tabular}{ccccccccccc}
    \toprule\hline
& $\rho$     & $C_{11}$   & $C_{12}$   & $C_{44}$ &$C^0_{11}$ & $C^0_{44}$ & $\nu$  & $A$  & $\epsilon_L$  & $\epsilon_T$  \\\hline
    Al    & 2700&  103.4& 57.1& 28.6 & 107.76 & 26.42 & -10.9& 1.23 & 0.0088 & 0.0270 \\
    Fe & 7860&219.2&136.8&109.2 & 273.6 & 82 &-136&2.65& 0.0434 & 0.1086\\
    Li  & 534 &13.4&11.3&9.6& 20.24 & 6.18 &-17.1&9.14& 0.0737 & 0.1811\\
    \hline
    \bottomrule
    \end{tabular}
  \label{tab1}
\end{table}

\par The current model introduces a quadrilocal approximation to the diagrammatic Dyson equation in Equation (\ref{diag_Dyson}) by extending from the existing bilocal approximation shown in Equation (\ref{bilocalDyson}) to

\begin{tikzpicture}[scale=1.5,>=latex]


\draw[line width=2.5pt] (4.5,-3) -- (5.5,-3) node[right] {$=$};
\draw (6,-3) --  (7,-3)  node[right] {$+$};
\draw (7.5,-3) --  (8,-3); 
\draw (8,-3) --  (8.5,-3);
\draw[densely dashed] (8,-3) arc [start angle=180, end angle=0, x radius=0.25, y radius=0.25];
\draw[densely dashed] (8,-3) arc [start angle=180, end angle=0, x radius=0.25, y radius=0.25];
\draw [line width=2.5pt](8.5,-3) --  (9,-3) node[right] {$+$};
\draw (9.5,-3) --  (10,-3); 
\draw (10,-3) --  (10.25,-3);
\draw (10.25,-3) --  (10.75,-3);

\draw[densely dashed] (10,-3) arc [start angle=180, end angle=0, x radius=0.25, y radius=0.25];
\draw[densely dashed] (10.25,-3) arc [start angle=180, end angle=0, x radius=0.25, y radius=0.25];
\draw[densely dashed] (10,-3) arc [start angle=180, end angle=0, x radius=0.25, y radius=0.25];
\draw[densely dashed] (10.25,-3) arc [start angle=180, end angle=0, x radius=0.25, y radius=0.25];

\draw [line width=2.5pt](10.75,-3) --  (11.25,-3)node[right] {$+$};
\draw (11.75,-3) --  (12.25,-3); 
\draw (12.25,-3) --  (12.75,-3);
\draw (12.75,-3) --  (13.25,-3);

\draw[densely dashed] (12.25,-3) arc [start angle=180, end angle=0, x radius=0.5, y radius=0.5];
\draw[densely dashed] (12.5,-3) arc [start angle=180, end angle=0, x radius=0.25, y radius=0.25];
\draw[densely dashed] (12.25,-3) arc [start angle=180, end angle=0, x radius=0.5, y radius=0.5];
\draw[densely dashed] (12.5,-3) arc [start angle=180, end angle=0, x radius=0.25, y radius=0.25];

\draw [line width=2.5pt](13.25,-3) --  (13.75,-3);
\filldraw[black] (8,-3) circle (0.03);
\filldraw[black] (8.5,-3) circle (0.03);
\filldraw[black] (10.75,-3) circle (0.03);
\filldraw[black] (10.25,-3) circle (0.03);
\filldraw[black] (10.5,-3) circle (0.03);
\filldraw[black] (10,-3) circle (0.03);
\filldraw[black] (12.25,-3) circle (0.03);
\filldraw[black] (12.5,-3) circle (0.03);
\filldraw[black] (13,-3) circle (0.03);
\filldraw[black] (13.25,-3) circle (0.03);
\end{tikzpicture}\begin{align}\label{SOSADyson}
\end{align}
This quadrilocal or SOSA-based scattering theory involves the four-point statistics by including the terms $_1\textbf{m}^{(2)}$ and $_2\textbf{m}^{(2)}$, in addition to the FOSA-based $_1\textbf{m}^{(1)}$ term into the self-energy series shown in Equation (\ref{GausmassSeries}).
The SOSA-based self-energy can be analytically expressed as,

\begin{align}
\label{mspc31}
    _1m_{i_1i_2}^{(2)}(\textbf{x}_1,\textbf{x}_4)&=-\int\int d^3\textbf{x}_2d^3\textbf{x}_3\nabla_{x_{1_{j_1}}} \left[\vphantom{\frac{1}{2}}\right. \langle \gamma_{i_1j_1k_1l_1}\left(\textbf{x}_1\right)\gamma_{j_2k_4k_2l_2}\left(\textbf{x}_3\right)\rangle\langle\gamma_{j_3j_5k_3l_3}\left(\textbf{x}_2\right)\notag\\
    \quad &\gamma_{j_4j_6i_2l_4}\left(\textbf{x}_4\right)\rangle G^{0''}_{k_1j_2}\left(\textbf{x}_{1_{l_1}},\textbf{x}_{2_{k_4}}\right) G^{0''}_{k_2j_3}\left(\textbf{x}_{2_{l_2}},\textbf{x}_{3_{j_5}}\right)G^{0''}_{k_3j_4}\left(\textbf{x}_{3_{l_3}},\textbf{x}_{4_{j_6}}\right)\left.\vphantom{\frac{1}{2}}\right]\nabla_{x_{4_{l_4}}}\left[\vphantom{\frac{}{}}\right.,
\end{align}

and,
\begin{align}
\label{mspc32}
    _2m_{i_1i_2}^{(2)}(\textbf{x}_1,\textbf{x}_4)&=-\int\int d^3\textbf{x}_2d^3\textbf{x}_3\nabla_{x_{1_{j_1}}} \left[\vphantom{\frac{1}{2}}\right. \langle \gamma_{i_1j_1k_1l_1}\left(\textbf{x}_1\right)\gamma_{j_2k_4k_2l_2}\left(\textbf{x}_4\right)\rangle\langle\gamma_{j_3j_5k_3l_3}\left(\textbf{x}_2\right)\notag\\
    \quad &\gamma_{j_4j_6i_2l_4}\left(\textbf{x}_3\right)\rangle G^{0''}_{k_1j_2}\left(\textbf{x}_{1_{l_1}},\textbf{x}_{2_{k_4}}\right) G^{0''}_{k_2j_3}\left(\textbf{x}_{2_{l_2}},\textbf{x}_{3_{j_5}}\right)G^{0''}_{k_3j_4}\left(\textbf{x}_{3_{l_3}},\textbf{x}_{4_{j_6}}\right)\left.\vphantom{\frac{1}{2}}\right]\nabla_{x_{4_{l_4}}}\left[\vphantom{\frac{}{}}\right..
\end{align}
The FOSA and SOSA approximations invoke the requirement that perturbations must be small ($\epsilon^2$). To be shown in Section \ref{resultsSec}, the perturbation parameter $\epsilon$ can be calculated, and values for aluminum, iron, and lithium are shown in Table \ref{TabMat}. Lithium, in particular, is one of the most anisotropic crystalline elements in nature. Note that the $\epsilon$ parameter for lithium remains small for both longitudinal and transverse waves. FOSA is expected to be accurate when $\epsilon^2\ll 1$; for comparison, $\epsilon_T^2=0.033$ for transverse waves in lithium. For SOSA, it is $\epsilon^4$ that is of importance and $\epsilon^4\ll 1$; for reference, $\epsilon_T^4=0.0011$ for transverse waves in lithium. Clearly, the conditions for both FOSA and SOSA are easily satisfied even in the extreme case of lithium. If we truncated at even higher orders, the self-energy would include even higher-order powers of $\epsilon$, which also indicates a convergent series for the self-energy when considering polycrystalline random media. Furthermore, small or negligible higher-order powers of $\epsilon$ will tend to be very small, indicating that even higher-order scattering effects would likely be negligible. Now, the attention turns to evaluating Equations (\ref{mspc31}) and (\ref{mspc32}) and including their influence on the SOSA-based model predictions of attenuation and velocity dispersion in polycrystals.

\section{Dyson equation solutions in the wavenumber domain}
\label{SecSFTDomain}

Solutions to the Dyson equation are most readily obtained in the wave vector domain. In the wave vector domain, the Dyson equation is
\begin{align}
\label{DysonP}
    \langle \textbf{G}(\textbf{p})\rangle=\textbf{G}^{0}(\textbf{p}) - \textbf{G}^{0}(\textbf{p})\boldsymbol{\cdot}\boldsymbol{\sigma}(\textbf{p})\boldsymbol{\cdot}\langle\textbf{G}(\textbf{p})\rangle,
\end{align}
 and its solution is,
\begin{align}
\label{DysonPfin}
    \langle \textbf{G}(\textbf{p})\rangle=\left[\left(\textbf{G}^{0}(\textbf{p})\right)^{-1} + \boldsymbol{\sigma}(\textbf{p})\right]^{-1},
\end{align}
where $\boldsymbol{\sigma}$ is the self-energy in the wave vector domain. Evaluation then turns to solving for the poles of Equation (\ref{DysonPfin}), which in turn allows us to find the attenuation and velocity dispersion. 
Clearly, Equation (\ref{DysonPfin}) requires finding the Fourier transforms of both the homogeneous Green's function and self-energy. These are
\begin{align}
\label{GpDef}
\textbf{G}^0(\textbf{p})\delta^3(\textbf{p}-\textbf{q})=\frac{1}{(2\pi)^3}\int\int d^3\textbf{x}d^3\textbf{x}'e^{-i\textbf{p}\boldsymbol{\cdot}\textbf{x}}\textbf{G}^0(\textbf{x},\textbf{x}')e^{i\textbf{q}\boldsymbol{\cdot}\textbf{x}'},
\end{align}
\begin{align}
\label{sigmaDef}
    -\boldsymbol{\sigma}(\textbf{p})\delta^3(\textbf{p}-\textbf{q})=\frac{1}{(2\pi)^3}\int\int d^3\textbf{x}d^3\textbf{x}'e^{-i\textbf{p}\boldsymbol{\cdot}\textbf{x}}\textbf{m}(\textbf{x},\textbf{x}')e^{i\textbf{q}\boldsymbol{\cdot}\textbf{x}'}.
\end{align} 
where $\textbf{p}$ and $\textbf{q}$ are wave vectors. As the homogeneous reference medium and the scattering medium are both isotropic, $\left(\textbf{G}^0(\textbf{p})\right)^{-1}$ and $\boldsymbol{\sigma}$ can be written in isotropic dyadic form,
\begin{align}
\label{G_decomp}
\textbf{G}^0(\textbf{p})=g_{0L}(p_L) \hat{\textbf{p}}\hat{\textbf{p}} + g_{0T}(p_T) \left(\textbf{I}-\hat{\textbf{p}}\hat{\textbf{p}} \right),
\end{align}
\begin{align}
\label{Gm1_decomp}
\left(\textbf{G}^0\left(\textbf{p}\right)\right)^{-1}=g_{0L}^{-1}(p_L) \hat{\textbf{p}}\hat{\textbf{p}} + g_{0T}^{-1}\left(p_T\right) \left(\textbf{I}-\hat{\textbf{p}}\hat{\textbf{p}} \right),
\end{align}
\begin{align}
\label{sigLT_decomp}
\boldsymbol{\sigma}\left(\textbf{p}\right)=\sigma^{L}(\textbf{p}) \hat{\textbf{p}}\hat{\textbf{p}} + \sigma^{T}\left(\textbf{p}\right) \left(\textbf{I}-\hat{\textbf{p}}\hat{\textbf{p}} \right),
\end{align}
where $p_L$ and $p_T$ are the complex wavenumbers of the two possible longitudinal and transverse wave modes in the infinite and statistically isotropic polycrystal. The components $\sigma^L$ and $\sigma^T$ depend on whether FOSA or SOSA is being applied and will be shown in the following two subsections. Note that while $\textbf{G}^0$ represents the Green's function in the non-scattering reference medium, Equation (\ref{G_decomp}) contains propagators that are a function of the effective wavenumbers for the scattering medium. This is because the Dyson equation involves the spatial Fourier transform to the $\textbf{p}$ domain to arrive at Equation (\ref{DysonP}).
 In Equation (\ref{G_decomp}), the homogeneous propagators $g_{0L}$ and $g_{0T}$ must satisfy \cite{weaver1990diffusivity}
\begin{align}
\label{homoGFourier}
\left[\rho\omega^2\delta_{ik}-C_{ijkl}^0p_jp_l\right]G_{km}^0(\textbf{p}) = \delta_{im},
\end{align}
where $\omega=2\pi f$ is the angular frequency. Explicitly, $g_{0L}$ and $g_{0T}$ are 
\begin{align}
\label{propLong}
g_{0L}\left(p_L\right) = \frac{1}{\rho\left(\omega^2-p_L^2c^2_{0L}\right)} = \frac{1}{\rho c^2_{0L}\left(p_{0L}^2-p_L^2\right)}, 
\end{align}
and,
\begin{align}
\label{propShear}
g_{0T}\left(p_T\right) = \frac{1}{\rho\left(\omega^2-p_T^2c^2_{0T}\right)} = \frac{1}{\rho c^2_{0T}\left(p_{0T}^2-p_T^2\right)},  
\end{align}
where $c_{0L}=\omega/p_{0L}$ and $c_{0T}=\omega/p_{0T}$ are the respective longitudinal and transverse phase velocities in the non-scattering (and non-dispersive) reference medium. It is noted that the reference medium has a stiffness represented by, $\textbf{C}^0=\left\langle\textbf{C}\right\rangle$, and thus $\rho c_{0L}^2=\left\langle\textbf{C}_{11}\right\rangle=C_{12}+2C_{44}+3\nu/5$ and $\rho c_{0T}^2=\left\langle\textbf{C}_{44}\right\rangle=C_{44}+\nu/5$. The Green's function $\textbf{G}^0$ is complex with the real part $\Re[{g_{0M}}]$ involving the Cauchy principal value (CPV),
\begin{align}
\label{Reg0}
\Re\left[g_{0M}\left(p_M\right)\right] \quad & =\frac{-1}{\rho c^2_{0M} }\mathcal{P}\left( \frac{1}{p_M^2-p_{0M}^2}\right),
\end{align}
where a short form is introduced in which $M$ can be $L$ or $T$. Similarly, the imaginary part $\Im\left[g_{0M}\right]$, involves a delta function \cite{weaver1990diffusivity} as,
\begin{align}
\label{ImG0MDef}
\Im\left[g_{0M}\left(p_M\right)\right] \quad & =\frac{-\pi}{2\rho c^2_{0M} p_{0M}}\delta\left(p_M-p_{0M}\right).
\end{align}
\par We return to the Dyson equation solution seen in Equation (\ref{DysonPfin}). Note that the real and complex parts of $\textbf{p}$ can be directly solved from the poles of Equation (\ref{DysonPfin}) when $\left(\textbf{G}^{0}(\textbf{p})\right)^{-1}+\boldsymbol{\sigma}(\textbf{p})=0$ or equivalently
\begin{align}
\label{ginvL}
    \left(g^{-1}_{0L}+\sigma^L\right)\hat {\textbf{p}}\hat {\textbf{p}} = 0
\end{align}
and
\begin{align}
\label{ginvT}
\left(g^{-1}_{0T}+\sigma^T\right)\left(\textbf{I}-\hat {\textbf{p}}\hat {\textbf{p}}\right) = 0.
\end{align}
We define
\begin{align}
\label{SigmasigmaL}
\Sigma^L\left(\textbf{p} \right)=\frac{\sigma^L\left(\textbf{p}\right)}{\rho(p_Lc_{0L})^2},
\end{align}
and
\begin{align}
\label{SigmasigmaT}
\Sigma^T\left(\textbf{p} \right)=\frac{\sigma^T\left(\textbf{p}\right)}{\rho(p_Tc_{0T})^2},
\end{align}
for which the effective wavenumbers are contained within. The terms in parenthesis in Equations (\ref{ginvL}) and (\ref{ginvT}) must be zero. Setting these terms to zero and combining with Equations (\ref{SigmasigmaL}) and (\ref{SigmasigmaT}) leads to the general dispersion equations,
\begin{align}
\label{dispL_org}
p^2_L= p^2_{0L}\left[1-\Sigma^L\left(\textbf{p}\right)\right]^{-1},
\end{align}
and
\begin{align}
\label{dispT_org}
p^2_T = p^2_{0T}\left[1-\Sigma^T\left(\textbf{p}\right)\right]^{-1}.
\end{align}
Equations (\ref{dispL_org}) and (\ref{dispT_org}) give the squared effective wavenumbers $p^2$ and are equal to the square of the non-scattering wavenumbers $p_0^2$ modified by a factor that contains \textbf{all scattering effects}. Note also that the respective effective wavenumbers make Equations (\ref{dispL_org}) and (\ref{dispT_org}) transcendental. When appropriate, the Born approximation is sometimes used to deal with transcendental equations in scattering theory \cite{weaver1990diffusivity}.
Various orders of Born approximation can be obtained when the approximations  $\textbf{p}_L\approx \textbf{p}_{0L}$ and $\textbf{p}_T\approx \textbf{p}_{0T}$ are used in Equations (\ref{dispL_org}) and (\ref{dispT_org}) (including their unit vectors). It is emphasized that Equations (\ref{dispL_org}) and (\ref{dispT_org}) are general since no smoothing approximations have been applied yet.  The attenuations and phase velocity dispersion can be directly obtained from Equations (\ref{dispL_org}) and (\ref{dispT_org}) by finding the imaginary and real parts of the effective wavenumbers,
\begin{align}
\label{att_DEFL}
\alpha=\Im(p).
\end{align}
and,
\begin{align}
\label{ws_DEFT}
c=\frac{\omega}{\Re(p)}.
\end{align}
Explicitly, the real and imaginary parts of $p$ are
\begin{align}
\label{reX}
\Re(p)&=p_{0}\Re\left[1/\sqrt{1-\Re_\Sigma-i\Im_\Sigma}\right]\notag\\ 
    \quad &=\frac{p_0}{\sqrt{2}}\left[\frac{\left(\sqrt{\Im_\Sigma^2 + (1-\Re_\Sigma)^2}\right)+1-\Re_\Sigma}{\Im_\Sigma^2 + (1-\Re_\Sigma)^2}\right]^{1/2},
\end{align}
\begin{align}
\label{imX}
\Im(p)= &p_{0}\Im\left[1/\sqrt{1-\Re_\Sigma-i\Im_\Sigma}\right]\notag\\ 
    \quad & =\frac{p_0}{\sqrt{2}}\left[\frac{\left(\sqrt{\Im_\Sigma^2 + (1-\Re_\Sigma)^2}\right)-1+\Re_\Sigma}{\Im_\Sigma^2 + (1-\Re_\Sigma)^2}\right]^{1/2},
\end{align}
where the real part of the self-energy $\Re\left(\Sigma\right)$ is denoted by $\Re_\Sigma$ and the imaginary part $\Im\left(\Sigma\right)$ is denoted by $\Im_\Sigma$. Thus, the attenuations and phase velocities follow from obtaining the real and imaginary part of the scaled self-energy where
\begin{align}
\label{SigLFi}
    \Sigma^L\left(\textbf{p}\right) = \frac{\sigma_{ij}\left(\textbf{p}\right)}{\rho\left(p_L c_{0L}\right)^2}\hat{p}_i\hat{p}_j
\end{align}
and
\begin{align}
\label{SigTFi}
    \Sigma^T\left(\textbf{p}\right) =\frac{\sigma_{ij}\left(\textbf{p}\right)}{2\rho\left(p_T c_{0T}\right)^2}\left(\delta_{ij}-\hat{p}_i\hat{p}_j\right).
\end{align}
The general framework for calculating the effective wavenumbers is now provided. In principle, this framework could be used to find the effective wave numbers for all orders of multiple scattering if the self-energies can be determined (analytically or numerically), which would lead to the exact mean Green's function. Here, we proceed analytically and focus on the approximations of FOSA and SOSA. Thus, the next step is to obtain the real and imaginary parts of the scaled self-energy terms seen in Equations (\ref{reX}) and (\ref{imX}) through the FOSA in Section \ref{FOSA_Sec} and the SOSA in Section \ref{SOSA_Sec}.

\subsection{First-order smoothing approximation}
\label{FOSA_Sec}

Weaver \cite{weaver1990diffusivity} has shown the self-energy $\boldsymbol{\sigma}$ in the FOSA is 
\begin{align}
\label{convfirstapp}
    \sigma_{i_1i_2}^{FOSA}(\textbf{p})=-p_{j_1}p_{j_2}\int d^3\textbf{s}\Xi^{i_1j_1k_1l_1}_{i_2j_2k_2l_2}s_{k_1}s_{k_2}G_{l_1l_2}^0(\textbf{s})\Tilde{\eta}(\textbf{p}-\textbf{s}).
\end{align}
The components of the scaled self-energies follow from Equations (\ref{SigLFi}) and (\ref{SigTFi}),
\begin{align}
\label{SigprojLongF}
    \Sigma^L_{FOSA}(\textbf{p})=-\frac{\hat{p}_{i_1}\hat{p}_{i_2}\hat{p}_{j_1}\hat{p}_{j_2}}{\rho c_{0L}^2}\int d^3\textbf{s}\Xi^{i_1j_1k_1l_1}_{i_2j_2k_2l_2}s_{k_1}s_{k_2}G_{l_1l_2}^0(\textbf{s})\Tilde{\eta}(\textbf{p}-\textbf{s}),
\end{align}
and,
\begin{align}
\label{SigprojShearF}
    \Sigma^T_{FOSA}(\textbf{p})=-\frac{\left(\delta_{i_1i_2}-\hat{p}_{i_1}\hat{p}_{i_2}\right)\hat{p}_{j_1}\hat{p}_{j_2}}{2\rho c_{0T}^2}\int d^3\textbf{s}\Xi^{i_1j_1k_1l_1}_{i_2j_2k_2l_2}s_{k_1}s_{k_2}G_{l_1l_2}^0(\textbf{s})\Tilde{\eta}(\textbf{p}-\textbf{s}).
\end{align}
Moreover, $\textbf{G}^0$ shown in Equation (\ref{G_decomp}) can be used to decompose the scaled self-energies into
\begin{align}
\label{SigLFdefn}
\Sigma^L_{FOSA}=2\left[\Sigma^{LL}_{FOSA}+\Sigma^{LT}_{FOSA}\right],
\end{align}
and,
\begin{align}
\label{SigTFdefn}
\Sigma^T_{FOSA}=\left[\Sigma^{TL}_{FOSA}+\Sigma^{TT}_{FOSA}\right].
\end{align}
Equations (\ref{SigLFdefn}) and (\ref{SigTFdefn}) are decompositions of all possible scattering modes, including longitudinal to longitudinal, transverse to transverse, and mode-converted scattering of longitudinal to transverse and transverse to longitudinal. Each component on the right-hand side of Equations (\ref{SigLFdefn}) and (\ref{SigTFdefn}) are of the form
\begin{align}
\label{GenSigLLFOSA}
    \Sigma^{PQ}_{FOSA}(\textbf{p})=\int d^3 \textbf{s} \mathcal{Q}^P_{FOSA}\left(p_P\hat{\textbf{p}},\textbf{s}\right) \Xi^{PQ}_{FOSA} g_{0Q}\left(s\right) ,
\end{align}
where $P$ and $Q$ can take the values of $L$ or $T$ and the common scalar $\mathcal{Q}^P_{FOSA}\left(\textbf{p},\textbf{s}\right)$ arrives from the Fourier transform of the TPC seen in Equation (\ref{TPC_fosatransform}),
\begin{align}
    \mathcal{Q}^P_{FOSA}\left(\textbf{p},\textbf{s}\right)=-\frac{\ell^3}{2\pi^2\rho c_{0P}^2}\left[\frac{\left|\textbf{s}\right|}{\left(1+\ell^2\left|\textbf{p}-\textbf{s}\right|^2\right)}\right]^2.
\end{align}
The inner products denoted as $\Xi_{FOSA}^{PQ}$ are
\begin{align}
\label{GenSigLLFOSA22}
    \Xi^{LL}_{FOSA}=L,~\Xi^{LT}_{FOSA}=\Xi^{TL}_{FOSA}=M-L,~ \Xi^{TT}_{FOSA}=N-2M+L,
\end{align}
where
\begin{flalign}
\label{LFOSADEF}
L&=\Xi^{i_1j_1k_2l_1}_{j_2k_3i_2l_3}\hat{p}_{i_1}\hat{p}_{i_2}\hat{p}_{j_1}\hat{p}_{l_3}\hat{s}_{k_2}\hat{s}_{j_2}\hat{s}_{l_1}\hat{s}_{k_3}=\frac{\nu^2}{525}\left(3+\cos^2\theta\right)^2,
\notag\\
\quad 
M&=\Xi^{i_1j_1k_2l_1}_{j_2k_3i_2l_3}\delta_{k_2j_2}\hat{p}_{i_1}\hat{p}_{i_2}\hat{p}_{j_1}\hat{p}_{l_3}\hat{s}_{l_1}\hat{s}_{k_3}=\frac{12\nu^2}{525}\left(2+\cos^2\theta\right),\notag\\
\quad  N&=\Xi^{i_1j_1k_2l_1}_{j_2k_3i_2l_3}\delta_{i_1i_2}\delta_{k_2j_2}\hat{p}_{j_1}\hat{p}_{l_3}\hat{s}_{l_1}\hat{s}_{k_3}=\frac{21\nu^2}{525}\left(3+\cos^2\theta\right).
\end{flalign}
The angle $\theta$ in Equation (\ref{LFOSADEF}) is the scattering angle between $\hat{\textbf{p}}$ and $\hat{\textbf{s}}$. The eighth rank covariance tensor needed to obtain the right-hand side of Equation (\ref{LFOSADEF}) is found in Appendix \ref{covTrans}. The remaining step to obtain the scaled self-energies is to complete the integration, which is described in the results section of Section \ref{SecResultsAll}.

\subsection{Second-order smoothing approximation}
\label{SOSA_Sec}
The SOSA involves the process of the scattering of an incident wave vector $\textbf{p}$ into wave vector $\textbf{a}$, which in turn scatters into the wave vector $\textbf{b}$, which scatters into $\textbf{c}$  [note that $\textbf{a}$ should not be confused with the rotation matrices seen in Equations (\ref{intGrain}) and (\ref{latingreekEq})]. The Fourier transform of the two SOSA contributions in Equations (\ref{mspc31}) and (\ref{mspc32}) reduces to a single expression for the self-energy involving three convolutions,
\begin{align}
\label{sigFOSAeq}
    \sigma_{i_1i_2}^{SOSA}(\textbf{p})=&-p_{j_1}p_{j_2}\int \int \int d^3\textbf{a}d^3\textbf{b}d^3\textbf{c} \Xi^{i_1j_1k_1l_1}_{i_3j_3k_3l_3}\Xi^{i_4j_4k_4l_4}_{i_2j_2k_2l_2}a_{k_1}a_{i_4}b_{i_3}b_{k_4}c_{k_2}c_{k_3}G_{l_1j_4}^0(\textbf{a})\tilde{\eta}\left(\textbf{p}-\textbf{a}\right)\notag\\
    \quad & G_{j_3l_4}^0(\textbf{b})\tilde{\eta}'\left(\textbf{p}-\textbf{b}\right) G_{l_2l_3}^0(\textbf{c})\tilde{\eta}\left(\textbf{p}-\textbf{c}\right),
\end{align}
 where $\tilde{\eta}'\left(\textbf{p}\right)$ comes from Fourier transform on the spatial TPC, $\eta'\left(\textbf{s}\right)=\eta\left(2\textbf{s}\right)$ following Equation (\ref{TPC_fosatransform}). It is noted that the reduction of the two SOSA terms into a single term in the wave vector domain is a result of the exponential TPC and the exponential identity $e^ae^b=e^{a+b}$. Other forms of TPC could require evaluation of both expressions in Equations (\ref{mspc31}) and (\ref{mspc32}). The components of the scaled self-energies follow from Equations (\ref{SigLFi}) and (\ref{SigTFi}),
\begin{align}
\label{SigTLMod1}
    \Sigma_{SOSA}^L(\textbf{p})=&-\frac{\hat{p}_{i_1}\hat{p}_{i_2}\hat{p}_{j_1}\hat{p}_{j_2}}{\rho c_{0L}^2}\int \int \int d^3\textbf{a}d^3\textbf{b}d^3\textbf{c} \Xi^{i_1j_1k_1l_1}_{i_3j_3k_3l_3}\Xi^{i_4j_4k_4l_4}_{i_2j_2k_2l_2}a_{k_1}a_{i_4}b_{i_3}b_{k_4}c_{k_2}c_{k_3}G_{l_1j_4}^0(\textbf{a})\notag\\
    \quad & \tilde{\eta}\left(\textbf{p}-\textbf{a}\right)G_{j_3l_4}^0(\textbf{b})\tilde{\eta}'\left(\textbf{p}-\textbf{b}\right) G_{l_2l_3}^0(\textbf{c})\tilde{\eta}\left(\textbf{p}-\textbf{c}\right),
\end{align}
and
\begin{align}
\label{SigTTMod1}
    \Sigma_{SOSA}^T(\textbf{p})=&-\frac{\left(\delta_{i_1i_2}-\hat{p}_{i_1}\hat{p}_{i_2}\right)\hat{p}_{j_1}\hat{p}_{j_2}}{2\rho c_{0T}^2} \int \int \int d^3\textbf{a}d^3\textbf{b}d^3\textbf{c} \Xi^{i_1j_1k_1l_1}_{i_3j_3k_3l_3}\Xi^{i_4j_4k_4l_4}_{i_2j_2k_2l_2}a_{k_1}a_{i_4}b_{i_3}b_{k_4}c_{k_2}c_{k_3}\notag\\
    \quad & G_{l_1j_4}^0(\textbf{a}) \tilde{\eta}\left(\textbf{p}-\textbf{a}\right)G_{j_3l_4}^0(\textbf{b})\tilde{\eta}'\left(\textbf{p}-\textbf{b}\right) G_{l_2l_3}^0(\textbf{c})\tilde{\eta}\left(\textbf{p}-\textbf{c}\right).
\end{align}
Moreover, $\textbf{G}^0$ shown in Equation (\ref{G_decomp}) decomposes the scaled self-energy into
\begin{align}
\label{SigLdefn}
\Sigma_{SOSA}^L=2\left[\Sigma_{SOSA}^{LLLL}+2\Sigma_{SOSA}^{LLLT}+\Sigma_{SOSA}^{LLTL}+2\Sigma_{SOSA}^{LLTT}+\Sigma_{SOSA}^{LTLT}+\Sigma_{SOSA}^{LTTT}\right],
\end{align}
and,
\begin{align}
\label{SigTdefn}
\Sigma^T_{SOSA}=\left[\Sigma_{SOSA}^{TLLL}+2\Sigma_{SOSA}^{TLLT}+\Sigma_{SOSA}^{TLTL}+2\Sigma_{SOSA}^{TLTT}+\Sigma_{SOSA}^{TTLT}+\Sigma_{SOSA}^{TTTT}\right],
\end{align}
where the superscripts in the SOSA-based self-energy components $\Sigma_{SOSA}^{PQUV}$ correspond to the scattering of wave vector $\textbf{p}$ of mode $P$ into scattered wave vectors $\textbf{a}$, $\textbf{b}$, and $\textbf{c}$ having the modes $Q, U,$ and $V$, respectively. Analytical expressions are of the form
\begin{align}
\label{SigmaFullT_LLLL}
\Sigma_{SOSA}^{PQUV} &=\int \int \int d^3\textbf{a}d^3\textbf{b}d^3\textbf{c}\mathcal{Q}^P_{SOSA}\left(\textbf{p},\textbf{a},\textbf{b},\textbf{c}\right)\Xi_{SOSA}^{PQUV}g_{0Q}\left(a\right)g_{0U}\left(b\right)g_{0V}\left(c\right),
\end{align}
where $\mathcal{Q}^P_{SOSA}\left(\textbf{p},\textbf{a},\textbf{b},\textbf{c}\right)$ is
\begin{align}
\label{QSOSA}
\mathcal{Q}^P_{SOSA}\left(\textbf{p},\textbf{a},\textbf{b},\textbf{c}\right)=-\frac{\ell^9}{\pi^6\rho c_{0P}^2}\left[\frac{\left|\textbf{a}\right|\left|\textbf{b}\right|\left|\textbf{c}\right|}{\left(1+\ell^2\left|\textbf{p}-\textbf{a}\right|^2\right)\left(4+\ell^2\left|\textbf{p}-\textbf{b}\right|^2\right)\left(1+\ell^2\left|\textbf{p}-\textbf{c}\right|^2\right)}\right]^2,
\end{align}
and the inner products denoted as $\Xi^{PQUV}_{SOSA}$ are 
\begin{flalign}
\label{SigmaFullT_LLLL2}
&\Xi_{SOSA}^{LLLL} =L,\notag\\
\quad
&\Xi_{SOSA}^{LLLT}=M_1-L,\notag\\
\quad&
\Xi_{SOSA}^{LLTL}=M_2-L,\notag\\
\quad&
 \Xi_{SOSA}^{LLTT}= M_3-M_2-M_1+L ,\notag\\
\quad
 & \Xi_{SOSA}^{LTLT}= M_4-2M_1+L,\notag\\
\quad 
&\Xi_{SOSA}^{LTTT}=M_5-2M_3-M_4+2M_1+M_2-L,\notag\\ \quad
&\Xi_{SOSA}^{TLLL}=N_1-L,\notag\\ \quad
&\Xi_{SOSA}^{TLLT}=N_2-N_1-M_1+L,\notag\\
\quad
&\Xi_{SOSA}^{TLTL}=N_3-N_1-M_2+L,\notag\\ \quad
&\Xi_{SOSA}^{TLTT}=N_4-N_3-N_2+N_1-M_3+M_2 +M_1-L,\notag\\ \quad
&\Xi_{SOSA}^{TTLT}=N_5-2N_2+N_1-M_4+2M_1-L,\notag\\ \quad
&\Xi_{SOSA}^{TTTT}=N_6-2N_4-N_5+2N_2+N_3-N_1-M_5+2M_3+M_4 -2M_1-M_2+L,
\end{flalign}
which are the complete contracted inner products involving two eighth rank tensors $\boldsymbol{\Xi}$ and up to sixteen unit wave vector directions, 
\begin{flalign}
\label{LSOSADef}
L&=\Xi^{i_1j_1k_1l_1}_{\alpha_1\beta_1\gamma_1\kappa_1}\Xi^{i_2j_2k_2l_2}_{\alpha_2\beta_2\gamma_2\kappa_2}\hat{p}_{i_1}\hat{p}_{j_1}\hat{p}_{i_2}\hat{p}_{j_2}\hat{a}_{k_1}\hat{a}_{l_1}\hat{a}_{\gamma_2}\hat{a}_{\kappa_2}\hat{b}_{\alpha_1}\hat{b}_{\beta_1}\hat{b}_{\alpha_2}\hat{b}_{\beta_2}\hat{c}_{\gamma_1}\hat{c}_{\kappa_1}\hat{c}_{k_2}\hat{c}_{l_2},
\notag\\
\quad
M_1&=\Xi^{i_1j_1k_1l_1}_{\alpha_1\beta_1\gamma_1\kappa_1}\Xi^{i_2j_2k_2l_2}_{\alpha_2\beta_2\gamma_2\kappa_2}\hat{p}_{i_1}\hat{p}_{j_1}\hat{p}_{i_2}\hat{p}_{j_2}\hat{a}_{k_1}\hat{a}_{l_1}\hat{a}_{\gamma_2}\hat{a}_{\kappa_2}\hat{b}_{\alpha_1}\hat{b}_{\beta_1}\hat{b}_{\alpha_2}\hat{b}_{\beta_2}\hat{c}_{\kappa_1}\hat{c}_{l_2}\delta_{k_2\gamma_1},
\notag\\
\quad
M_2&=\Xi^{i_1j_1k_1l_1}_{\alpha_1\beta_1\gamma_1\kappa_1}\Xi^{i_2j_2k_2l_2}_{\alpha_2\beta_2\gamma_2\kappa_2}\hat{p}_{i_1}\hat{p}_{j_1}\hat{p}_{i_2}\hat{p}_{j_2}\hat{a}_{k_1}\hat{a}_{l_1}\hat{a}_{\gamma_2}\hat{a}_{\kappa_2}\hat{b}_{\beta_1}\hat{b}_{\beta_2}\hat{c}_{\kappa_1}\hat{c}_{l_2}\hat{c}_{k_2}\hat{c}_{\gamma_1}\delta_{\alpha_2\alpha_1},
\notag\\
\quad
M_3&=\Xi^{i_1j_1k_1l_1}_{\alpha_1\beta_1\gamma_1\kappa_1}\Xi^{i_2j_2k_2l_2}_{\alpha_2\beta_2\gamma_2\kappa_2}\hat{p}_{i_1}\hat{p}_{j_1}\hat{p}_{i_2}\hat{p}_{j_2}\hat{a}_{k_1}\hat{a}_{l_1}\hat{a}_{\gamma_2}\hat{a}_{\kappa_2}\hat{b}_{\beta_1}\hat{b}_{\beta_2}\hat{c}_{\kappa_1}\hat{c}_{l_2}\delta_{k_2\gamma_1}\delta_{\alpha_2\alpha_1},
\notag\\
\quad M_4&=\Xi^{i_1j_1k_1l_1}_{\alpha_1\beta_1\gamma_1\kappa_1}\Xi^{i_2j_2k_2l_2}_{\alpha_2\beta_2\gamma_2\kappa_2}\hat{p}_{i_1}\hat{p}_{j_1}\hat{p}_{i_2}\hat{p}_{j_2}\hat{a}_{l_1}\hat{a}_{\kappa_2}\hat{b}_{\alpha_1}\hat{b}_{\beta_1}\hat{b}_{\alpha_2}\hat{b}_{\beta_2}\hat{c}_{\kappa_1}\hat{c}_{l_2}\delta_{k_2\gamma_1}\delta_{k_1\gamma_2},
\notag\\
\quad
M_5&=\Xi^{i_1j_1k_1l_1}_{\alpha_1\beta_1\gamma_1\kappa_1}\Xi^{i_2j_2k_2l_2}_{\alpha_2\beta_2\gamma_2\kappa_2}\hat{p}_{i_1}\hat{p}_{j_1}\hat{p}_{i_2}\hat{p}_{j_2}\hat{a}_{l_1}\hat{a}_{\kappa_2}\hat{b}_{\beta_1}\hat{b}_{\beta_2}\hat{c}_{\kappa_1}\hat{c}_{l_2}\delta_{k_2\gamma_1}\delta_{k_1\gamma_2}\delta_{\alpha_2\alpha_1},\notag\\
\quad
N_1&=\Xi^{i_1j_1k_1l_1}_{\alpha_1\beta_1\gamma_1\kappa_1}\Xi^{i_2j_2k_2l_2}_{\alpha_2\beta_2\gamma_2\kappa_2}\hat{p}_{j_1}\hat{p}_{j_2}\hat{a}_{k_1}\hat{a}_{l_1}\hat{a}_{\gamma_2}\hat{a}_{\kappa_2}\hat{b}_{\alpha_1}\hat{b}_{\beta_1}\hat{b}_{\alpha_2}\hat{b}_{\beta_2}\hat{c}_{\gamma_1}\hat{c}_{\kappa_1}\hat{c}_{k_2}\hat{c}_{l_2}\delta_{i_1i_2},
\notag\\
\quad
N_2&=\Xi^{i_1j_1k_1l_1}_{\alpha_1\beta_1\gamma_1\kappa_1}\Xi^{i_2j_2k_2l_2}_{\alpha_2\beta_2\gamma_2\kappa_2}\hat{p}_{j_1}\hat{p}_{j_2}\hat{a}_{k_1}\hat{a}_{l_1}\hat{a}_{\gamma_2}\hat{a}_{\kappa_2}\hat{b}_{\alpha_1}\hat{b}_{\beta_1}\hat{b}_{\alpha_2}\hat{b}_{\beta_2}\hat{c}_{\kappa_1}\hat{c}_{l_2}\delta_{i_1i_2}\delta_{k_2\gamma_1},
\notag\\
\quad
N_3&=\Xi^{i_1j_1k_1l_1}_{\alpha_1\beta_1\gamma_1\kappa_1}\Xi^{i_2j_2k_2l_2}_{\alpha_2\beta_2\gamma_2\kappa_2}\hat{p}_{j_1}\hat{p}_{j_2}\hat{a}_{k_1}\hat{a}_{l_1}\hat{a}_{\gamma_2}\hat{a}_{\kappa_2}\hat{b}_{\beta_1}\hat{b}_{\beta_2}\hat{c}_{\kappa_1}\hat{c}_{l_2}\hat{c}_{k_2}\hat{c}_{\gamma_1}\delta_{i_1i_2}\delta_{\alpha_2\alpha_1},
\notag\\
\quad
N_4&=\Xi^{i_1j_1k_1l_1}_{\alpha_1\beta_1\gamma_1\kappa_1}\Xi^{i_2j_2k_2l_2}_{\alpha_2\beta_2\gamma_2\kappa_2}\hat{p}_{j_1}\hat{p}_{j_2}\hat{a}_{k_1}\hat{a}_{l_1}\hat{a}_{\gamma_2}\hat{a}_{\kappa_2}\hat{b}_{\beta_1}\hat{b}_{\beta_2}\hat{c}_{\kappa_1}\hat{c}_{l_2}\delta_{i_1i_2}\delta_{k_2\gamma_1}\delta_{\alpha_2\alpha_1},
\notag\\
\quad N_5&=\Xi^{i_1j_1k_1l_1}_{\alpha_1\beta_1\gamma_1\kappa_1}\Xi^{i_2j_2k_2l_2}_{\alpha_2\beta_2\gamma_2\kappa_2}\hat{p}_{j_1}\hat{p}_{j_2}\hat{a}_{l_1}\hat{a}_{\kappa_2}\hat{b}_{\alpha_1}\hat{b}_{\beta_1}\hat{b}_{\alpha_2}\hat{b}_{\beta_2}\hat{c}_{\kappa_1}\hat{c}_{l_2}\delta_{i_1i_2}\delta_{k_2\gamma_1}\delta_{k_1\gamma_2},
\notag\\
\quad
N_6&=\Xi^{i_1j_1k_1l_1}_{\alpha_1\beta_1\gamma_1\kappa_1}\Xi^{i_2j_2k_2l_2}_{\alpha_2\beta_2\gamma_2\kappa_2}\hat{p}_{j_1}\hat{p}_{j_2}\hat{a}_{l_1}\hat{a}_{\kappa_2}\hat{b}_{\beta_1}\hat{b}_{\beta_2}\hat{c}_{\kappa_1}\hat{c}_{l_2}\delta_{i_1i_2}\delta_{k_2\gamma_1}\delta_{k_1\gamma_2}\delta_{\alpha_2\alpha_1}.
\end{flalign}
The SOSA includes higher-order scattering effects in polycrystals by including the two equally contributing terms \cite{roy2023elastic} for exponential TPC functions. Thus, the total scaled self-energies within the SOSA are
\begin{flalign}
\label{massopCur}
\Sigma^L&=\Sigma^L_{FOSA}+2\Sigma^L_{SOSA},\\
    \label{massopCurT}\quad \Sigma^T&=\Sigma^T_{FOSA}+2\Sigma^T_{SOSA},
\end{flalign}
which can be used with Equations (\ref{reX}) and (\ref{imX}) to find the real and imaginary parts of the effective wavenumbers and, therefore, the attenuations and phase velocity dispersion as seen in Equations (\ref{att_DEFL}) and (\ref{ws_DEFT}). Section \ref{SecResultsAllSOSA} describes the integration steps for the scaled self-energies in SOSA, followed by evaluations for two specific polycrystalline materials, iron and lithium.

\section{Results}
\label{resultsSec}

Results in this section will include the evaluation of FOSA in Section \ref{SecResultsAll} and SOSA in Section \ref{SecResultsAllSOSA} for polycrystalline iron and lithium. The evaluations will be performed for all frequencies spanning the low-frequency Rayleigh regime to the geometric scattering regime to be defined. As the stochastic and geometric scattering regimes are to be used in the foregoing discussion, we note the definitions of the three scattering regimes. To do so, we note the perturbation parameters known as degrees of inhomogeneity \cite{stanke1984unified}, 
\begin{align}
\label{epLeq}
\epsilon_L=\frac{1}{C^0_{11}}\sqrt{\frac{4\nu^2}{525}},\end{align}
\begin{align}
\label{epTeq}
\epsilon_T=\frac{1}{C^0_{44}}\sqrt{\frac{3\nu^2}{700}}.
\end{align}
The degree of inhomogeneity clearly depends on the elastic anisotropy factor $\nu=C_{11}-C_{12}-2C_{44}$, suggesting that the degree of inhomogeneity increases with increases in anisotropy of the independent stiffness constants inherent to the crystal structure of the material (cubic structure in the present case). As a reminder, $C_{11}^0=C_{12}+2C_{44}+3\nu/5$ and $C_{44}^0=C_{44}+\nu/5$. Furthermore, the degree of inhomogeneity also corresponds to scattering strength in the polycrystal. Then, the three scattering regimes are

1. the \textit{Rayleigh} regime, for $p_0\ell <1/2$,

2. the \textit{stochastic} regime, for $1/2\leq p_0\ell<1/2\epsilon$, and,

3. the \textit{geometric} regime, for $p_0\ell\geq 1/2\epsilon$,

where $p_{0}\ell$ are normalized wavenumbers. $p_{0}$ is proportional to frequency and, thus, the quantity $p_0\ell$ describes both frequency and correlation length $\ell$ (which is on the order of the grain size). In Secs. \ref{SecResultsAll} and \ref{SecResultsAllSOSA}, results will be provided at \textit{all frequencies} that at least span the Rayleigh, stochastic, and geometric regions as defined above. Of course, a minimum $p_0\ell$ and maximum $p_0\ell$ are required in evaluation. The minimum $p_0\ell$ will be shown to be asymptotic (Rayleigh limit), and convergence between FOSA and SOSA will be shown. At large $p_0\ell$ in the geometric regime, we truncate results where we felt the numerical solver for $p$ becomes unstable while still going well beyond $p_0\ell=1/2\epsilon$ at the start of the geometric scattering regime. 
\subsection{First-order smoothing approximation}
\label{SecResultsAll}
The FOSA for describing elastic wave propagation in polycrystals has been evaluated within the Born approximation by Weaver \cite{weaver1990diffusivity}. The Born approximation is made when the mean Green's function $\langle\textbf{G}(\textbf{p})\rangle$ is replaced by $\textbf{G}^0(\textbf{p}_0)$ on the right-hand side of Equation (\ref{DysonP}). Note that the wave vector and wave number components of the non-scattering reference medium are used within $\textbf{G}^0(\textbf{p}_0)$ while retaining the diagonal form of Equation (\ref{G_decomp}). While this approximation can be made in Equation (\ref{G_decomp}), Weaver applies the Born approximation directly within the dispersion equations for the squared effective wavenumbers by letting $\Sigma(\textbf{p}=\textbf{p}_0)$. The Born approximation from Ref. \cite{weaver1990diffusivity} predicts attenuations that closely follow predictions from UT of Stanke and Kino \cite{stanke1984unified} to within experimental resolution for moderate frequencies and begin to diverge in the stochastic scattering regime. 
\par
\begin{figure}[H]
\includegraphics[scale=0.58]{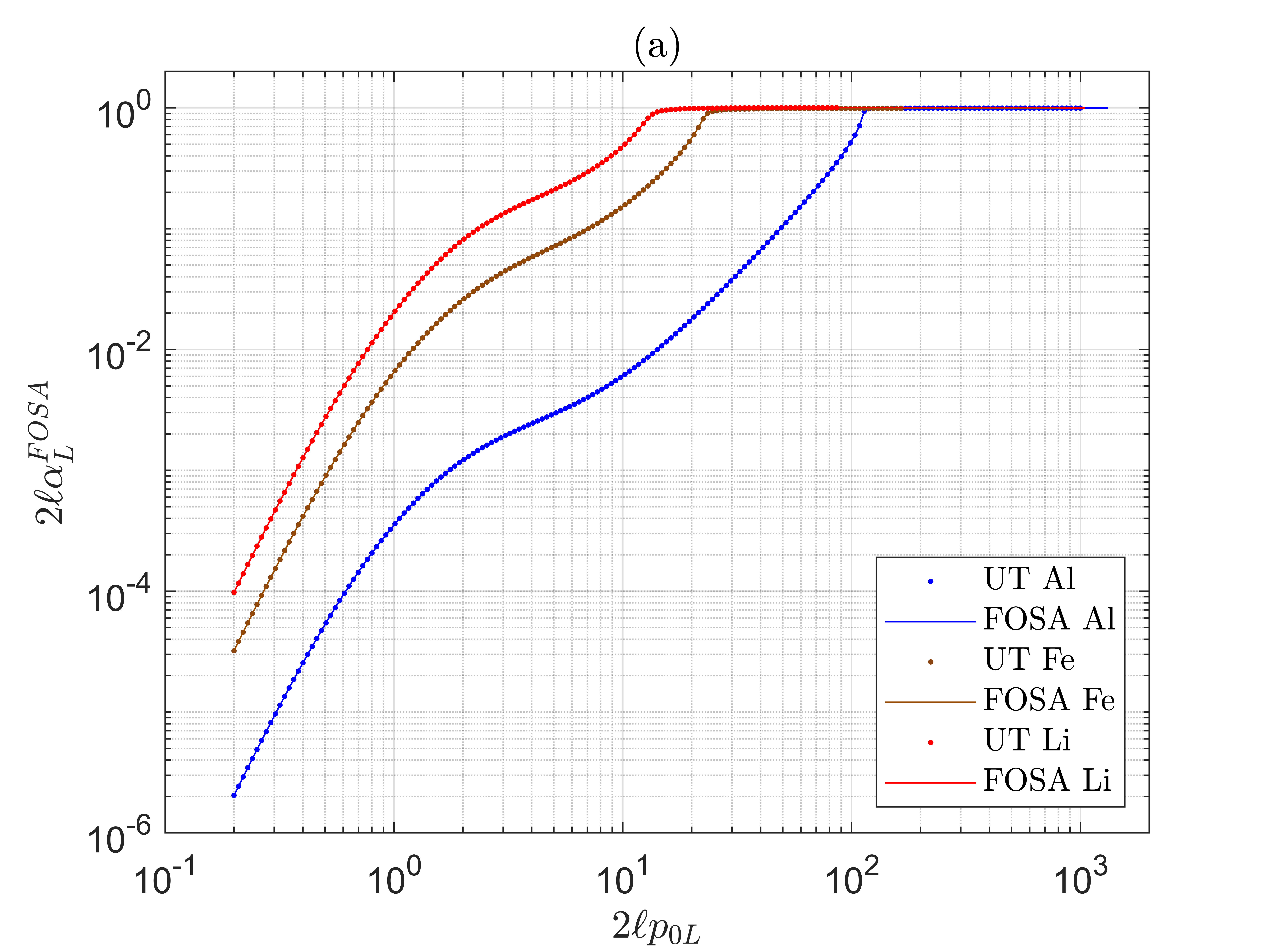}
\includegraphics[scale=0.58]{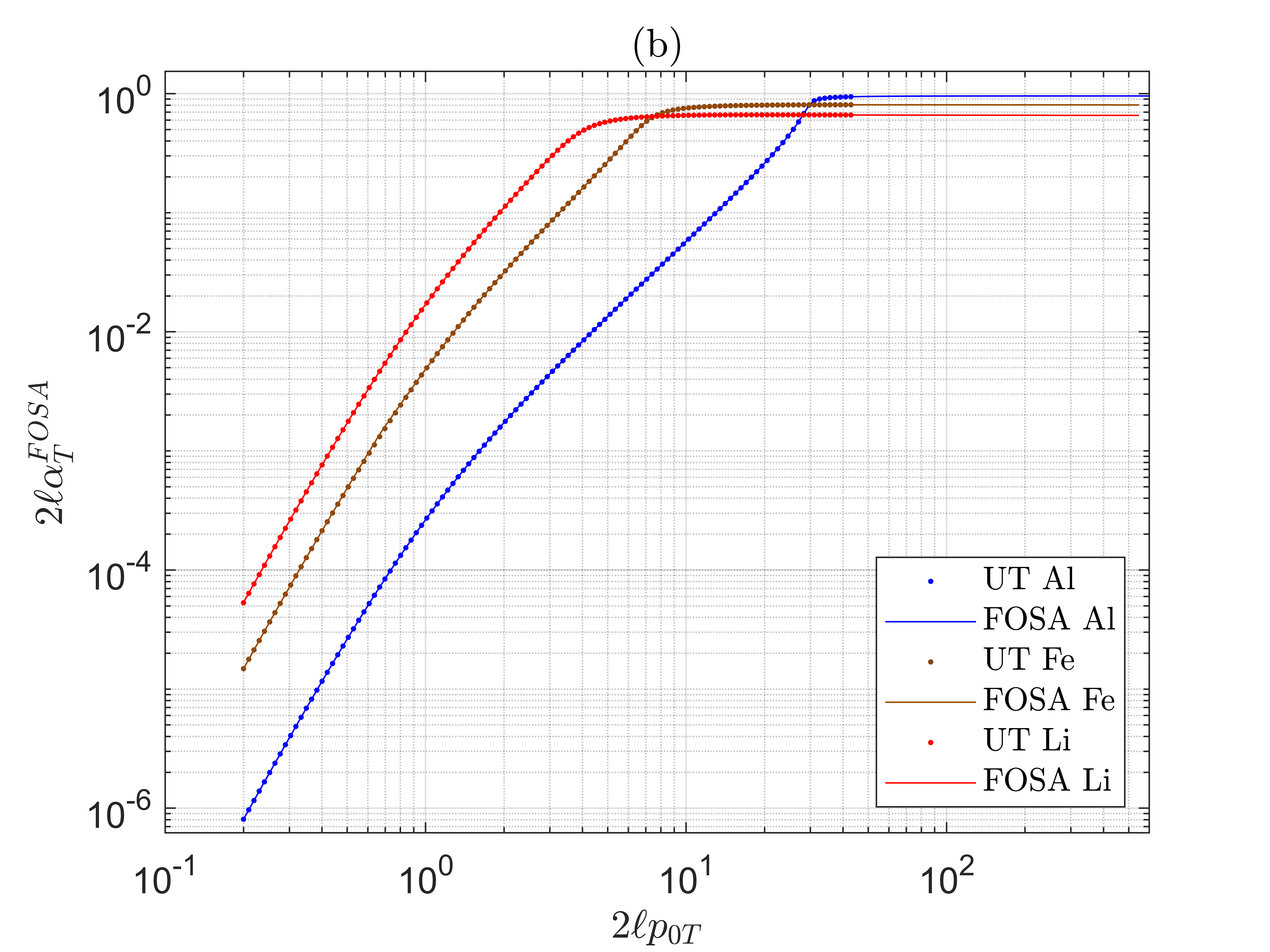}
\includegraphics[scale=0.58]{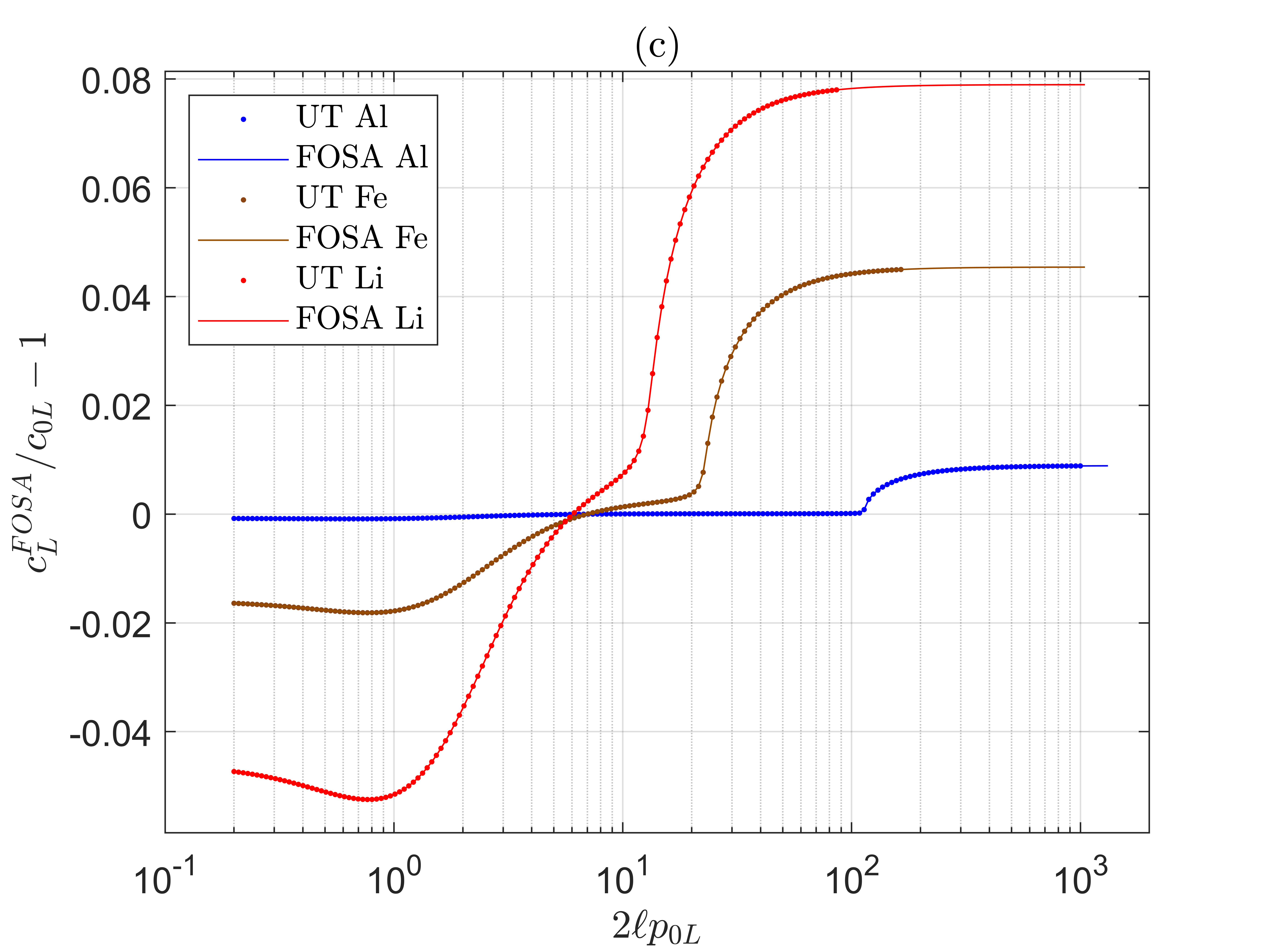}
\includegraphics[scale=0.58]{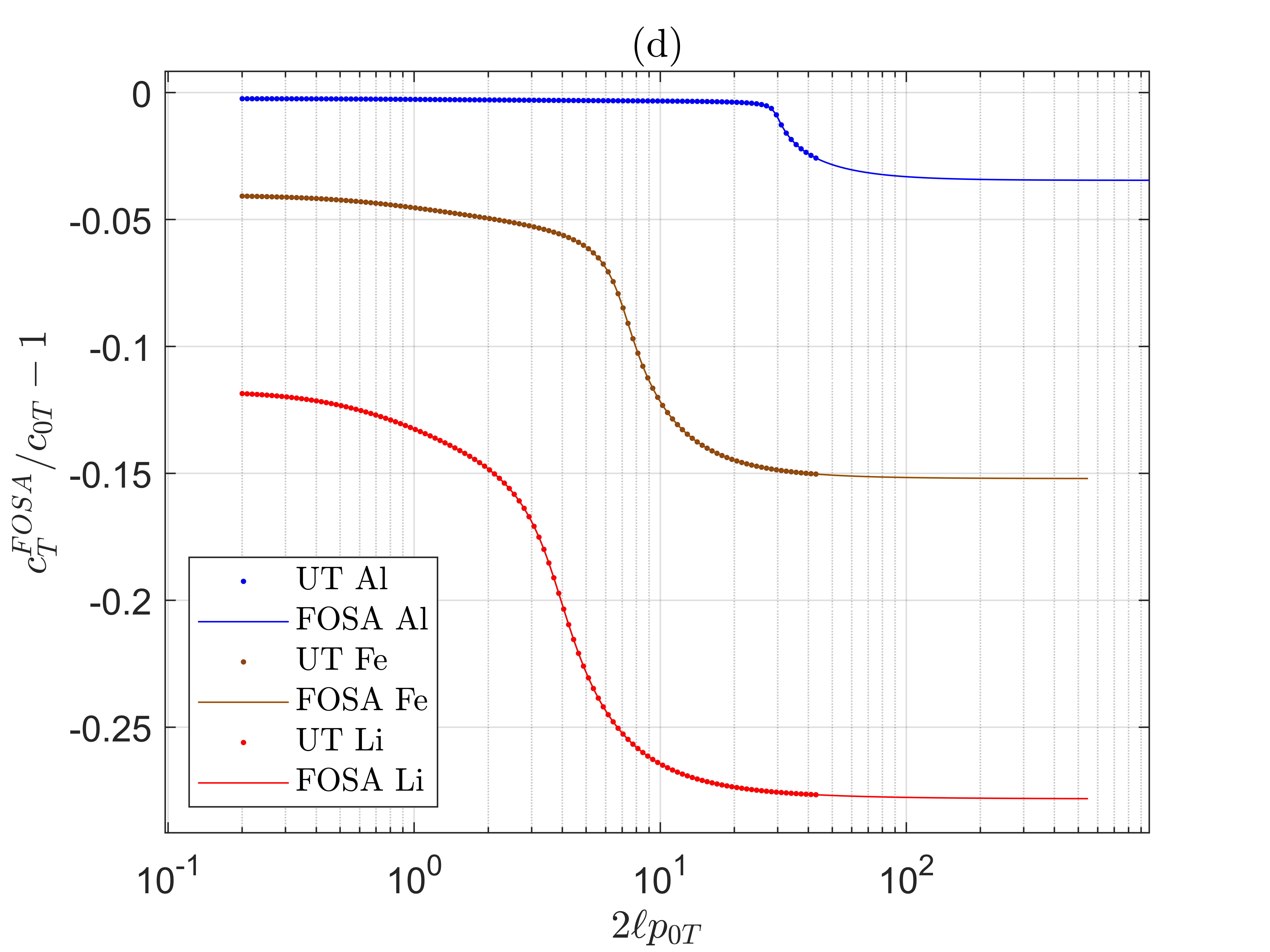}
\caption{Equivalence of the UT of Stanke and Kino and FOSA for all frequencies in the three scattering regimes in which the Rayleigh and geometric scattering display asymptotic behavior at low and high $2\ell p_0$ values, with the stochastic regime being between. Results are shown for low, moderate, and strongly scattering materials of aluminum, iron, and lithium, respectively.}
\label{fig2}
\end{figure}

 While Refs. \cite{weaver1990diffusivity} and \cite{stanke1984unified} provide the exact same dispersion equations (see Equation 94 from \cite{stanke1984unified} and Equation 3.15 from \cite{weaver1990diffusivity}), the FOSA from Ref. \cite{weaver1990diffusivity} has not been evaluated at all frequencies outside of the Born approximation. The present article aims at evaluating FOSA and SOSA at all frequencies because the importance of multiple scattering is most impactful at higher frequency ranges likely beyond where the Born approximation fails. To this end, this section describes the evaluation of FOSA for all frequencies and shows the exact equivalence to the UT model of Stanke and Kino \cite{stanke1984unified}, with results shown in Fig. \ref{fig2}.
\par To evaluate FOSA, we must evaluate the integrations of the scaled self-energies seen in Equation (\ref{GenSigLLFOSA}). It is convenient to transform the integrands of Equation (\ref{GenSigLLFOSA}) to incorporate normalized wavenumbers $x=p\ell$, $x_s=s\ell$ where $p$ is the effective wave number and $s$ is the wave number that is part of the integration domain. Furthermore, Equation (\ref{GenSigLLFOSA}) is evaluated in spherical coordinates over the domain $s$, $\varphi$, and $\theta$. Additionally, it is convenient to introduce $\mu=\cos\theta$ to handle the angular integration. As the domain is statistically isotropic, the integration over the angle $\varphi$ is trivial and provides a $2\pi$ contribution. With these considerations, Equation (\ref{GenSigLLFOSA}) can be written as
\begin{align}
\label{GenSigNRFOSA}
    \Sigma^{PQ}_{FOSA}(\textbf{p})=\frac{1}{\pi\rho^2 c_{0P}^2c_{0Q}^2}\int \limits_{-1}^{1} \int \limits_0^{\infty} dx_s d\mu \frac{\Xi^{PQ}_{FOSA}\left(\mu \right) x_s^4\left(x_s^2-x^2_{0Q}\right)^{-1}}{\left(1+x_P^2+x_s^2-2x_P x_s\mu\right)^2},
\end{align}
where $x_{0Q}=p_{0Q}\ell$, and $x_P=p_P\ell$ are normalized wavenumbers that include the indicators where $P,Q$ can represent either longitudinal ($L$) or transverse ($T$) modes. The indicators $P,Q$ are also used within the phase velocities of the reference medium $c_{0P}$ and $c_{0Q}$. In Equation (\ref{GenSigNRFOSA}), the inner products $\Xi_{FOSA}^{PQ}$ are functions of even powers of $\mu=\cos\theta$ as seen in Equation (\ref{LFOSADEF}). The following four scaled self-energies can be obtained from Equation (\ref{GenSigNRFOSA}), 
\begin{align}
 \label{sec5sigLL}
    \Sigma_{FOSA}^{LL}\left(\textbf{p}\right)=\frac{\nu^2}{525\pi\rho^2 c_{0L}^4}\left(9\mathcal{Z}^{LL}_0+6\mathcal{Z}^{LL}_2+\mathcal{Z}^{LL}_4\right),
\end{align}
\begin{align}
    \Sigma_{FOSA}^{LT}\left(\textbf{p}\right)=\frac{\nu^2}{525\pi\rho^2 c_{0L}^2c_{0T}^2}\left(15\mathcal{Z}^{LT}_0+6\mathcal{Z}^{LT}_2-\mathcal{Z}^{LT}_4\right),
\end{align}
\begin{align}
    \Sigma_{FOSA}^{TL}\left(\textbf{p}\right)=\frac{\nu^2}{525\pi\rho^2 c_{0L}^2c_{0T}^2}\left(15\mathcal{Z}^{TL}_0+6\mathcal{Z}^{TL}_2-\mathcal{Z}^{TL}_4\right),
\end{align}
\begin{align}
    \label{sec5sigTT}
    \Sigma_{FOSA}^{TT}\left(\textbf{p}\right)=\frac{\nu^2}{525\pi\rho^2 c_{0T}^4}\left(24\mathcal{Z}^{TT}_0+3\mathcal{Z}^{TT}_2+\mathcal{Z}^{TT}_4\right),
\end{align}
where
\begin{align}
\label{ZMassF}
\mathcal{Z}^{PQ}_n=\left[\Re\left(\mathcal{R}_n^{PQ}\right)-\Im\left(\mathcal{I}_n^{PQ}\right)\right]+i\left[\Im\left(\mathcal{R}_n^{PQ}\right)+\Re\left(\mathcal{I}_n^{PQ}\right)\right],
\end{align} 
and $n$ is 0, 2, or 4 as seen in Equations (\ref{sec5sigLL}-\ref{sec5sigTT}). Inside Equation (\ref{ZMassF}) are the real and imaginary parts of the integrals $\mathcal{R}$ and $\mathcal{I}$, which arrive from the real and imaginary parts of $g^0$ as seen in Equations (\ref{Reg0}) and (\ref{ImG0MDef}), respectively. The $\mathcal{R}$ integrals involve the Cauchy principle value, $\mathcal{P}^{PQ}$, 
\begin{align}
\label{SigRs}
   \mathcal{R}^{PQ}_{n}=\int \limits_{-1}^{1}d\mu\mathcal{P}^{PQ}\left(x_P,x_{0Q},\mu\right)\mu^{n}.
\end{align}
Specifically, the Cauchy principal value integral, provided in Appendix \ref{appCont}, is
\begin{align}
\label{SigPu}
   \mathcal{P}^{PQ}&=\int \limits_0^{\infty} \frac{x_s^4  dx_s}{\left(x_s - z_P\right)^2\left(x_s - \bar{z}_P\right)^2\left(x_s^2-x_{0Q}^2\right)},
\end{align}

where the conjugates $z_P=x_P\mu+i\sqrt{1+x_P^2(1-\mu^2)}$ and $\bar{z}_P=x_P\mu-i\sqrt{1+x_P^2(1-\mu^2)}$ are used.

The $\mathcal{I}$ integrals can be evaluated analytically,
\begin{align}
\label{SigIs0}
\mathcal{I}^{PQ}_0=\frac{\pi x_{0Q}}{8x_{P}^2}  \left(\frac{2}{\phi_{PQ}^2 - 1}\right),
\end{align}
\begin{align}
\label{SigIs2}
\mathcal{I}^{PQ}_2=\frac{\pi x_{0Q}}{8x_{P}^2} \left[\frac{2 \phi_{PQ}^2}{\phi_{PQ}^2 - 1}-2\phi_{PQ}\log{\left(\frac{\phi_{PQ} + 1}{\phi_{PQ}- 1}\right)}+2\right],
\end{align}
and,
\begin{align}
\label{SigIs4}
\mathcal{I}^{PQ}_4=\frac{\pi x_{0Q}}{8x_{P}^2} \left[\frac{2\phi_{PQ}^4}{\phi_{PQ}^2 - 1}-4\phi_{PQ}^3\log{\left(\frac{\phi_{PQ} + 1}{\phi_{PQ}-1}\right)}+6\phi_{PQ}^2+\frac{2}{3}\right],
\end{align}
where, unlike Weaver's model \cite{weaver1990diffusivity}, the $\phi_{PQ}$ is complex-valued and is,
\begin{align}
\label{phiDef}
\phi_{PQ}=\frac{1+x_{P}^2+x_{0Q}^2}{2x_{P}x_{0Q}}.
\end{align}

\par Once Equations (\ref{sec5sigLL}-\ref{sec5sigTT}) are evaluated, their real and imaginary parts can be utilized in Equations (\ref{reX}) and (\ref{imX}) to find the attenuations and phase velocity dispersion seen in Equations (\ref{att_DEFL}) and (\ref{ws_DEFT}), respectively. This evaluation procedure was performed, and the results were evaluated for polycrystalline aluminum, iron, and lithium. The material parameters used are seen in Table \ref{TabMat}.

Figure \ref{fig2} shows the equivalence of the attenuations and phase velocity dispersion between solutions arriving from FOSA and the UT of Stanke and Kino \cite{stanke1984unified}. The results from the UT utilize an iterative numerical solver \cite{kube2017iterative} for Equations (101) and (102) of Stanke and Kino \cite{stanke1984unified}. Now that the equivalence of the two models is established, we will turn to SOSA in Section \ref{SecResultsAllSOSA}, in which the behavior of the attenuation and phase velocity curves will be analyzed closely.

\subsection{Second-order smoothing approximation}
\label{SecResultsAllSOSA}

The evaluation of attenuation and phase velocity dispersion within the SOSA follows the same procedure as FOSA and requires integrations of similar functional forms. For SOSA, our first objective is to evaluate the needed integrations required in the scaled-self energies as seen in Equations (\ref{SigLdefn}) - (\ref{SigmaFullT_LLLL}), which are specific to SOSA. Once obtained, the scaled self-energies for FOSA and SOSA are combined as seen in Equations (\ref{massopCur}) and (\ref{massopCurT}). The resulting longitudinal and transverse scaled self-energies from Equations (\ref{massopCur}) and (\ref{massopCurT}) can then be used to calculate attenuations and phase velocity dispersion as described in Equations (\ref{att_DEFL})-(\ref{imX}).

While the scaled self-energies seen in FOSA require integration over a single scattered wave vector $\textbf{s}$, SOSA has integrations over the wave vectors $\textbf{a}$, $\textbf{b}$, and $\textbf{c}$ as seen in Equation (\ref{SigmaFullT_LLLL}). The three wave vectors are a result of the multiple scattering process in which an incident wave with wave vector $\textbf{p}$ scatters into the wave vector $\textbf{c}$, then into $\textbf{b}$, and finally $\textbf{a}$. All wave vectors in this process are complex, and the wave vector $\textbf{p}$ can be interpreted as the primary wave for which the effective wave number $p$ contains the desired attenuation and phase velocity quantities of the polycrystal. For SOSA, the scaled self-energies can be written in compact form,
\begin{align}
\label{SOSAMod}
\Sigma_{SOSA}^{PQUV}\left(\textbf{p}\right) = \frac{1}{4\pi^3}\left(\frac{\nu^2}{525\rho^2c_{0P}c_{0Q}c_{0U}c_{0V}}\right)^2\sum^{4}_{n_1=0,2}\sum^{4}_{n_2=0,2}\sum^{4}_{n_3=0,2} \Omega^{PQUV}_{n_1n_2n_3}\mathcal{Z}_{n_1}^{PQ}\left(a\right)\mathcal{Z}_{n_2}^{PU}\left(b\right)\mathcal{Z}_{n_3}^{PV}\left(c\right),
\end{align}

where the $\mathcal{Z}$ terms are integrations in the same form as in Equation (\ref{ZMassF}), except for $\mathcal{Z}(b,\mu_2)$ which involves the TPC function of the form $\eta'(\textbf{b})=\eta(2\textbf{b})$ as described after Equation (\ref{sigFOSAeq}) and results in the middle denominator term seen in Equation (\ref{QSOSA}). For this specific term, the poles are $z^b_P=x_P\mu+i\sqrt{4+x_P^2(1-\mu^2)}$ and $\bar{z}^b_P=x_P\mu-i\sqrt{4+x_P^2(1-\mu^2)}$ and $\phi_{PQ}^b$ is
\begin{align}
\label{phiDefb}
\phi^{b}_{PQ}=\frac{4+x_{P}^2+x_{0Q}^2}{2x_{P}x_{0Q}}.
\end{align}
The other difference in the scaled self-energies seen in Equation (\ref{SOSAMod}) is in how the azimuthal angles are handled in the spherical coordinate integration. In FOSA, the azimuthal angle $\varphi$ contributed a trivial factor of $2\pi$ because the scattering into the single direction $\hat{\textbf{s}}$ only requires evaluation in a single plane over a single scattering angle $\theta$ where $\mu=\cos\theta$. However, in SOSA all six angular variables in the three spherical coordinate integrations must be evaluated. With that said, the three integrations over $\varphi_1$ (for $\textbf{c}$), $\varphi_2$ (for $\textbf{b}$), and $\varphi_3$ (for $\textbf{a}$) are pre-integrated and do not appear in Equation (\ref{SOSAMod}). The integrations over $\varphi_1$, $\varphi_2$, and $\varphi_3$ are directly obtained from integrating all of the individual inner products seen in Equation (\ref{LSOSADef}) over the respective  $\varphi_1$, $\varphi_2$, and $\varphi_3$ each spanning from 0 to $2\pi$ and substituting the result into the terms in Equation (\ref{SigmaFullT_LLLL2}). The integrals involving a product of four unit vectors can be written in tensorial form,
\begin{align}
\label{fourasphi}
&\int_0^{2\pi}d\varphi\hat{a}_i\hat{a}_j\hat{a}_k\hat{a}_l=\frac{\pi}{4}\left(1-\mu_3^2\right)\left(\delta_{ij}\delta_{kl}+\delta_{ik}\delta_{jl}+\delta_{il}\delta_{jk}\right)\notag\\
\quad &-\frac{\pi}{4}\left(1-6\mu_3^2+5\mu_3^4\right)\left(\delta_{ij}\delta_{k3}\delta_{l3}+\delta_{kl}\delta_{i3}\delta_{j3}+\delta_{ik}\delta_{j3}\delta_{l3}+\delta_{jk}\delta_{i3}\delta_{l3}+\delta_{il}\delta_{j3}\delta_{k3}+\delta_{jl}\delta_{i3}\delta_{k3}\right)\notag\\
\quad &+\frac{\pi}{4}\left(3-30\mu_3^2+35\mu_3^4\right)\delta_{i3}\delta_{j3}\delta_{k3}\delta_{l3},
\end{align}
and an integral of a product of two unit vectors can be written as,
\begin{align}
\label{twoasphi}
    \int_0^{2\pi}d\varphi\hat{a}_i\hat{a}_j = \pi\left(1-\mu_3^2\right)\delta_{ij} - \pi\left(1-3\mu_3^2\right)\delta_{i3}\delta_{j3},
\end{align}
The same procedure can be used for integrating the products of any four unit vectors over the azimuthal angle from 0 to $2\pi$ in spherical coordinates. For integrating products of $\hat{\textbf{b}}$, we will have the same tensors as Equations (\ref{fourasphi}) and (\ref{twoasphi}) with the replacement of $\mu_3$ with $\mu_2$. For $\hat{\textbf{c}}$, we would use $\mu_1$ instead of $\mu_3$ in Equation (\ref{fourasphi}) and (\ref{twoasphi}).
Since these specific integrations are performed analytically, Equation (\ref{SOSAMod}) is an expression involving six integrals over the complex wavenumbers $a$, $b$, and $c$ and their corresponding angles $\mu_3$, $\mu_2$, and $\mu_1$, respectively. Like in FOSA, each $\mathcal{Z}$ factor in Equation (\ref{SOSAMod}) contains two integrations. Furthermore, the values of $n_1$, $n_2$, and $n_3$ can each take the values of 0, 2, or 4. These values correspond to $\mu_1$, $\mu_2$, or $\mu_3$ integrals that involve powers of $n_3$, $n_2$, or $n_1$ respectively following Equation (\ref{SigRs}). The coefficients in $\Omega_{n_1n_2n_3}^{PQUV}$ are tabulated in Appendix \ref{coeffTable}. 

Table \ref{tabAllCoeffs} in Appendix \ref{coeffTable} and the integration procedure described in Equations (\ref{ZMassF})-(\ref{phiDef}) enable the computation of inner products in Equation (\ref{SigmaFullT_LLLL2}) and then the scaled self-energies seen in Equations (\ref{massopCur}) and (\ref{massopCurT}), including both FOSA and SOSA scattering contributions. Finally, with the scaled self-energies, Equations (\ref{att_DEFL})-(\ref{imX}) may be used to calculate the attenuations and phase velocity dispersion relations influenced by the scattering in the polycrystal.

\begin{figure}[t]
\includegraphics[scale=0.46]{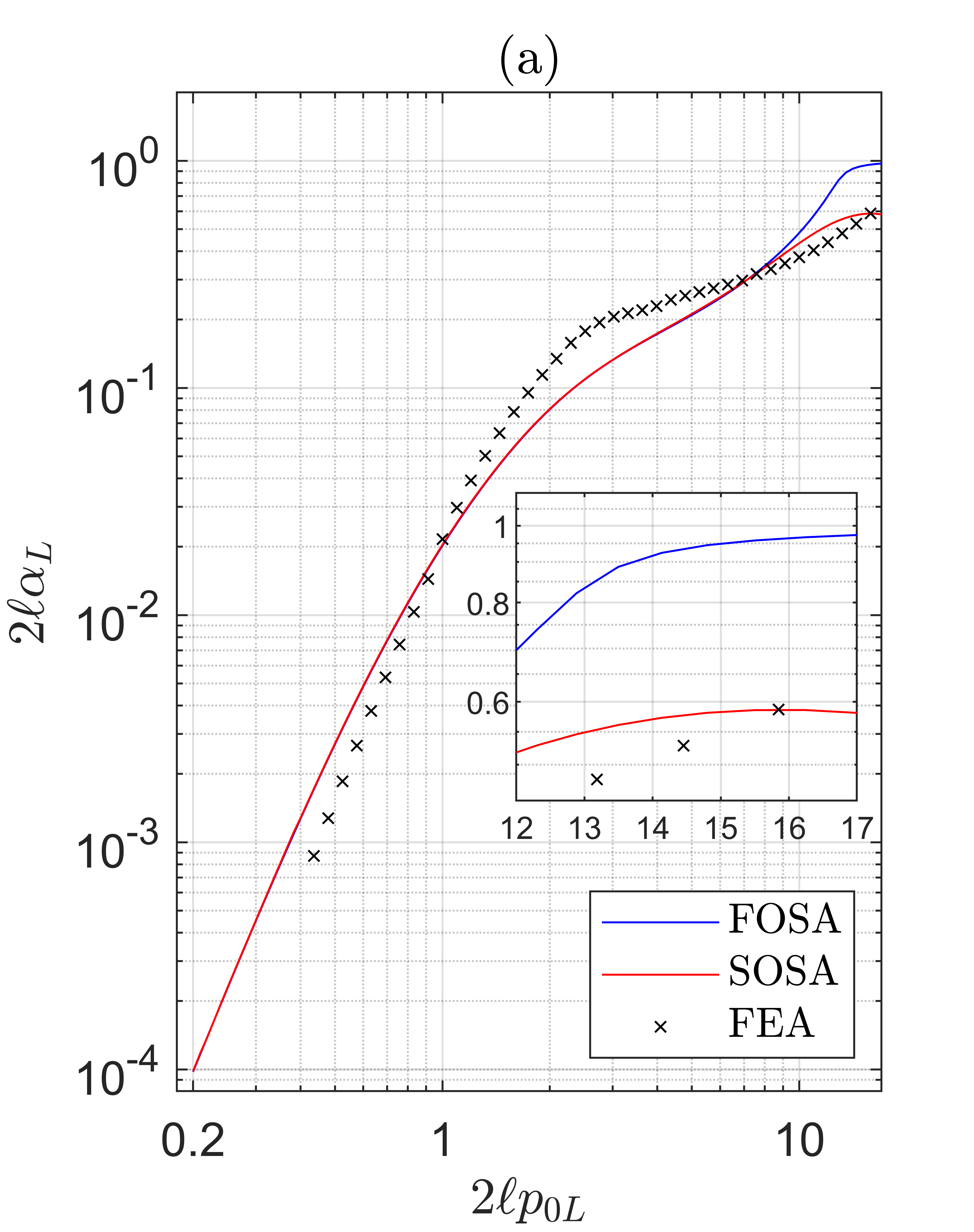}
\includegraphics[scale=0.46]{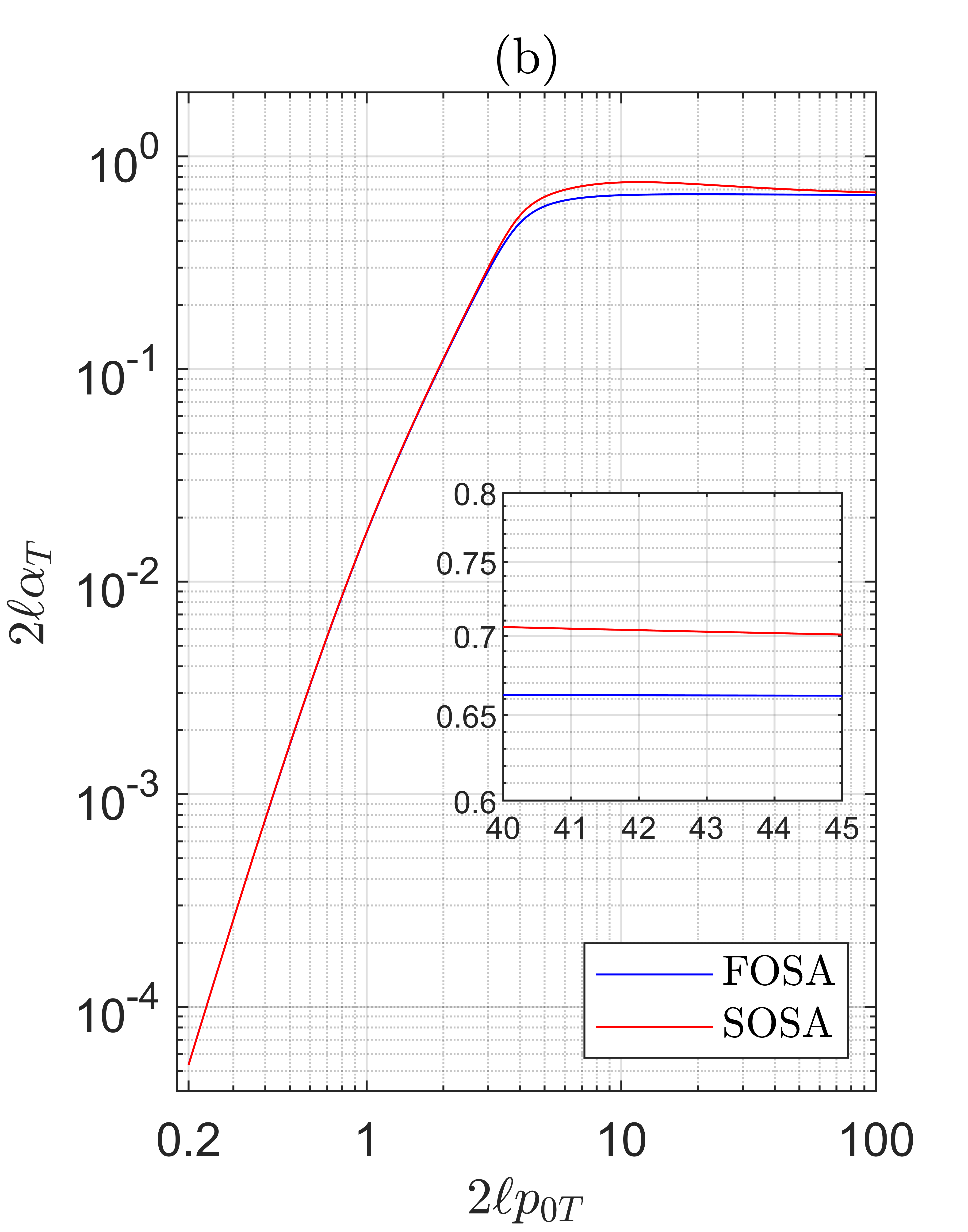}
\includegraphics[scale=0.46]{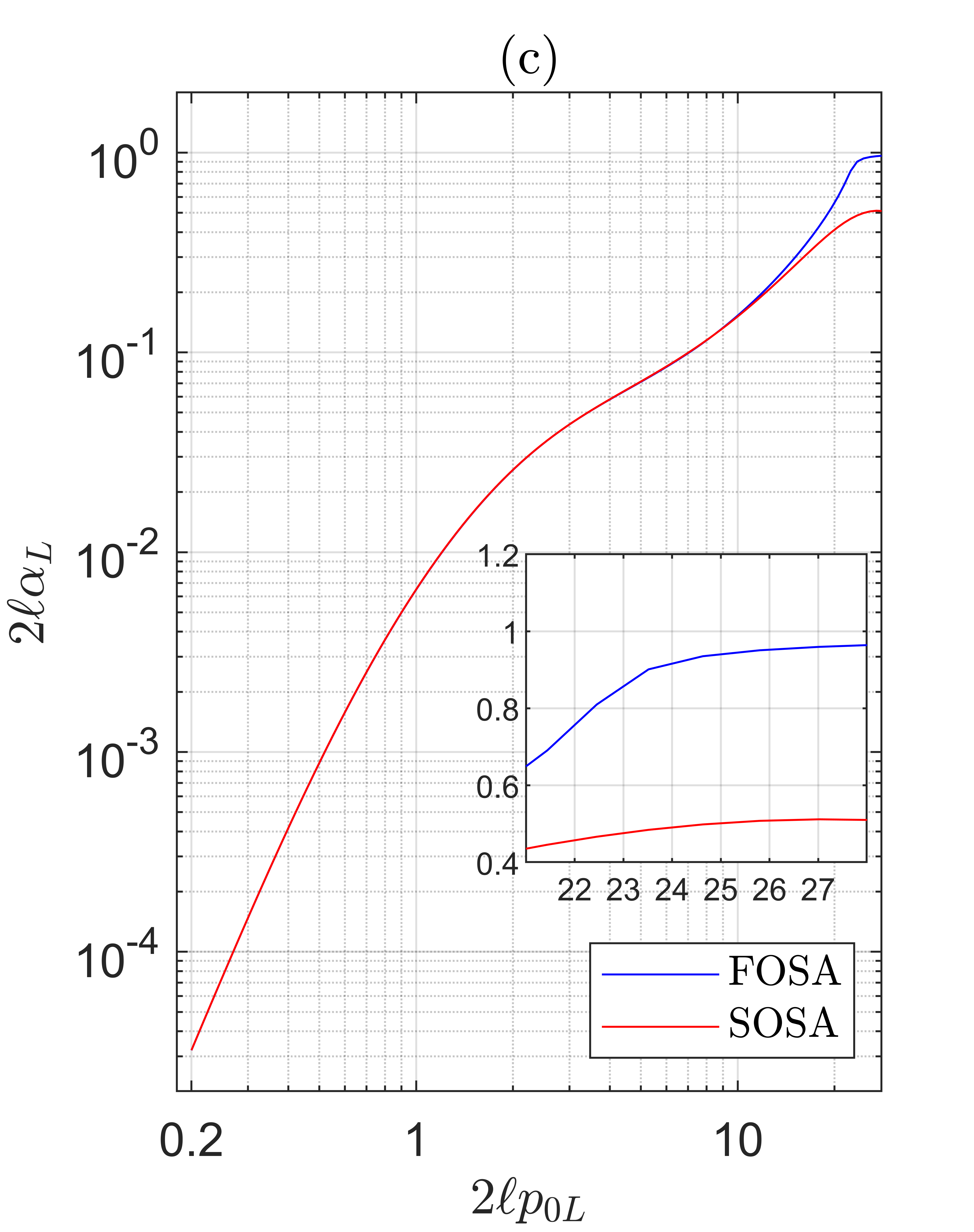}
\includegraphics[scale=0.46]{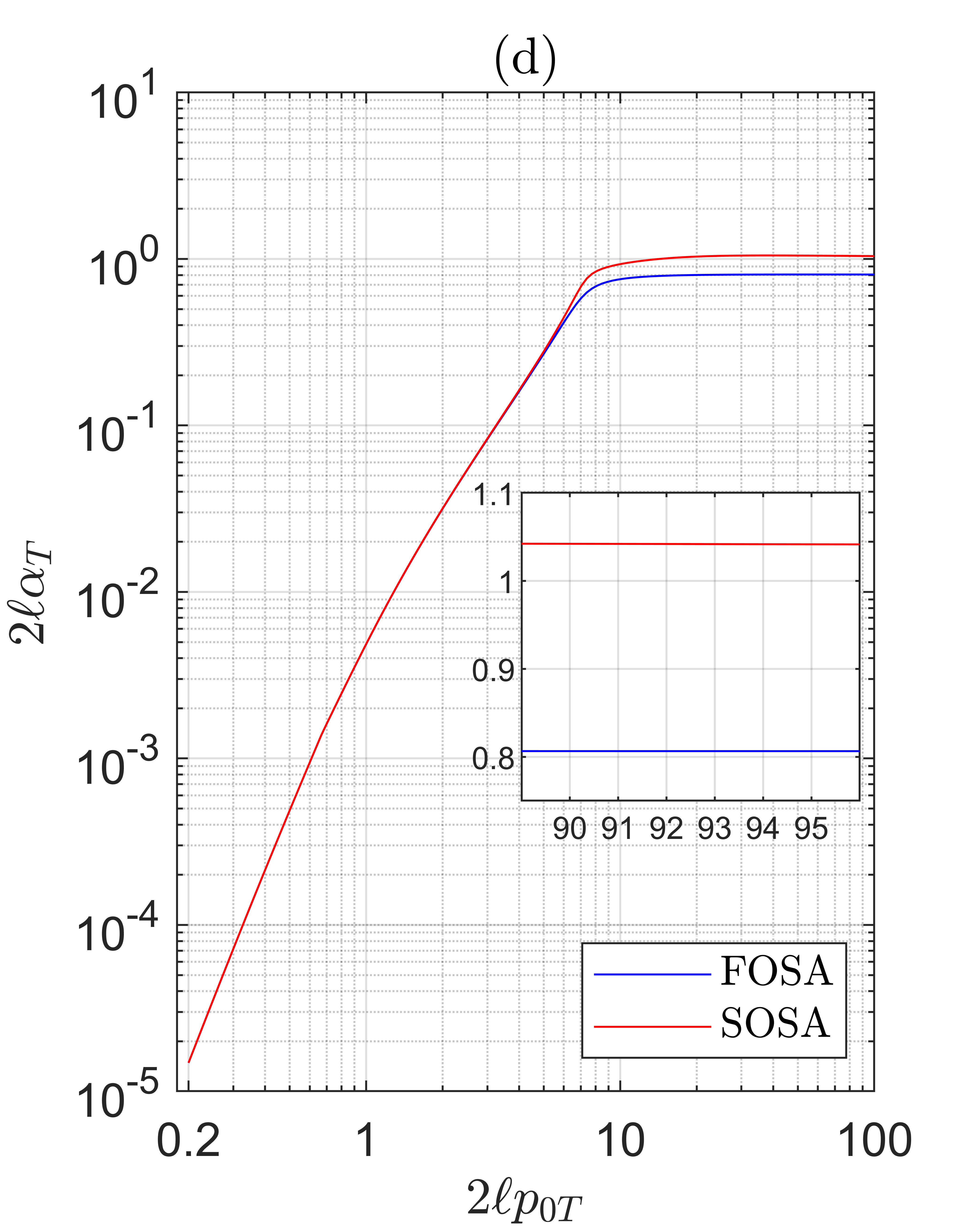}
\includegraphics[scale=0.465]{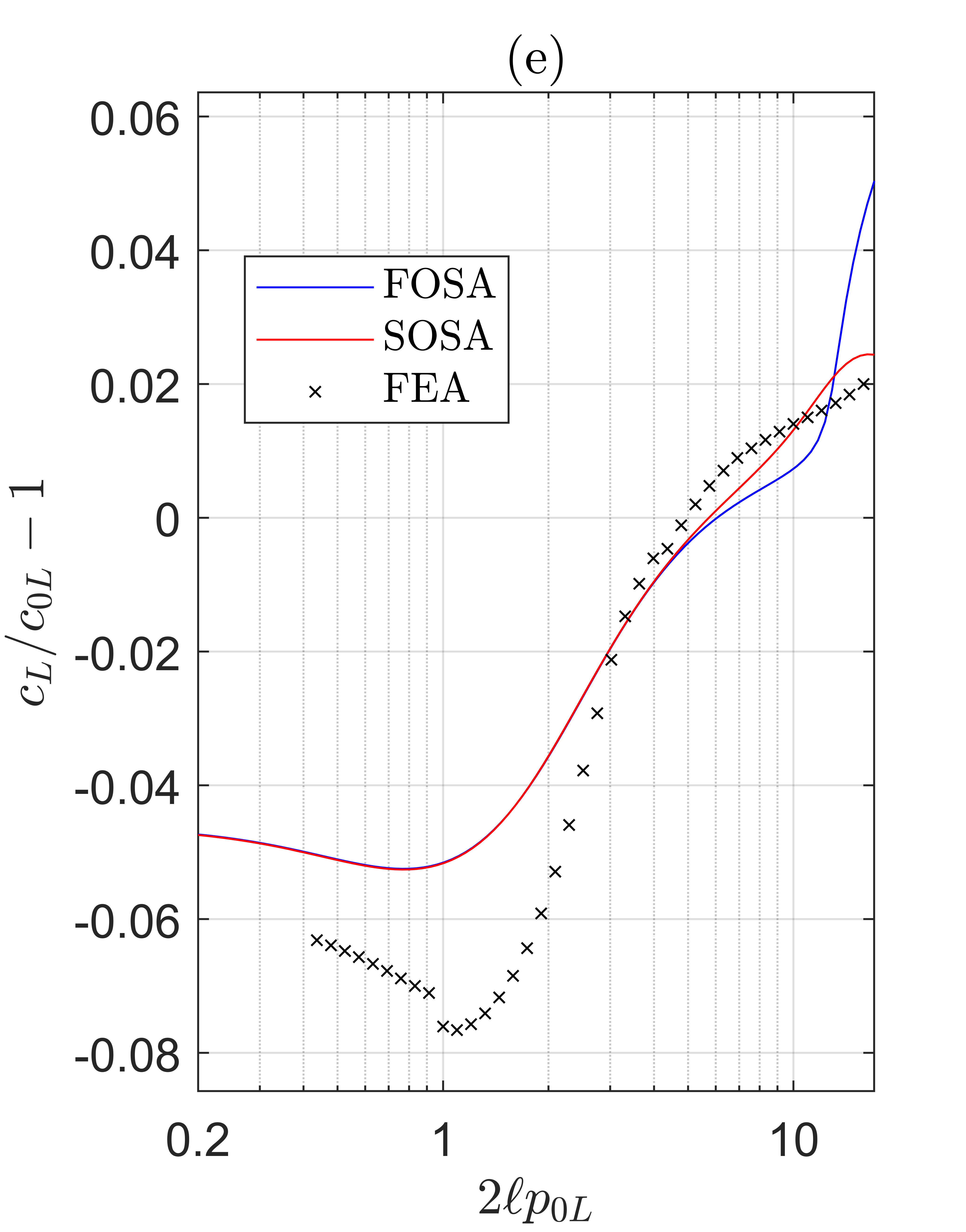}
\includegraphics[scale=0.465]{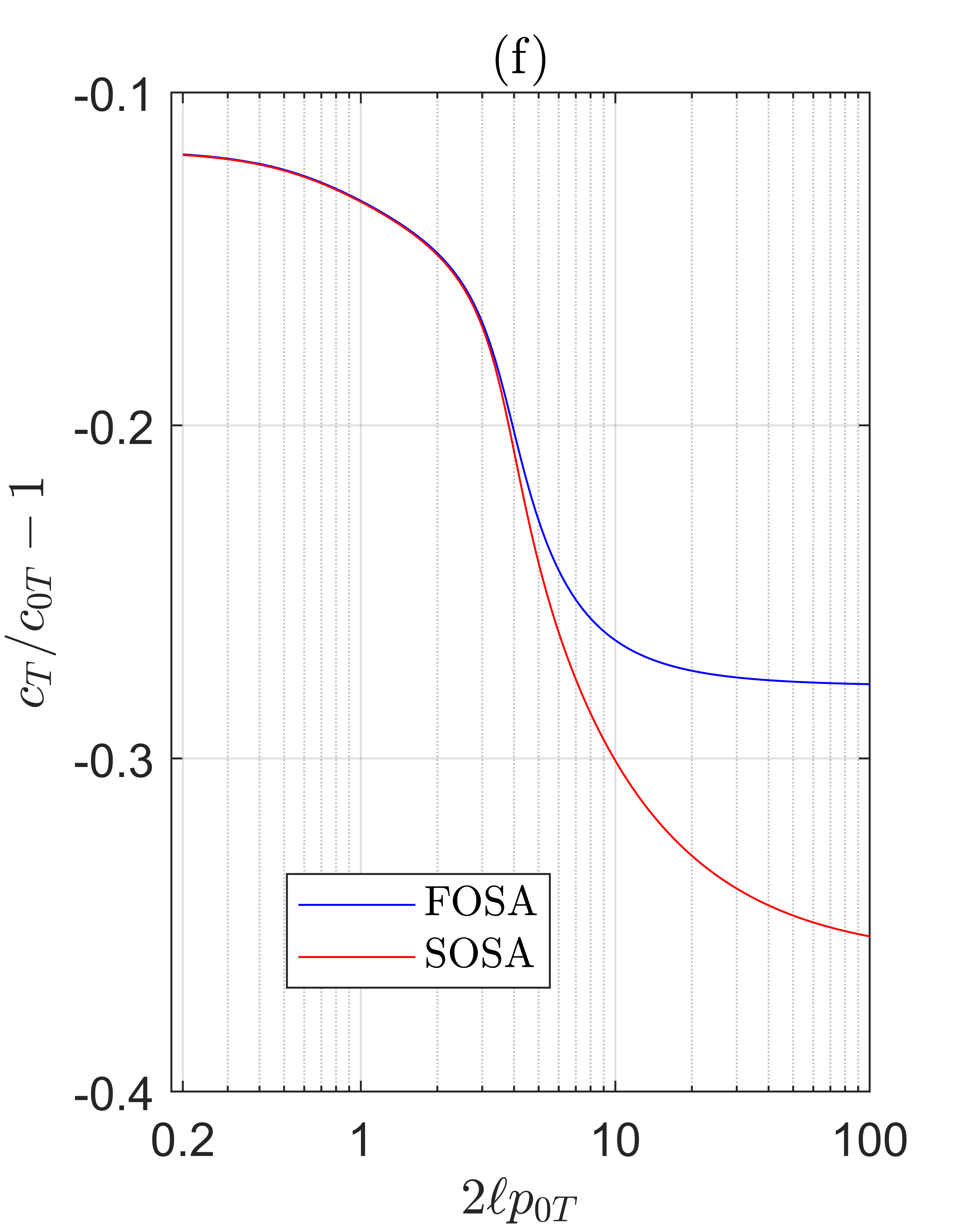}
\includegraphics[scale=0.465]{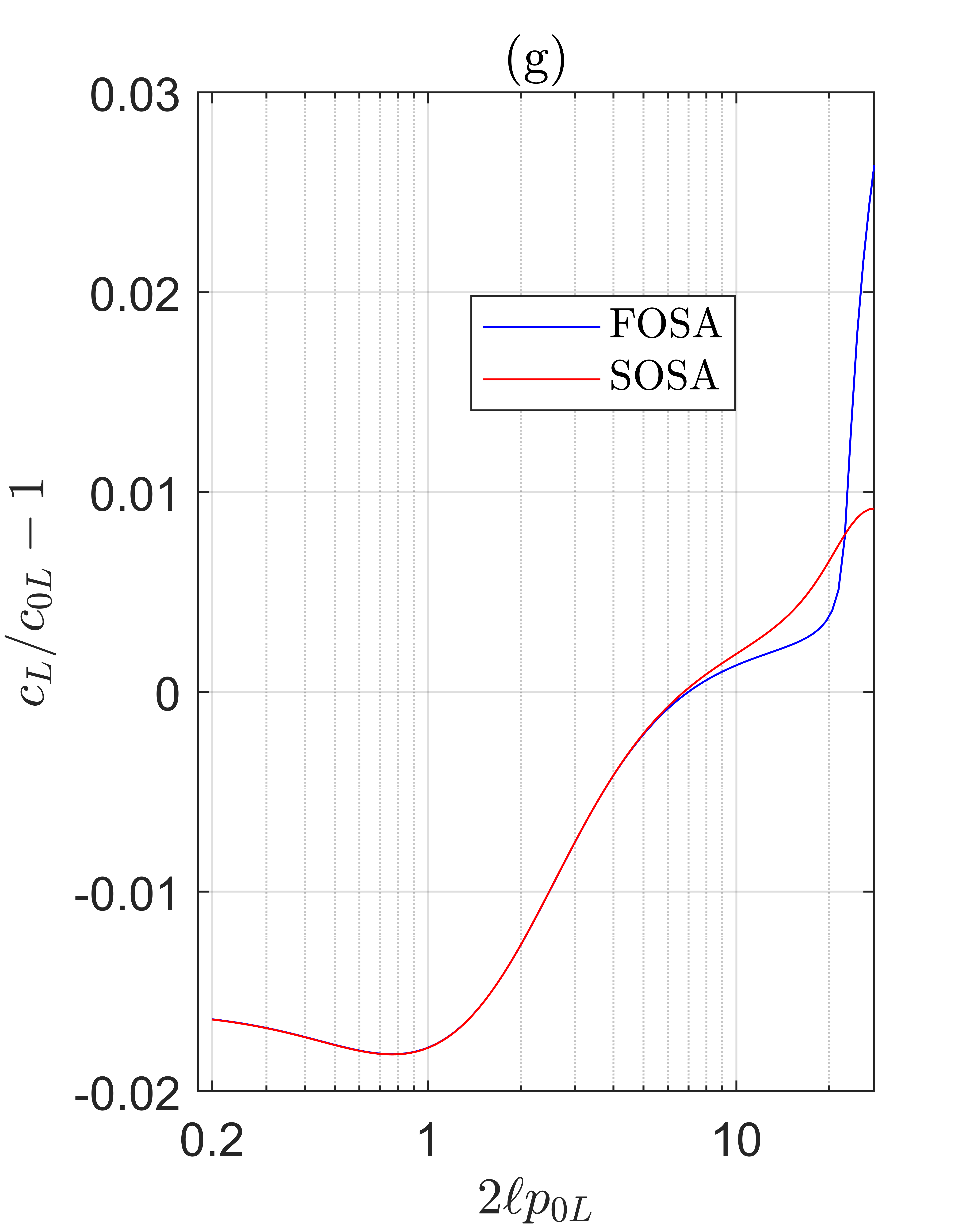}
\includegraphics[scale=0.465]{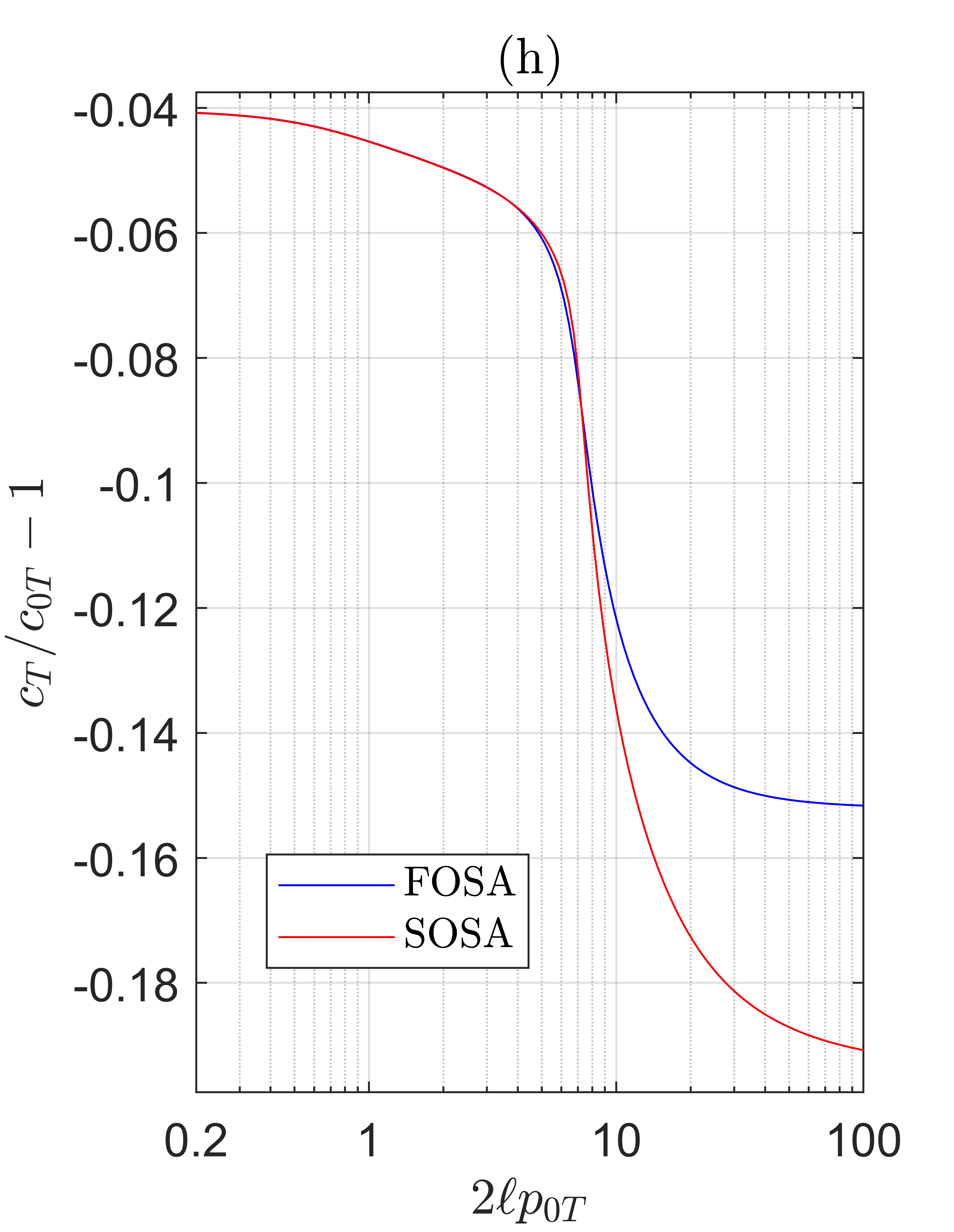}
\caption{Attenuation results for (a) longitudinal waves in lithium, (b) transverse waves in lithium, (c) longitudinal waves in iron, and (d) transverse waves in iron. Phase velocity dispersion for (e) longitudinal waves in lithium, (f) transverse waves in lithium, (g) longitudinal waves in iron, and (h) transverse waves in iron. The phase velocity dispersion is normalized with respect to the phase velocities of a non-scattering reference medium to isolate the influence of scattering.}
\label{fig5}
\end{figure}

At this point, we turn to quantitative attenuation and phase velocity results for polycrystalline lithium and iron for both longitudinal and transverse wave modes as seen in Figure \ref{fig5}. These two materials were chosen as two materials that are moderately (iron) and strongly (lithium) scattering, which is a function of elastic anisotropy or degree of heterogeneity as seen in Table \ref{TabMat}.

The attenuation results shown in Fig. \ref{fig5}(a-d) are normalized with respect to a factor of $2\ell$, which is twice the correlation length present in the TPC. Additionally, the factor $2\ell$ makes the results consistent with Fig. \ref{fig2} and the results in Stanke and Kino \cite{stanke1984unified}, which suggests $2\ell$ is related to the mean grain diameter. In this work, the correlation length $\ell$ is a measure of the average characteristic distance a wave travels between two points such that the two points are located in grains with different stiffness values. Thus, $\ell$ is related to an arithmetic mean grain radius, but is a distinct statistical quantity. The normalization by $2\ell$ allows one to observe the change in attenuation or phase velocity as a function of either $\ell$ or $p_0$ (proportional to frequency) by fixing one or the other. For example, the geometric scattering regime is located in the far right region in Fig. \ref{fig5}(a-d) where the curve begins to flatten out. The factor of $2\ell p_0$ in the geometric scattering regime suggests this regime is achieved in polycrystalline materials when the grain size and/or wave number (proportional to frequency) is sufficiently large or high. The stochastic scattering regime is also known as the transition regime and is the portion of the attenuation curves that display a nonlinear behavior (in logspace) while the Rayleigh regime is on the lower end of $2\ell p_0$ values and displays a near constant trend (but is an artifact of plotting in logspace and actually has a fourth-order $p_0$ or frequency dependence). 
\par
\begin{figure}[t]
\includegraphics[scale=0.55]{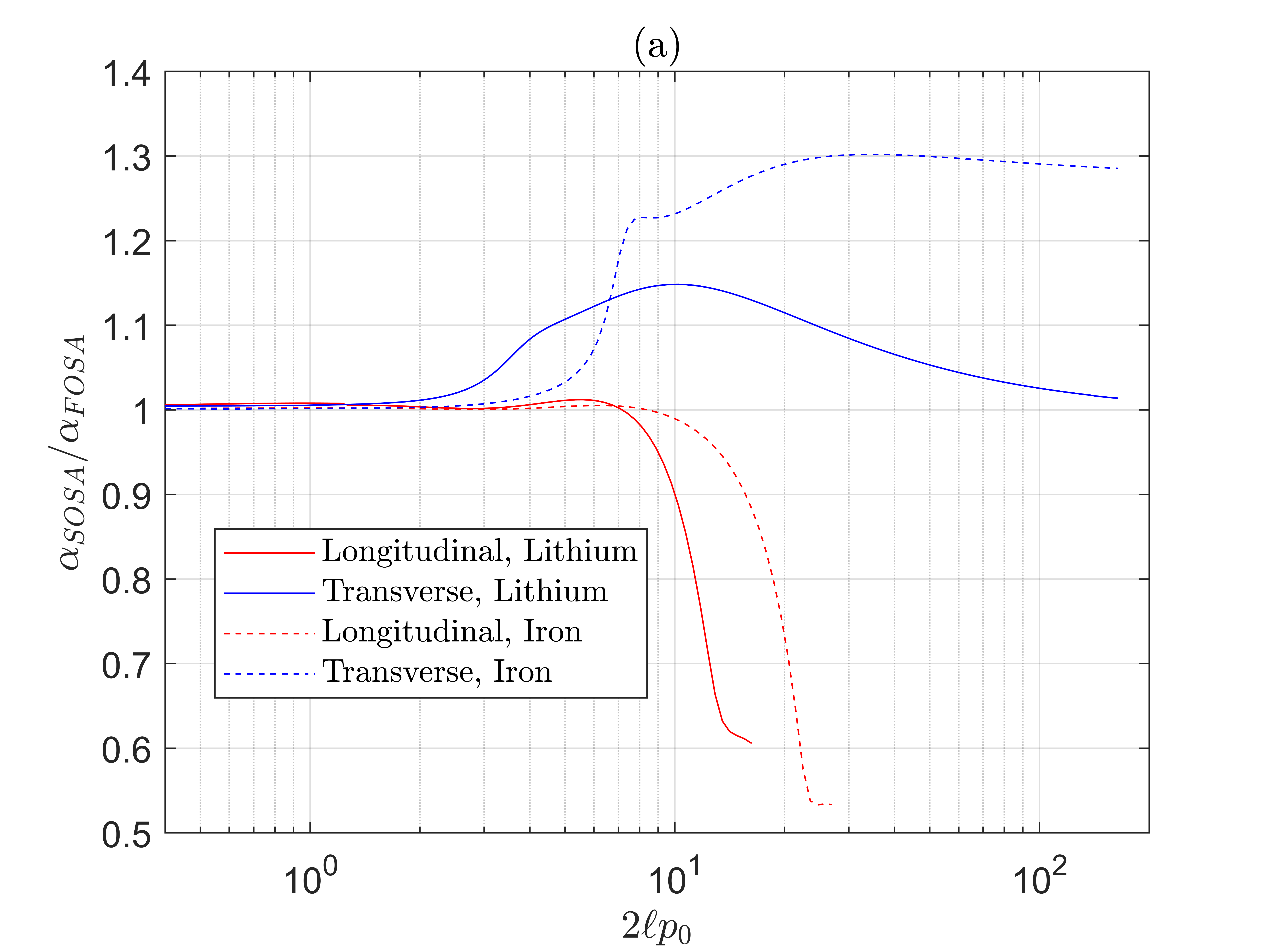}
\includegraphics[scale=0.55]{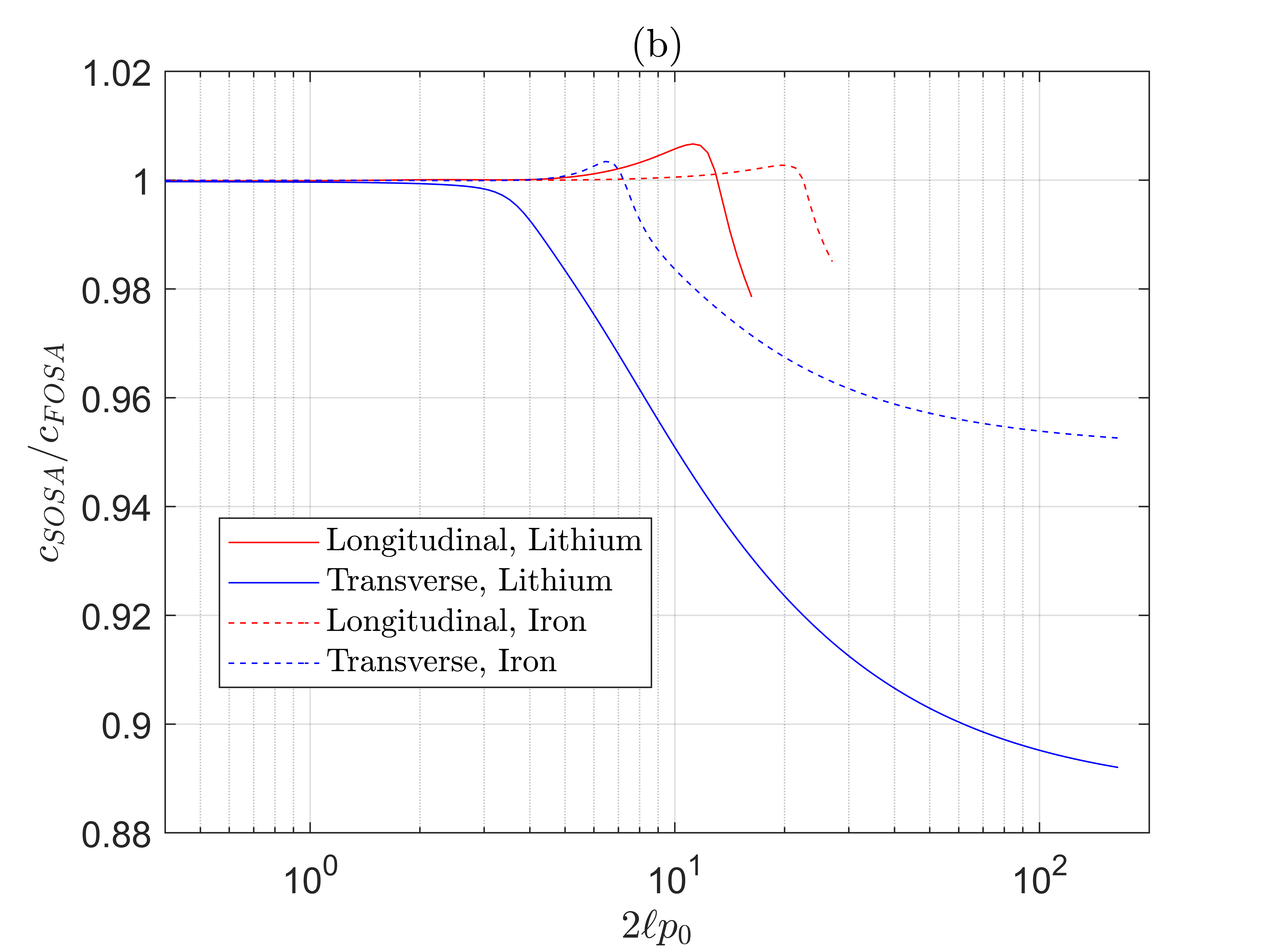}
\caption{The effects of including the SOSA terms is studied as a ratio to the corresponding FOSA estimates of attenuation and wavespeed for longitudinal and transverse wave propagation in lithium and iron.}
\label{fig4New}
\end{figure}

 In each of Figs. \ref{fig5}(a-d), the attenuations with first- and second-order scattering effects are shown in the red SOSA lines, while only first-order scattering effects are shown in the blue FOSA lines. The reader is reminded that second-order scattering allows each grain to be revisited in the scattering process, whereas FOSA does not include this possibility. In all attenuation results, the second-order scattering effects become more pronounced toward the higher end of the stochastic scattering regime and into the geometric scattering regime. The same statement can be said about the phase velocity dispersion. For both iron and lithium, the SOSA attenuations are larger for transverse waves and smaller for longitudinal waves. Thus, the primary longitudinal wave field is predicted to be enhanced due to the second-order scattering effects, suggesting a forward scattering contribution that constructively supports the primary wave. The situation for transverse waves suggests that more energy is scattered out of the forward direction and never returns when including second-order scattering effects, leading to a decrease in the amplitude of the primary wave field. These observations hint at the possibility of enhanced backscatter for transverse waves. Indeed, the addition of recurrent scattering events can begin to be captured in SOSA. However, the present model is concerned only with $\langle\textbf{G}\rangle$ through the Dyson equation and does not provide information on the differential scattering cross-section. Fig. \ref{fig4New} presents the ratio of SOSA and FOSA predictions as a function of $2\ell p_0$ to clearly show the prominence of the second-order scattering effects in the upper stochastic and geometric scattering regimes.

\par Also in Figs. \ref{fig5}(e-h) and Fig. \ref{fig4New}(b) are the corresponding results for phase velocity dispersion. The phase velocity dispersion is normalized by the respective phase velocity values for the non-scattering reference medium (using $c_{0L}=\sqrt{C_{11}^0/\rho}$ and $c_{0T}=\sqrt{C_{44}^0/\rho}$). Since the non-scattering medium is non-dispersive, the normalization explicitly shows the dispersive behavior from the scattering. As can be seen in Figs. \ref{fig5}(f) and \ref{fig5}(h) for transverse waves, the second-order scattering effects cause the phase velocity to notably decrease. In Figs. \ref{fig5}(e) and \ref{fig5}(g), the phase velocity for longitudinal waves is shown to reach a peak prior to FOSA in the geometric scattering regime. The phase velocity results in Figs. \ref{fig5}(e-h) plot $c/c_0-1$ versus the normalized wave number $p_0\ell$. As the group velocity $c_G$ is usually easier to measure experimentally, we note that if we let $f=c_G/c_0-1$ and $g=c/c_0-1$ then it can be shown that $f\approx g+p_0\ell dg/d(p_0\ell)$ where the derivative is exactly the slope evaluated at various points in Figs. \ref{fig5}(e-h). Thus, the corresponding plots for $f$ are easily accessible and will indicate that the group velocity dispersion will be largest when $p_0\ell$ multiplied by the slope in Figs. \ref{fig5}(e-h) is largest.

\par Finite element (FE) results of attenuation and phase velocity dispersion are also included in Figs. \ref{fig5}(a) and \ref{fig5}(e) from Ref. \cite{huang2022finite}. Modeling elastic waves using FE is beneficial since multiple scattering effects are naturally included. In both Figs. \ref{fig5}(a) and \ref{fig5}(e), the SOSA predictions are shown to be closer to the FE predictions for both attenuation and phase velocity. Note that the digitized microstructure used in the simulations toward the FE results in Ref. \cite{huang2022finite} is not fully represented by a single exponential TPC. Rather, the authors of Ref. \cite{huang2022finite} use a multi-term exponential. Additional details on the finite element analysis results can be found in Ref. \cite{huang2022finite}. Thus, agreement between SOSA and FE predictions could be improved if the generated microstructure had statistics describable by a single exponential TPC or the SOSA theory was extended to accommodate more general TPC functions, which will be a topic for further investigation.

\section{Conclusions and Discussion}
\label{SecConc}
This work presents a comprehensive theoretical framework for modeling multiple scattering of elastic waves in polycrystalline materials across all frequency regimes. We have extended previous models by advancing beyond the first-order smoothing approximation (FOSA) to incorporate the second-order smoothing approximation (SOSA), which accounts for higher-order scattering effects where grains can be revisited during the scattering process.
 
\par Our analysis of polycrystalline iron and lithium demonstrates that second-order scattering enhances the primary longitudinal wave field through constructive forward scattering, while causing more energy to scatter out of the forward direction for transverse waves. These effects are particularly pronounced in materials with higher elastic anisotropy.
While the current model employs a single-term exponential two-point correlation function with a single characteristic length scale, real polycrystalline materials often exhibit broad grain size distributions that cannot be adequately represented by this simplified approach. A promising direction for future work is to incorporate more general correlation functions, such as the Von Kármán spatial correlation function proposed by Argüelles \cite{arguelles2022karman}, which can accommodate wider distributions of grain sizes. Such extensions would likely reveal cases where SOSA becomes even more critical, as diverse grain size distributions create more varied scattering paths where multiple-visit scattering events become increasingly significant.
Another important extension would be to apply the SOSA framework to multiphase polycrystals or crystals embedded in a matrix, as studied by Hirsekorn \cite{hirsekorn1988scattering} and by Calvet and Margerin \cite{calvet2012velocity}. In these heterogeneous systems, the acoustic impedance contrast between phases creates stronger scattering interfaces, likely enhancing multiple scattering effects and making the SOSA approach particularly valuable for accurate predictions. The additional complexity of multiple phase boundaries would create more opportunities for waves to revisit scattering regions, precisely the phenomenon that SOSA is designed to capture. 
\par Experimental validation of the present SOSA-based model through attenuation and velocity measurements (phase or group) is viewed as feasible with transverse waves being a promising direction. Experimental efforts should painstakingly confirm that specimens have correlation functions that closely match the model assumptions. Additionally, the experiments themselves must minimize sources of experimental error by carefully isolating the scattering-based effects on measured or perceived attenuation or dispersion from other factors. This includes accounting for items such as diffraction, coupling conditions, attenuation due to absorption, sample flatness and parallelism etc. Readers are encouraged to consult the chapters of Papadakis \cite{papadakis1990measurementV, papadakis1990measurementA} for experimental considerations of velocity and attenuation measurements. Additionally, SOSA effects will be probed most readily at sufficiently high frequencies or large $p_0\ell$ values. Thus, a challenge exists between dealing with high attenuations in this regime, propagation distances, and having a sufficient number of scatterers in the path of the ultrasonic wave to be representative of the mean displacement fields modeled here (through the mean Green's function). Interestingly, Rayleigh waves have been used to explore geometric scattering regimes \cite{grabec2022surface}. As Rayleigh waves contain a combination of longitudinal and transverse components impacted by scattering, this further supports directions involving transverse waves. 
\par The challenges associated with experiments at high frequencies could somewhat be alleviated through experiments on samples with more complicated microstructures that do not strictly follow the simple single exponential TPC. In these settings, the strong scattering effects seen in SOSA could be realized at lower and more experimentally accessible frequencies. This would require similar model extensions as discussed a couple of paragraphs ago.

\par It should be emphasized that the present work is concerned with the mean Green's function available from the Dyson equation, thus capturing scattering effects on the mean field. The improved solution of the Dyson equation using SOSA represents a crucial prerequisite for future work on the Bethe-Salpeter equation in terms of the scattering intensity $\textbf{K}$, 
\begin{flalign}
    \langle \textbf{G}\left(\textbf{x}_1,\textbf{x}_1'\right)\textbf{G}^{*}\left(\textbf{x}_2,\textbf{x}_2'\right)\rangle &= \langle \textbf{G}\left(\textbf{x}_1,\textbf{x}_1'\right)\rangle\langle\textbf{G}^{*}\left(\textbf{x}_2,\textbf{x}_2'\right) \rangle \notag\\
   \quad & + \int \langle \textbf{G}\left(\textbf{x}_1,\textbf{x}_3'\right)\rangle\langle\textbf{G}^{*}\left(\textbf{x}_2,\textbf{x}_4'\right) \rangle \textbf{K}\left(\textbf{x}_3',\textbf{x}_4';\textbf{x}_3,\textbf{x}_4\right)\notag\\
   \quad & \times \langle\textbf{G}\left(\textbf{x}_3,\textbf{x}_1'\right)\textbf{G}^{*}\left(\textbf{x}_4,\textbf{x}_2'\right)\rangle
\end{flalign}
which provides access to the mean of the squared Green's function and would enable the calculation of differential scattering cross-sections and more detailed analysis of scattered field directionality. As demonstrated by Barabanenkov et al. \cite{barabanenkov1971status} and Barabanenkov \cite{barabanenkov1974energy}, there exists an energy equivalence between the Dyson and Bethe-Salpeter equations in multiple scattering theory, suggesting that our SOSA improvements to the Dyson equation solution provide a necessary foundation for subsequently enhancing the modeling of scattered fields through the Bethe-Salpeter equation. Furthermore, it is the Bethe-Salpeter equation that forms the starting point for modeling ultrasonic backscatter in applications for nondestructive evaluation and characterization \cite{ghoshal2007wigner, hu2015contribution, arguelles2016mode, kube2018ultrasonic, arguelles2019generalized}. In Ref. \cite{ghoshal2007wigner}, the first-order ladder approximation to the Bethe-Salpeter equation is used to model singly-scattered backscatter which is described by the coherent propagation of wave vector $\textbf{q}$ that is then scattered into a wave vector $\textbf{p}$ into the opposite direction. Both $\textbf{p}$ and $\textbf{q}$ in Ref. \cite{ghoshal2007wigner} are effective wave vectors defined through the Born approximation of the mean Green's function derived in the context of FOSA. Thus, $\textbf{p}$ and $\textbf{q}$ are attenuating and include all orders of multiple scattering events involving no revisiting of grains (no recurrent scattering). The Born approximation removes the dispersive character of $\textbf{p}$ and $\textbf{q}$, which relegates the backscatter model to weak scattering.

\par Hu and Turner \cite{hu2015contribution} incorporated the next order term of the Bethe-Salpeter equation to arrive at a doubly-scattered theory, which is the same order of approximation as our present treatment of SOSA. The intent of the doubly-scattered theory was to accommodate more strongly scattering materials. However, the wave vectors contributing to the double scattering process remain defined by the Born approximation of FOSA and still place a potentially severe limitation on how strongly scattering the polycrystal can be. Thus, improvements in grain scattering models \cite{ghoshal2007wigner, hu2015contribution, arguelles2016mode, kube2018ultrasonic, arguelles2019generalized} could be realized by consistently incorporating the same scattering contributions in the Bethe-Salpeter equation as the Dyson equation. This would be achieved by being consistent with including the analogous multi-point diagrams in both settings. 
Furthermore, our model addresses a significant limitation identified by Weaver \cite{weaver1990diffusivity}, who noted that previous Born approximation calculations could not be safely applied in the geometric optics high-frequency regime, which has been termed the ``diffusion regime" by some authors. By calculating attenuations at all frequencies beyond the Born approximation, including first-order and second-order effects, our work provides values that can enable a meaningful study of diffusion in the geometric scattering regime—an area previously identified as lacking adequate theoretical treatment. This advancement opens new possibilities for investigating multiple scattering in polycrystals.


\section*{Acknowledgements}
This work was done as part of the first author’s doctoral research at the Pennsylvania State University’s Engineering Science and Mechanics Department. This material is based upon work supported by the National Science Foundation under Grant Number \hyperlink{https://www.nsf.gov/awardsearch/showAward?AWD_ID=2225215&HistoricalAwards=false}{2225215}. Any opinions, findings, and conclusions or recommendations expressed in this material are those of the author(s) and do not necessarily reflect the views of the National Science Foundation. The authors acknowledge the encouragement and support from the American Society for Nondestructive Testing for awarding the 2022 ASNT Student Fellowship. CMK would like to dedicate this work to the late Lucas W. Koester, who helped spark his interest in these directions. CMK would like to thank Joseph A. Turner for ongoing support over many years and Richard L. Weaver for useful discussions and initiating the foundation for this work. AR and CMK would like to thank Jacob L. Bourjaily, Akhlesh Lakhtakia, and 
Nikhil Kalyanapuram for useful discussions surrounding this work.

\appendix
\section*{Appendices}
\section{Expression for the covariance tensor}
\label{covTrans}

\begin{align}
    \Xi_{ijkl}^{\alpha\beta\gamma\delta} = \langle\left(C_{ijkl}-C_{ijkl}^0\right)\left(C_{\alpha\beta\gamma\delta}-C_{\alpha\beta\gamma\delta}^0\right)\rangle
= \langle C_{ijkl}C_{\alpha\beta\gamma\delta}\rangle - \langle C_{ijkl}\rangle\langle C_{\alpha\beta\gamma\delta}\rangle
\end{align}

where $\textbf{C}^0=\langle\textbf{C}\rangle$ and
\begin{align}
    \langle\textbf{C}\rangle=(C_{12}+\nu/5)\delta_{ij}\delta_{kl}+(C_{44}+\nu/5)(\delta_{ik}\delta_{jl}+\delta_{il}\delta_{jk})
\end{align}
with $\nu=C_{11}-C_{12}-2C_{44}$

Then, $\boldsymbol{\Xi}$ can be written as an isotropic eighth-rank isotropic Cartesian tensor.

\begin{flalign}
\label{covfin2}
&\Xi_{ijkl}^{\alpha \beta \kappa \delta }=BI_{LLLLGGGG}+DI_{LLGGLGLG}+HI_{LGLGLGLG}\notag \\ \quad &=B\left(\delta_{ij}\delta_{kl}+\delta_{ik}\delta_{jl}+\delta_{il}\delta_{jk}\right)\left(\delta_{\alpha \beta }\delta_{\kappa \delta }+\delta_{\alpha \kappa }\delta_{\beta \delta }+\delta_{\alpha \delta }\delta_{\beta \kappa }\right) \notag \\ \quad &
+D\left(\phantom {\frac{}{}}\right.\delta_{\alpha i}\delta_{\beta j}\delta_{\kappa \delta }\delta_{kl}+\delta_{\alpha i}\delta_{\kappa j}\delta_{\beta \delta }\delta_{kl}+\delta_{\alpha i}\delta_{\delta j}\delta_{\beta \kappa }\delta_{kl}+\delta_{\beta i}\delta_{\kappa j}\delta_{\alpha \delta }\delta_{kl}\notag\\ \quad &
+\delta_{\beta i}\delta_{\delta j}\delta_{\alpha \kappa }\delta_{kl}+\delta_{\kappa i}\delta_{\delta j}\delta_{\alpha \beta }\delta_{kl}+\delta_{\alpha j}\delta_{\beta i}\delta_{\kappa \delta }\delta_{kl}+\delta_{\alpha j}\delta_{\kappa i}\delta_{\beta \delta }\delta_{kl}\notag\\ \quad &
+\delta_{\alpha j}\delta_{\delta i}\delta_{\beta \kappa }\delta_{kl}+\delta_{\beta j}\delta_{\kappa i}\delta_{\alpha \delta }\delta_{kl}+\delta_{\beta j}\delta_{\delta i}\delta_{\alpha \kappa }\delta_{kl}+\delta_{\kappa j}\delta_{\delta i}\delta_{\alpha \beta }\delta_{kl}\notag\\ \quad &
+\delta_{\alpha k}\delta_{\beta l}\delta_{\kappa \delta }\delta_{ij}+\delta_{\alpha k}\delta_{\kappa l}\delta_{\beta \delta }\delta_{ij}+\delta_{\alpha k}\delta_{\delta l}\delta_{\beta \kappa }\delta_{ij}+\delta_{\beta k}\delta_{\kappa l}\delta_{\alpha \delta }\delta_{ij}\notag\\ \quad &
+\delta_{\beta k}\delta_{\delta l}\delta_{\alpha \kappa }\delta_{ij}+\delta_{\kappa k}\delta_{\delta l}\delta_{\alpha \beta }\delta_{ij}+\delta_{\alpha l}\delta_{\beta k}\delta_{\kappa \delta }\delta_{ij}+\delta_{\alpha l}\delta_{\kappa k}\delta_{\beta \delta }\delta_{ij}\notag\\ \quad &
+\delta_{\alpha l}\delta_{\delta k}\delta_{\beta \kappa }\delta_{ij}+\delta_{\beta l}\delta_{\kappa k}\delta_{\alpha \delta }\delta_{ij}+\delta_{\beta l}\delta_{\delta k}\delta_{\alpha \kappa }\delta_{ij}+\delta_{\kappa l}\delta_{\delta k}\delta_{\alpha \beta }\delta_{ij}\notag\\ \quad &
+\delta_{\alpha j}\delta_{\beta l}\delta_{\kappa \delta }\delta_{ik}+\delta_{\alpha j}\delta_{\kappa l}\delta_{\beta \delta }\delta_{ik}+\delta_{\alpha j}\delta_{\delta l}\delta_{\beta \kappa }\delta_{ik}+\delta_{\beta j}\delta_{\kappa l}\delta_{\alpha \delta }\delta_{ik}\notag\\ \quad &
+\delta_{\beta j}\delta_{\delta l}\delta_{\alpha \kappa }\delta_{ik}+\delta_{\kappa j}\delta_{\delta l}\delta_{\alpha \beta }\delta_{ik}+\delta_{\alpha l}\delta_{\beta j}\delta_{\kappa \delta }\delta_{ik}+\delta_{\alpha l}\delta_{\kappa j}\delta_{\beta \delta }\delta_{ik}\notag\\ \quad &
+\delta_{\alpha l}\delta_{\delta j}\delta_{\beta \kappa }\delta_{ik}+\delta_{\beta l}\delta_{\kappa j}\delta_{\alpha \delta }\delta_{ik}+\delta_{\beta l}\delta_{\delta j}\delta_{\alpha \kappa }\delta_{ik}+\delta_{\kappa l}\delta_{\delta j}\delta_{\alpha \beta }\delta_{ik}\notag\\ \quad &
+\delta_{\alpha j}\delta_{\beta k}\delta_{\kappa \delta }\delta_{il}+\delta_{\alpha j}\delta_{\kappa k}\delta_{\beta \delta }\delta_{il}+\delta_{\alpha j}\delta_{\delta k}\delta_{\beta \kappa }\delta_{il}+\delta_{\beta j}\delta_{\kappa k}\delta_{\alpha \delta }\delta_{il}\notag\\ \quad &
+\delta_{\beta j}\delta_{\delta k}\delta_{\alpha \kappa }\delta_{il}+\delta_{\kappa j}\delta_{\delta k}\delta_{\alpha \beta }\delta_{il}+\delta_{\alpha k}\delta_{\beta j}\delta_{\kappa \delta }\delta_{il}+\delta_{\alpha k}\delta_{\kappa j}\delta_{\beta \delta }\delta_{il}\notag\\ \quad &
+\delta_{\alpha k}\delta_{\delta j}\delta_{\beta \kappa }\delta_{il}+\delta_{\beta k}\delta_{\kappa j}\delta_{\alpha \delta }\delta_{il}+\delta_{\beta k}\delta_{\delta j}\delta_{\alpha \kappa }\delta_{il}+\delta_{\kappa k}\delta_{\delta j}\delta_{\alpha \beta }\delta_{il}\notag\\ \quad &
+\delta_{\alpha i}\delta_{\beta l}\delta_{\kappa \delta }\delta_{jk}+\delta_{\alpha i}\delta_{\kappa l}\delta_{\beta \delta }\delta_{jk}+\delta_{\alpha i}\delta_{\delta l}\delta_{\beta \kappa }\delta_{jk}+\delta_{\beta i}\delta_{\kappa l}\delta_{\alpha \delta }\delta_{jk}\notag\\ \quad &
+\delta_{\beta i}\delta_{\delta l}\delta_{\alpha \kappa }\delta_{jk}+\delta_{\kappa i}\delta_{\delta l}\delta_{\alpha \beta }\delta_{jk}+\delta_{\alpha l}\delta_{\beta i}\delta_{\kappa \delta }\delta_{jk}+\delta_{\alpha l}\delta_{\kappa i}\delta_{\beta \delta }\delta_{jk}\notag\\ \quad &
+\delta_{\alpha l}\delta_{\delta i}\delta_{\beta \kappa }\delta_{jk}+\delta_{\beta l}\delta_{\kappa i}\delta_{\alpha \delta }\delta_{jk}+\delta_{\beta l}\delta_{\delta i}\delta_{\alpha \kappa }\delta_{jk}+\delta_{\kappa l}\delta_{\delta i}\delta_{\alpha \beta }\delta_{jk}\notag\\ \quad &
+\delta_{\alpha i}\delta_{\beta k}\delta_{\kappa \delta }\delta_{jl}+\delta_{\alpha i}\delta_{\kappa k}\delta_{\beta \delta }\delta_{jl}+\delta_{\alpha i}\delta_{\delta k}\delta_{\beta \kappa }\delta_{jl}+\delta_{\beta i}\delta_{\kappa k}\delta_{\alpha \delta }\delta_{jl}\notag\\ \quad &
+\delta_{\beta i}\delta_{\delta k}\delta_{\alpha \kappa }\delta_{jl}+\delta_{\kappa i}\delta_{\delta k}\delta_{\alpha \beta }\delta_{jl}+\delta_{\alpha k}\delta_{\beta i}\delta_{\kappa \delta }\delta_{jl}+\delta_{\alpha k}\delta_{\kappa i}\delta_{\beta \delta }\delta_{jl}\notag\\ \quad &
+\delta_{\alpha k}\delta_{\delta i}\delta_{\beta \kappa }\delta_{jl}+\delta_{\beta k}\delta_{\kappa i}\delta_{\alpha \delta }\delta_{jl}+\delta_{\beta k}\delta_{\delta i}\delta_{\alpha \kappa }\delta_{jl}+\delta_{\kappa k}\delta_{\delta i}\delta_{\alpha \beta }\delta_{jl}\left . \phantom {\frac{}{}}\right)\notag\\ \quad &
+H\left(\phantom {\frac{}{}}\right . \delta_{\alpha i}\delta_{\beta j}\delta_{\kappa k}\delta_{\delta l}+\delta_{\alpha i}\delta_{\beta j}\delta_{\kappa l}\delta_{\delta k}+\delta_{\alpha j}\delta_{\beta i}\delta_{\kappa k}\delta_{\delta l}+\delta_{\alpha j}\delta_{\beta i}\delta_{\kappa l}\delta_{\delta k}\notag\\ \quad &
+\delta_{\alpha i}\delta_{\beta k}\delta_{\kappa j}\delta_{\delta l}+\delta_{\alpha i}\delta_{\beta k}\delta_{\kappa l}\delta_{\delta j}+\delta_{\alpha k}\delta_{\beta i}\delta_{\kappa j}\delta_{\delta l}+\delta_{\alpha k}\delta_{\beta i}\delta_{\kappa l}\delta_{\delta j}\notag\\ \quad &
+\delta_{\alpha i}\delta_{\beta l}\delta_{\kappa j}\delta_{\delta k}+\delta_{\alpha i}\delta_{\beta l}\delta_{\kappa k}\delta_{\delta j}+\delta_{\alpha l}\delta_{\beta i}\delta_{\kappa j}\delta_{\delta k}+\delta_{\alpha l}\delta_{\beta i}\delta_{\kappa k}\delta_{\delta j}\notag\\ \quad &
+\delta_{\alpha j}\delta_{\beta k}\delta_{\kappa i}\delta_{\delta l}+\delta_{\alpha j}\delta_{\beta k}\delta_{\kappa l}\delta_{\delta i}+\delta_{\alpha k}\delta_{\beta j}\delta_{\kappa i}\delta_{\delta l}+\delta_{\alpha k}\delta_{\beta j}\delta_{\kappa l}\delta_{\delta i}\notag\\ \quad &
+\delta_{\alpha j}\delta_{\beta l}\delta_{\kappa i}\delta_{\delta k}+\delta_{\alpha j}\delta_{\beta l}\delta_{\kappa k}\delta_{\delta i}+\delta_{\alpha l}\delta_{\beta j}\delta_{\kappa i}\delta_{\delta k}+\delta_{\alpha l}\delta_{\beta j}\delta_{\kappa k}\delta_{\delta i}\notag\\ \quad &
+\delta_{\alpha k}\delta_{\beta l}\delta_{\kappa i}\delta_{\delta j}+\delta_{\alpha k}\delta_{\beta l}\delta_{\kappa j}\delta_{\delta i}+\delta_{\alpha l}\delta_{\beta k}\delta_{\kappa i}\delta_{\delta j}+\delta_{\alpha l}\delta_{\beta k}\delta_{\kappa j}\delta_{\delta i}\left . \phantom {\frac{}{}}\right)
\end{flalign}
where, $B$, $D$ and $H$ are \cite{weaver1990diffusivity}, 
\begin{align}
\label{bdef}
B = 2\nu^2/1575, D = -\nu^2/630, H = \nu^2/180. 
\end{align} 
The eighth rank tensors on the right-hand side of Equation (\ref{latingreekEq}) or the terms seen in Equation (\ref{covfin2}) are given here in descriptive form following Ref. \cite{kube2015stress} as,
\begin{flalign}
\label{ILG1}
I_{LLLLGGGG}&=\delta_{ij}\delta_{kl}\delta_{\alpha\beta}\delta_{\gamma\delta}+\text{all permutations of two pairs of Latin and
Latin} \notag\\\quad &\text{indices and two pairs of Greek and Greek indices}\left(9~\text{terms}\right)
\end{flalign}
\begin{flalign}
\label{ILG2}
I_{LGLGLGLG}&=\delta_{i\alpha}\delta_{j\beta}\delta_{k\gamma}\delta_{l\delta}+\text{all permutations of four pairs of Latin and
Greek} \notag\\\quad &\text{indices}\left(24~\text{terms}\right)
\end{flalign}
\begin{flalign}
\label{ILG3}
I_{LLGGLGLG}&=\delta_{ij}\delta_{\alpha\beta}\delta_{k\gamma}\delta_{l\delta}+\text{all permutations of two pairs of Latin and
Greek indices} \notag\\\quad &\text{and one Latin and Latin indices and one Greek and Greek indices}\left(72~\text{terms}\right)
\end{flalign}

\section{CPV integrals}
\label{appCont}
Toward computing $\mathcal{P}$, in terms of partial fractions, the Equation (\ref{SigPu}) can be rewritten as,
\begin{align}
\label{partFracDecom}
\mathcal{P}^{PQ} &= \frac{1}{2}\left[A_1\left(\mathcal{P}_1+\mathcal{P}_5\right)+A_2\left(\mathcal{P}_2+\mathcal{P}_6\right)+B_1\left(\mathcal{P}_3+\mathcal{P}_7\right)+B_2\left(\mathcal{P}_4+\mathcal{P}_8\right)\right],
\end{align}
where each of these $8$ CPV integrals take the form,
\begin{flalign}
\label{partFrac1}
\mathcal{P}_1 &= \int \limits_0^{\infty} dx_s \frac{1}{(x_s-z)^2(x_s-x_0)}\notag\\
\quad &=\frac{\log(x_0) + 2\pi i}{4(z - x_0)^2} - \frac{1}{z(z - x_0)} + \frac{\log(z)}{(z - x_0)^2} + \frac{1}{2x_0(z - x_0)},
\\
\quad 
\mathcal{P}_2 &= \int \limits_0^{\infty} dx_s \frac{1}{(x_s-z)(x_s-x_0)}=\frac{\log(x_0)-\log(z) + 2\pi i}{z - x_0} ,
\\ \quad
\mathcal{P}_3 &= \int \limits_0^{\infty} dx_s \frac{1}{(x_s-\bar{z})^2(x_s-x_0)}\notag\\
\quad &=\frac{\log(x_0) + 2\pi i}{4(\bar{z} - x_0)^2} - \frac{1}{\bar{z}(\bar{z} - x_0)} + \frac{\log(\bar{z})}{(\bar{z} - x_0)^2} + \frac{1}{2x_0(\bar{z} - x_0)},
\\ \quad
\mathcal{P}_4 &= \int \limits_0^{\infty} dx_s \frac{1}{(x_s-\bar{z})(x_s-x_0)} =\frac{\log(x_0)-\log(\bar{z})}{\bar{z} - x_0},
\\ \quad \mathcal{P}_5 &= \int \limits_0^{\infty} dx_s \frac{1}{(x_s-z)^2(x_s+x_0)} \notag\\
\quad &=\frac{\log(-x_0) + 2\pi i}{4(z + x_0)^2} - \frac{1}{z(z + x_0)} + \frac{\log(z)}{(z + x_0)^2} - \frac{1}{2x_0(z + x_0)},
\\ \quad
\mathcal{P}_6 &= \int \limits_0^{\infty} dx_s \frac{1}{(x_s-z)(x_s+x_0)}=\frac{\log(-x_0)-\log(z)}{z + x_0},
\end{flalign}
\begin{flalign}
\mathcal{P}_7 &= \int \limits_0^{\infty} dx_s \frac{1}{(x_s-\bar{z})^2(x_s+x_0)}\notag\\
\quad &=\frac{\log(-x_0) + 2\pi i}{4(\bar{z} + x_0)^2} - \frac{1}{\bar{z}(\bar{z} + x_0)} + \frac{\log(\bar{z})}{(\bar{z} + x_0)^2} - \frac{1}{2x_0(\bar{z} + x_0)},
\\ \quad \mathcal{P}_8 &= \int \limits_0^{\infty} dx_s \frac{1}{(x_s-\bar{z})(x_s+x_0)} =\frac{\log(-x_0)-\log(\bar{z}) - 2\pi i}{\bar{z} + x_0}
\end{flalign}  
with the coefficients of partial fractions being,
\begin{align}
\label{coeff1}
& A_1 = \frac{z^3}{\left(z - \bar{z}\right)^2},
\\
\quad &
A_2 = \frac{z^2\left(z - 3\bar{z}\right)}{\left(z - \bar{z}\right)^3},
\\
\quad &
B_1 = \frac{ \bar{z}^3}{\left(z - \bar{z}\right)^2},
\\
\quad &
B_2 = \frac{\bar{z}^2\left(3z - \bar{z}\right)}{\left(z - \bar{z}\right)^3}.
\end{align}
It is to be noted that since CPVs are generally complex, only the real component of the CPV is to be considered consistent with the propagators' definition. 

\section{Coefficients $\Omega^{PQUV}_{n_1n_2n_3}$}
\label{coeffTable}

\renewcommand{\thetable}{C.\arabic{table}}
\setcounter{table}{0}
\begin{table}[H]
  \centering
  \renewcommand{\arraystretch}{1.4}
  \caption{Coefficients for $n_1n_2n_3$ from $000$ to $444$ divided by 16}
  \resizebox{\textwidth}{!}{%
    \begin{tabular}{|c|c|c|c|c|c|c|c|c|c|c|c|c|}
      \hline
      $n_1n_2n_3$ & $LLLL$ & $LLLT$ & $LLTL$ & $LLTT$ & $LTLT$ & $LTTT$ & $TLLL$ & $TLLT$ & $TLTL$ & $TLTT$ & $TTLT$ & $TTTT$ \\
      \hline
      000 & 308 & 140  &  80 & -480  & 600 & -300  &   0 & -440 &  475 &   45  & -360 & 783 \\
      \hline
      002 & -1288 &  1512 & -2400 &  1920 & -1092 & -2460 & -1760 &  1320 & -2670 &  2790 & -1920 & -2442 \\
      \hline
      004 & 1652 & -1652 &  2480 & -2480 &  1652 &  2480 &  1760 & -1760 &  3195 & -3195 &  1760 &  3195  \\
      \hline
      020 & -1528 & -2760 &  2052 &  2460 & -4800 &  5700 &  1900 &  2820 & -1425 & -3015 &  5292 & -4629 \\
      \hline
      022 & -1872 & -1392 & -2472 &  5688 & -3288 & -1608 &  -120 &  3240 & -3510 &   270 &  1752 & -5394 \\
      \hline
      024 & 8 &    -8 &  5028 & -5028 &     8 &  5028 &  2220 & -2220 &  1935 & -1935 &  2220 &  1935 \\
      \hline
      040 & 2132 &  3020 & -2132 & -3020 &  6200 & -6200 & -1900 & -3420 &  1900 &  3420 & -5732 &  5732 \\
      \hline
      042 & -712 &  4488 &   712 & -4488 &   572 &  -572 & -2280 & -1440 &  2280 &  1440 & -4472 &  4472  \\
      \hline
      044 & 3348 & -3348 & -3348 &  3348 &  3348 & -3348 &   180 &  -180 &  -180 &   180 &   180 &  -180  \\
      \hline
      200 & -1288 &  1560 & -2400 &  1680 & -1188 & -1980 & -1760 &  1200 & -2670 &  2862 & -1680 & -2586 \\
      \hline
      202 & 13968 & -13632 &  21120 & -20400 &  14868 &  18180 &  15360 & -15120 &  27468 & -26796 &  14040 &  28512 \\
      \hline
      204 & -16072 &  16072 & -24160 &  24160 & -16072 & -24160 & -17760 &  17760 & -31598 &  31598 & -17760 & -31598  \\
      \hline
      220 & -1872 &  -2640 &  -2472 &   8040 &   -792 &  -6312 &   -120 &   4248 &  -3510 &   -426 &   -264 &  -4002 \\
      \hline
      222 & 672 &    192 &  46992 & -49968 & -16488 &  73872 &  17712 & -16464 &  19356 & -20412 &  29808 &   8424  \\
      \hline
      224 & -208 &     208 &  -56808 &   56808 &    -208 &  -56808 &  -19832 &   19832 &  -23286 &   23286 &  -19832 &  -23286 \\
      \hline
      240 & -712 &    6680 &     712 &   -6680 &   -3812 &    3812 &   -2280 &   -2408 &    2280 &    2408 &   -2536 &    2536  \\
      \hline
      242 & 30992 &  -33408 &  -30992 &   33408 &   57492 &  -57492 &    4048 &   -5376 &   -4048 &    5376 &   -9288 &    9288  \\
      \hline
      244 & -37768 &   37768 &   37768 &  -37768 &  -37768 &   37768 &   -5608 &    5608 &    5608 &   -5608 &   -5608 &    5608  \\
      \hline
      400 & 1652 &   -1540 &    2480 &   -2320 &    1428 &    2160 &    1760 &   -1800 &    3195 &   -3107 &    1840 &    3019 \\
      \hline
      402 & -16072 &   16728 &  -24160 &   22320 &  -17384 &  -20480 &  -17760 &   16920 &  -31598 &   32406 &  -16080 &  -33214  \\
      \hline
      404 & 19188 &  -19188 &   26960 &  -26960 &   19188 &   26960 &   20160 &  -20160 &   37403 &  -37403 &   20160 &   37403 \\
      \hline
      420 & 8 &    -40 &   5028 &  -6660 &     72 &   8292 &   2220 &  -1628 &   1935 &  -2919 &   1036 &   3903 \\
      \hline
      422 & -208 & -11088 & -56808 &  74232 &  22384 & -91656 & -19832 &  28104 & -23286 &  15822 & -36376 &  -8358 \\
      \hline
      424 & -7288 &   7288 &  77892 & -77892 &  -7288 &  77892 &  28652 & -28652 &  22191 & -22191 &  28652 &  22191  \\
      \hline
      440 & 3348 &  -4420 &  -3348 &   4420 &   5492 &  -5492 &    180 &  -1132 &   -180 &   1132 &   2084 &  -2084 \\
      \hline
      442 & -37768 &  55032 &  37768 & -55032 & -72296 &  72296 &  -5608 &  -3504 &   5608 &   3504 &  12616 & -12616  \\
      \hline
      444 & 55572 & -55572 & -55572 &  55572 &  55572 & -55572 &    468 &   -468 &   -468 &    468 &    468 &   -468  \\
      \hline
    \end{tabular}
}
  \label{tabAllCoeffs}
\end{table}




\end{document}